\documentclass[10.5pt,compsoc]{JSC}
\usepackage{chemformula}

\graphicspath{{figures/}}
\usepackage{graphicx}
\usepackage{footmisc}
\usepackage{subfigure}
\usepackage{url}
\usepackage{multirow}
\usepackage{multicol}
\usepackage[noadjust]{cite}
\usepackage{amsmath,amsthm}
\usepackage{amssymb,amsfonts}
\usepackage{booktabs}
\usepackage{color}
\usepackage{ccaption}
\usepackage{booktabs}
\usepackage{float}
\usepackage{fancyhdr}
\usepackage{caption}
\usepackage{xcolor,stfloats, colortbl}
\usepackage{comment}
\usepackage{cuted}  
\usepackage{captionhack}
\usepackage{epstopdf}
\usepackage{enumitem}
\usepackage[square,numbers]{natbib}

\headevenname{\zihao{-5}{\textbf{\emph{Journal of Social Computing, \quad \quad }}} 20??, ?(?): ???-???}%
\headoddname{{\sf  et al.:}\quad {\textbf{\emph{From Symbols to Embeddings: A Tale of Two Representations in Computational Social Science}}}}%

\newtheoremstyle{mystyle}{0pt}{0pt}{\normalfont}{1em}{\bf}{}{1em}{}
\theoremstyle{mystyle}
\renewcommand\figurename{Fig.~}

\newcommand{\nop}[1]{}

\makeatletter
\renewcommand{\@biblabel}[1]{[#1]\hfill}
\makeatother

\begin{document}

\thispagestyle{empty}

\clearpage

\hyphenpenalty=50000

\makeatletter
\newcommand\mysmall{\@setfontsize\mysmall{7}{9.5}}

\newenvironment{tablehere}
  {\def\@captype{table}}
  {}
\newenvironment{figurehere}
  {\def\@captype{figure}}
  {}

\thispagestyle{plain}%
\thispagestyle{empty}%

\let\temp\footnote
\renewcommand \footnote[1]{\temp{\zihao{-5}#1}}
{}
\vspace*{-40pt}
\noindent{\zihao{5-}\textbf{\scalebox{0.95}[1.0]{\makebox[5.93cm][s]
{JOURNAL\hfil OF\hfil SOCIAL\hfil COMPUTING}}}}

\vskip .2mm
{\zihao{5-}
\textbf{
\hspace{-5mm}
\scalebox{1}[1.0]{\makebox[5.6cm][s]{%
I\hspace{0.70pt}S\hspace{0.70pt}S\hspace{0.70pt}N\hspace{0.70pt}{\color{black}%
\hspace{2pt}\hspace{5.70pt}2\hspace{0.70pt}6\hspace{0.70pt}8\hspace{0.70pt}8\hspace{0.70pt}-\hspace{0.70pt%
}5\hspace{0.70pt}2\hspace{0.70pt}5\hspace{0.00pt}5\hspace{0.70pt}\hspace{0.70pt%
}\hspace{0.70pt}{\color{white}l\hspace{0.70pt}l\hspace{0.70pt}}0\hspace{0.70pt}%
?\hspace{0.70pt}/\hspace{0.70pt}?\hspace{0.70pt}?\hspace{0.70pt}{\color{white}%
l\hspace{0.70pt}l\hspace{0.70pt}}p\hspace{0.70pt}p\hspace{0.70pt}?\hspace{0.70pt}?\hspace{0.70pt}?%
--\hspace{ 0.70pt}?\hspace{0.70pt}?\hspace{0.70pt}?}}}}

\vskip .2mm\noindent
{\zihao{5-}\textbf{\scalebox{1}[1.0]{\makebox[5.6cm][s]{%
V\hspace{0.4pt}o\hspace{0.4pt}l\hspace{0.4pt}u\hspace{0.4pt}m\hspace{0.4pt}%
e\hspace{0.4em}?\hspace{0.4pt},\hspace{0.8em}N\hspace{0.4pt}u\hspace{0.4pt}%
m\hspace{0.4pt}b\hspace{0.4pt}e\hspace{0.4pt}r\hspace{0.4em}?,\hspace{0.8em}%
\color{white}{J\hspace{0.4pt}a\hspace{0.4pt}n\hspace{0.4pt}u\hspace{0.4pt}a\hspace{0.4pt}%
\hspace{0.4pt}r\hspace{0.4pt}y\hspace{0.4em}2\hspace{0.4pt}0\hspace{0.4pt}1\hspace{0.4pt}8}}}}}}

\vskip .2mm
{\zihao{5-}
\textbf{
\hspace{-5mm}
\scalebox{1}[1.0]{\makebox[5.6cm][s]{%
DOI:~\color{white}{\hfill1\hfill0\hfill.\hfill2\hfill6\hfill5\hfill9\hfill9\hfill/\hfill B\hfill D\hfill M\hfill A\hfill .\hfill2\hfill0\hfill1\hfill8\hfill.\hfill9\hfill0\hfill2\hfill0\hfill0\hfill0\hfill0}}}}}\\

\begin{strip}
{\center
{\zihao{3}\textbf{From Symbols to Embeddings: A Tale of Two Representations in Computational Social Science}}
\vskip 9mm}

{\center {\sf \zihao{5}
Huimin Chen$^*$, Cheng Yang$^*$, Xuanming Zhang, Zhiyuan Liu$^\dagger$, Maosong Sun, Jianbin Jin$^\dagger$}
\vskip 5mm}


\centering{
\begin{tabular}{p{160mm}}

{\zihao{-5}
\linespread{1.6667} %
\noindent
\bf{Abstract:} {\sf
Computational Social Science (CSS), aiming at utilizing computational methods to address social science problems, is a recent emerging and fast-developing field. The study of CSS is data-driven and significantly benefits from the availability of online user-generated contents and social networks, which contain rich text and network data for investigation. However, these large-scale and multi-modal data also present researchers with a great challenge: how to represent data effectively to mine the meanings we want in CSS? To explore the answer, we give a thorough review of data representations in CSS for both  text and network. Specifically, we summarize existing representations into two schemes, namely symbol-based and embedding-based representations, and introduce a series of typical methods for each scheme. Afterwards, we present the applications of the above representations based on the investigation of more than $400$ research articles from $6$ top venues involved with CSS. From the statistics of these applications, we unearth the strength of each kind of representations and discover the tendency that embedding-based representations are emerging and obtaining increasing attention over the last decade. Finally, we discuss several key challenges and open issues for future directions. This survey aims to provide a deeper understanding and more advisable applications of data representations for CSS researchers.}

}

\end{tabular}
}
\vskip 6mm

\vskip -3mm
\zihao{6}\end{strip}

\vskip -3mm
\begin{figure}[b]
\vskip -6mm
\begin{tabular}{p{44mm}}
\toprule\\
\end{tabular}
\vskip -4.5mm
\noindent
\setlength{\tabcolsep}{1pt}
\begin{tabular}{p{1.5mm}p{79.5mm}}
&
 \\
$\bullet$&  Huimin Chen and Jianbin Jin are with the School of Journalism and Communication, Tsinghua University, Beijing 100084, China. E-mail: {huimchen1994}@gmail.com, jinjb@tsinghua.edu.cn \\
$\bullet$& Cheng Yang is with the School of Computer Science, Beijing University of Posts and Telecommunications, Beijing 100876, China. E-mail: {albertyang33}@gmail.com \\
$\bullet$& Xuanming Zhang, Zhiyuan Liu, and Maosong Sun are with the Department of Computer Science and Technology, Tsinghua University, Beijing 100084, China. E-mail: {billyzhang07}@outlook.com, \{liuzy, sms\}@mail.tsinghua.edu.cn \\
$\bullet$& $*$ indicates equal contribution.\\
$\bullet$& $\dagger$ indicates corresponding author.\\

\end{tabular}
\end{figure}\zihao{5}
\vskip -3mm

\thispagestyle{plain}%
\thispagestyle{empty}%
\makeatother
\pagestyle{tstheadings}

\section{Introduction}
\label{s:introduction}



\noindent
Computational Social Science (CSS) refers to the fields that utilizes computational approaches to model, simulate, and analyze social phenomena. CSS has received widespread attention and undergone rapid development over the past decade~\cite{lazer2020computational,zhang2020data}. It now includes numerous sub-fields such as computational sociology, computational politics, and computational communication.

\begin{figure*}
    \noindent
    \centering
    \includegraphics[width=\textwidth]{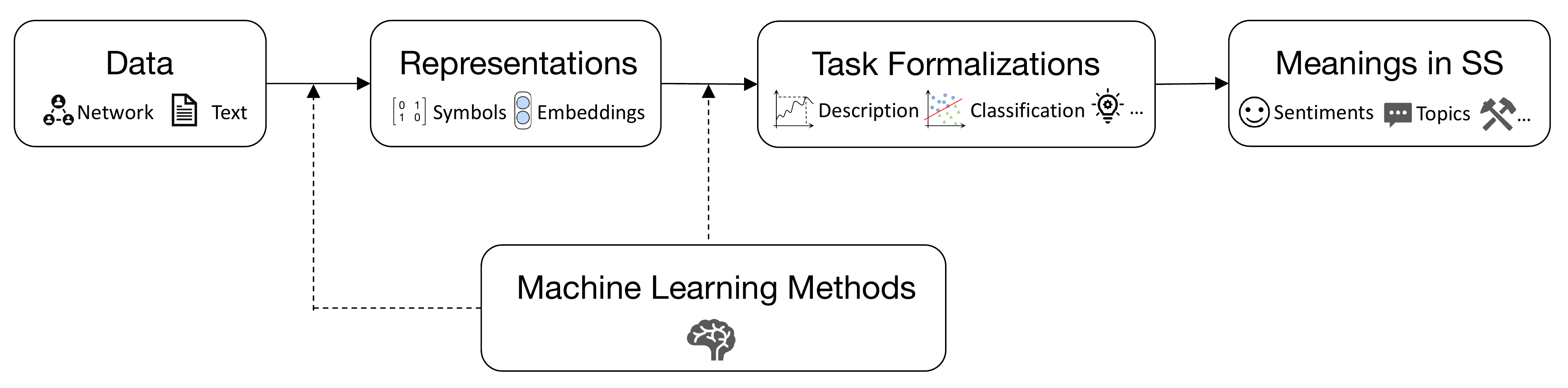}
    \caption{The operational framework in CSS, where SS denotes social science. }
    \noindent
    \label{fig:framework}
\end{figure*}


CSS is a data-driven field that was born due to the accessibility and analyzability of massive amounts of data~\cite{lazer2009social}. With the fast development of Internet technology and mobile devices, large-scale multi-modal data have been produced and digitally recorded, such as friendship and posts on online social networks, purchase behaviour on e-commerce websites, and movement trajectories recorded by mobile devices. These data provide us with an opportunity to mine meanings in social science directly and comprehensively from data, which include discovering the social phenomenon such as news framing and public opinion, explaining the phenomena, and finding the causal relations, etc. 

In general, we can summarize the operational framework of CSS: from data to meanings, as shown in Fig.~\ref{fig:framework}. Note that the operational process is different from the general research flow which can be problem-driven, followed by the selection of the required data, and then the identification of the task and the corresponding data representation. The operational framework we introduce here focuses on the implementation process of the study. Specifically, supposing we are conducting research in CSS, we first need to collect enough relevant data, which could be text or networks for our study. Afterwards, we need to transform the data into computationally processable representations, which are discrete or continuous numerals. Further, the representations of data are employed in practical applications, namely social issues we study. For each application, we formalize it into one of task prototypes, which commonly include data description, uncovering relationships between objects, clustering, classification, etc. Finally, the desired meaning in social science can be extracted based on the observation and analysis of the task results. Notably, the process from data to representations or representations to task formalizations usually requires the involvement of machine learning methods.  


In the framework, we can find that the module of representations is not only the foundation, but also the key component since the increasing scale of data in CSS requires more efficient and effective representations. According to statistics~\cite{nadhom2018survey}, there are now nearly 5 billion Internet users worldwide, who post hundreds of millions of tweets, view thousands of millions of videos on YouTube, and make billions of searches on the Google search engine every day. These massive data present us with a great challenge: how can we, the researchers in CSS, represent data effectively from such a large amount of multi-modal data? 

Recently, the rapid development of data representation in computer science has nourished a large amount of successes both in academia and in industry~\cite{bengio2013representation}. Therefore, in this paper, we provide a systematic introduction for data representations that are divided into two schemes: symbol-based and embedding-based representations, as well as their existing applications in CSS to explore the effective and desirable data representations for different types of applications. We focus on the introduction of two most commonly used data, namely text and network, since they not only contain rich meanings but also are harder to represent owing to the diverse expressions of text and complex structures of network.


To summarize, we make the following contributions in this survey:
\begin{itemize}
    \item We provide a thorough review of data representations in two schemes: symbol-based and embedding-based representations, both for text and network. Researchers major in CSS can obtain a deep perception of these representations and distinguish them from each other clearly. 
    
    \item We conduct a comprehensive survey on the applications utilizing the above representations, through investigating more than $400$ top-cited articles from $6$ representative publications over ten years. Base on the survey, we summarize the tasks in which each of the two representations excels, which can prompt the awareness of their expert areas and make advisable choices between them.
    
    \item We discover the trend that embedding-based representations are gaining growing attention, based on the statistics of their applications. This finding can encourage the usage of embedding-based representations in more relevant works and shed light on the future directions of CSS.
\end{itemize}

The rest of this survey is organized as follows. In Section~\ref{sec:a_tale}, we present in general terms the definitions of symbol-based and embedding-based representations and the differences between them. Afterwards, we meticulously introduce typical methods for constructing each kind of representations from text to network, in Section~\ref{sec:symbol_text} to Section~\ref{sec:emb_network}. In Section~\ref{sec:app}, we revisit the applications that use these representations and categorize them into different task prototypes in $6$ top venues over past ten years. Base on the well-organized applications, we examine the coverage of the two representations and present their skilled areas in Section~\ref{sec:etos}. In Section~\ref{sec:future}, we propose four open problems as well as future directions. Finally, we conclude the survey in Section~\ref{sec:conclusion}.
\section{A Tale of Two Representations}
\label{sec:a_tale}
\noindent
The representation indicated in this paper is behaved as computer-processable numerals, transformed from data in the real world. Each object (e.g., a word or a network node) in the real world can be assigned with a unique representation storing its characteristics. With the representation, we can conduct efficient analyses of large-scale data. It is the basis for data-driven CSS, since choosing an appropriate and exquisite representation will facilitate the subsequent analysis with fewer efforts. 

Traditional representations are based on symbols. Following the definition from Wikipedia\footnote{https://en.wikipedia.org/wiki/Symbol}, a symbol is ``a mark, sign, or word that indicates, signifies, or is understood as representing an idea, object, or relationship.'' Hence, in this article, we identify symbol-based representations as discrete or continuous numerals which characterize objects in real-world explicitly and recognizably, such as language and relationship. It generally relies on the manual definition from data, which greatly contributes to the interpretability of CSS. For example, the representation of a word can be defined as its frequency in the corpus or sentiment value, while the representation of a node in the network can be designed as its degree or centrality. 

Though symbol-based representation is explicit and human-readable, it suffers from several critical issues.~\footnote{Detailed issues of symbol-based representation are presented in Section~\ref{sec:etos}.} The most immediate shortcoming lies in heavy human efforts, since symbol-based representation is composed of manually defined features. To achieve a better performance, features should be elaborately designed. Besides, due to simple statistics and shallow combination of features, symbol-based representation usually fails to capture abstract semantics at a high level~\cite{bengio2013representation}. For example, humans can identify the similar semantic relation between ``king'' - ``queen'' and ``man'' - ``woman'', while it is hard to discover for symbol-based representation.

To overcome these issues, the embedding-based representation is proposed to encode an object into a low-dimensional continuous vector, with the rapid development of artificial intelligence and deep learning methods. The vector is learned automatically by optimization of a training objective instead of hand-crafted features. It is randomly initialized and updated during the training process just like climbing a mountain step by step. Once the training is finished, we can use the learned embeddings as object representations for downstream tasks. Learning representations in such an automatic way is very convenient without human efforts. Moreover, it usually behaves as a complex combination of shallow features, which can detect the high-level attributes from data such as the semantic relation mentioned above. But a shortcoming is that the interpretability of learned embeddings is poor, which means we usually have no idea about the exact meaning of embedding-based representations in each dimension.

In the following sections, we will introduce these two representations in detail, and further divide each kind of representation into text and network, namely symbol-based representations of text and network and embedding-based representations of text and network.

\section{Symbol-based Text Representation}
\label{sec:symbol_text}
\begin{figure*}
    \noindent
    \centering
    \includegraphics[width=\textwidth]{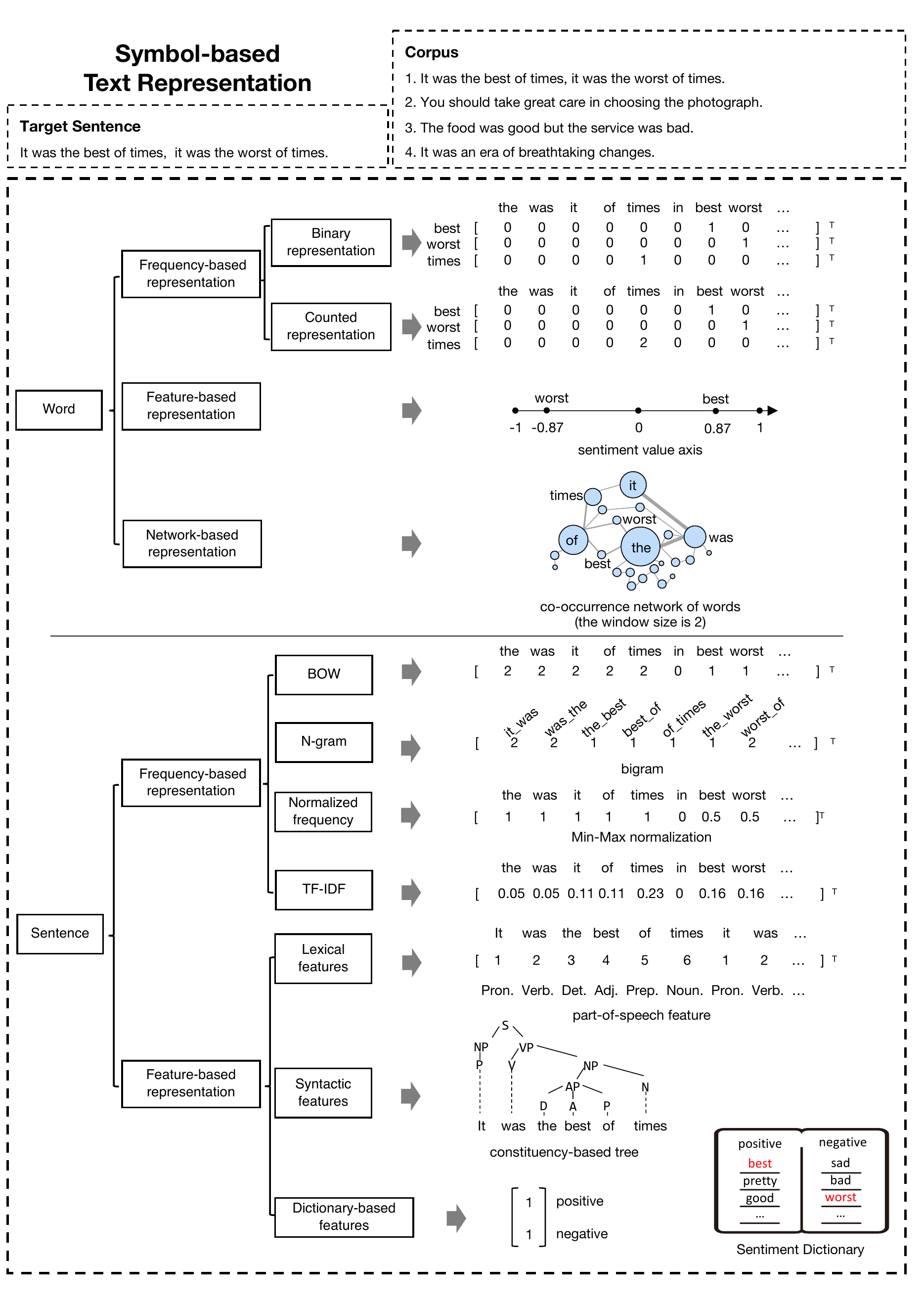}
    \caption{An illustration of symbol-based text representation.}
    \label{fig:reps_symbol_text}
    \noindent
\end{figure*}

\noindent
Text is the earliest and most common type of data we used. In linguistics, a word is the smallest unit of text that can be uttered in isolation with objective or practical meaning. Phrases, sentences, and documents are all compositions of words. Therefore, in this section, we will first introduce the word representation which is the basis of representing texts. Afterwards, we will delineate the sentence representation based on symbols. Note that the representation of a document is similar to a sentence, since it can be viewed as a longer sentence or multiple sentences composed together. An illustration of symbol-based text representation is shown in Fig.~\ref{fig:reps_symbol_text}.

\subsection{Symbol-based Word Representation}
\noindent

Existing symbol-based word representations can be divided into three brands, namely frequency-based, feature-based, and network-based representations. In the following, we will introduce each of them in detail.

\subsubsection{Frequency-based Representation}
Frequency is a basic statistic feature of words, reflecting the significance of words in the corpus. Frequency-based word representation transfers each word into a value or a vector based on its occurrence in the corpus. Specifically, it can be categorized into two settings:

 \textbf{Binary representation.} Each word is denoted as $0/1$ depending on whether it appears or not. Taking the word ``times'' in the target sentence in Fig.~\ref{fig:reps_symbol_text} as an example, it is represented with value $1$ as it appears in the corpus, while ``time'' is represented with value $0$ due to absence. Further, each word can also be indicated with a vector with its dimension size equal to the vocabulary size, \textit{i.e.,} the number of all words in the corpus. Each word is assigned with a unique index at first, then its vector behaves as that all elements are zeros except the only dimension of its index is one. As shown in Fig.\ref{fig:reps_symbol_text}, ``times'' is represented as a vector $[0,0,0,0,1,0,\dots]$, only the dimension indicating itself is $1$. Hence, it is also known as \textit{one-hot representation}, with its dimension probably tremendous if large vocabulary size. 
 
 \textbf{Counted representation.} Distinguished from binary representation, each word is expressed based on its number occurring in the corpus. For example, we can denote ``times" as its count: $2$, or a vector with the value in the dimension of its index to be the count: $[0,0,0,0,2,\dots]$. These two types of representations are corresponding to the value and vector in the binary representation, respectively. The difference is that they introduce information of the word's occurring number in this counted representation fashion. 

\subsubsection{Feature-based Representation}
Apart from the frequency-based approach, feature-based representation signifies each word with manual features defined depending on the research goal. For example, a word can be represented as a vector composed of its occurrences with designated words when measuring its semantics in some specific aspects. It also can be denoted as a human-defined sentiment value when considering its sentiment feature, as shown in Fig.\ref{fig:reps_symbol_text}. The ``worst'' is assigned to a sentiment value close to $-1$, while the ``best'' is arranged to be nearly $1$. 

\subsubsection{Network-based Representation}
Substituting for representing each word directly as a value or a vector, network-based representation maps each word into a node in the network, where each edge between two nodes is established based on defined relations, such as occurrences or semantic relations. In the light of the constructed network, we can represent each word with its degree, centrality, closeness, and neighboring nodes, etc. As shown in Fig.\ref{fig:reps_symbol_text}, each word in our example corpus is projected into a node in the word co-occurrence network. This representation manner allows for better modeling of the characteristics of words and the complex relationships between words through utilizing a range of network analysis algorithms.


\subsection{Symbol-based Sentence Representation}
Symbol-based sentence representation is usually built upon word representation and can be separated into two groups: frequency-based representation and feature-based representation. In this section, we will describe them in detail.

\subsubsection{Frequency-based Representation}
Frequency-based representation of sentences is constructed upon raw frequencies of words and phrases, as well as processed frequencies. As for the way based on raw frequencies, Bag-of-Words (BOW)~\cite{harris1954distributional} and N-gram representations~\cite{shannon1948mathematical} are widely utilized, while representation of normalized frequency and Term Frequency–Inverse Document Frequency (TF-IDF)~\cite{salton1983extended} are commonly used with regard to the way based on processed frequencies. Below we will present each of them, using the target sentence in Fig.\ref{fig:reps_symbol_text} as an illustrative example.

\textbf{BOW.} A sentence is represented as the bag of its words, where word order and grammar are disregarded and only word frequency is kept. For example, the bag of target sentence in Fig.\ref{fig:reps_symbol_text} contains $2$ "the", $2$ "was", and some other words with different numbers, which compose the representation vector of target sentence. We can see that it is a simple representation which is the sum of the one-hot representation of each word in the sentence. 

\textbf{N-gram.} Note that the word order information is disregarded in the BOW representation, resulting in that the two sentences ``Good, not bad'' and ``Bad, not good'' will have the same representation though they have completely different semantic meanings. Therefore, n-gram (\textit{i.e.,} $n$ consecutive words in a given sentence) count instead of word count is proposed. ``not bad'' and ``not good'' are two distinct 2-grams (or bigrams), so that the semantics of above sentences can be distinguished through 2-gram representations. 

\textbf{Normalized frequency.} Replacing raw frequency, each sentence is performed as a normalized version of its raw frequency. Two of the most prevailing normalization methods are Min-Max and Z-score normalization, with the first mapping the value of raw frequency into the range of $[0,1]$ and the latter transferring data into a standard normal distribution. Through this manner, representations of all sentences can be transformed into the same order of magnitude, enabling the comparison between sentences in different magnitudes, such as measuring semantics similarity between sentences with quite different lengths. It can benefit the efficient execution of downstream tasks as well.
 
\textbf{TF-IDF.} Frequencies of word and n-gram are the only considered features in the above representations. However, we can see that words with the most frequencies are not always the most important. For instance, ``a'', ``an'' and ``the'' are all frequent words but usually without meaning. Therefore, TF-IDF representation is proposed to further consider the document frequency, which is inspired by that a term's importance will decrease with the number of documents where it appears. Specifically, each value in BOW or N-gram representation is replaced with:
\begin{equation}
\text{tf-idf}(w,d) = \text{tf}(w,d) \times \text{idf}(w,D),
\end{equation}
where $\text{tf}(\cdot)$ denotes frequency of the term $w$ in document $d$ and $\text{idf}(\cdot)$ is inverse document frequency of the term in corpus $D$. It keeps a balance between the term frequency in a sentence and the document frequency of the term in a corpus. Through the processing of raw frequency, TF-IDF representation can re-weight words and catch the important ones of a sentence or a document.

\subsubsection{Feature-based Representation}
Feature-based representation is the most commonly utilized symbol-based sentence representation. It relies on artificially defined features, which can be divided into three main categories: lexical features, syntactic features, and dictionary-based features. The first two are based on features extracted from the text itself, and the last one depends on external dictionaries to obtain features. We then describe them in detail:

\textbf{Lexical features.} Specific words are distilled from the text as features, such as adjectives, adverbs, emoticons, and hashtags, which are informative lexicon features for downstream tasks. For example, adjectives and emoticons are central features for psychological studies, and verbs and nouns are particularly important when we intend to unearth topics from text. 

\textbf{Syntactic features.} Each sentence is equipped with a specific syntactic structure, which also plays a crucial role in the semantics of the sentence. For example, as for the sentence ``freedom is dearer than life'', its syntactic structure can inform that ``freedom'' is the nominal subject of ``dearer'' rather than ``life'', providing the key information of semantics. Hence, syntactic features of the sentence are extracted popularly, using syntactic analysis (i.e. parsing) or manual rules such as polarity shifts due to connectors and negations. Syntactic analysis is generally divided into constituency parsing and dependency parsing, with the former concentrating on breaking sentences into sub-components such as sub-phrases and the latter focusing on word connections based on their grammatical relations. Constituency parsing of part of the target sentence is shown in Fig.\ref{fig:reps_symbol_text}.

\textbf{Dictionary-based features.} Different from the above two kinds of features, dictionary-based features are recognized in the light of human-constructed dictionaries, such as Linguistic Inquiry and Word Count (LIWC)~\cite{pennebaker2001linguistic} and Language Assessment by Mechanical Turk (labMT)~\cite{dodds2015human}. Among these dictionaries, LIWC is most widely adopted, where each word falls into several pre-defined dimensions such as linguistic (e.g., person pronouns and conjunctions), psychological (e.g., anger and anxiety), cognitive dimension (e.g., insight and causation). It is worth mentioning that the difference from the lexical features lies in that dictionary-based features assimilate knowledge and wisdom summarized and accumulated in previous studies. Supposing there are two dimensions of words in a dictionary, i.e. positive and negative words, the target sentence can be represented in a two-dimensional vector $[1,1]$, with the first dimension indicating one positive word ``best'' and the second denoting one negative word ``worst'' occurring in the sentence, as shown in Fig.\ref{fig:reps_symbol_text}.

\begin{figure*}
    \noindent
    \centering
    \includegraphics[width=\textwidth]{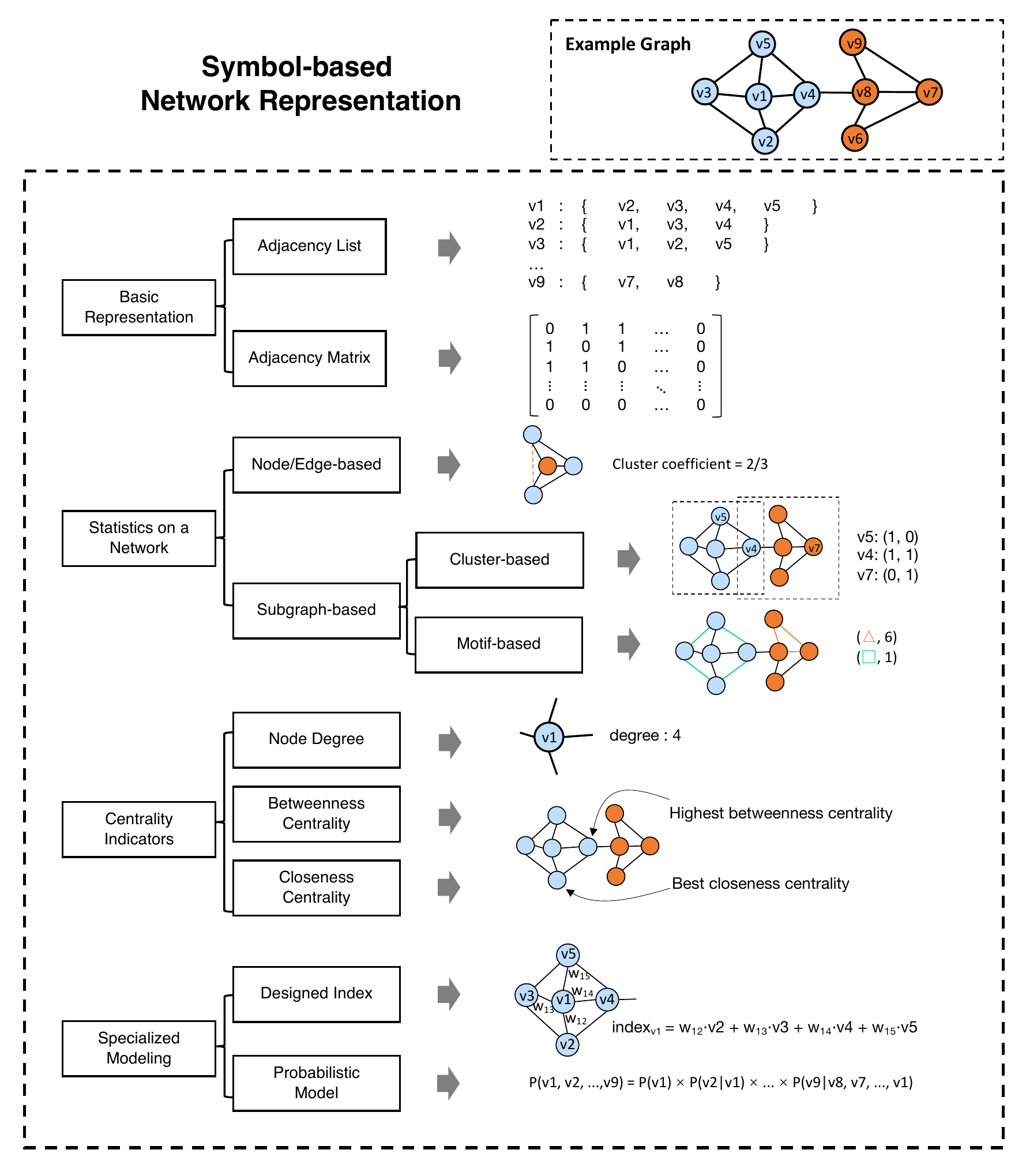}
    \caption{An illustration of symbol-based network representation.}
    \label{fig:reps_symbol_network}
    \noindent
\end{figure*}

\section{Symbol-based Network Representation}
\label{sec:symbol_network}
\noindent
A network (or graph) contains a set of objects and their relationships. An object is usually represented by a node (or vertex) and the relationship between two objects is represented by an edge between corresponding nodes. An edge can be directed to indicate an asymmetric relationship, weighted to emphasize the strength of a relationship, signed to represent a relationship is positive or negative, and etc. Most work will use adjacency list or adjacency matrix as the basic representations of a network. Then they will employ statistics or specialized modeling to build high-level representations.

\begin{figure}[!h]
    \noindent
    \centering
    \includegraphics[width=0.4\columnwidth]{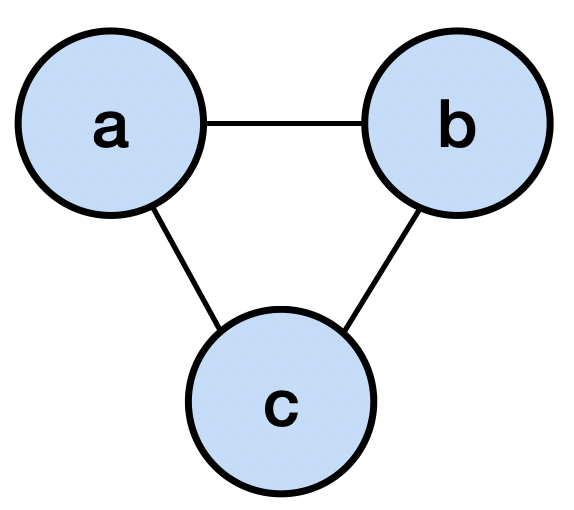}
    \caption{An example network for illustration.}
    \label{fig:net}
    \noindent
    \setlength{\abovecaptionskip}{0.cm}
    \setlength{\belowcaptionskip}{-0.cm}
\end{figure}

\subsection{Basic Representations}
Now we will start by presenting two basic representations of networks.

\subsection{Adjacency List}
Adjacency list is a collection of unordered lists where each list describes the set of neighbors of a node in the network. Taking the triangle structure in Fig.~\ref{fig:net} as an example, the corresponding adjacency list contains three lists: $a: \{b,c\}$, $b: \{a,c\}$ and $c: \{a,b\}$. The adjacency list representation can record all edges in a space-efficient manner and be suitable to describe an (un)directed graph structure.

\subsection{Adjacency Matrix}
Adjacency matrix is a square matrix whose dimension equals to the number of vertices. Each element of the adjacency matrix indicates a directed edge between corresponding nodes. The adjacency matrix representation of Fig.~\ref{fig:net} is

$$ \begin{pmatrix}

       0 & 1 & 1\\

       1 & 0 & 1\\
       
       1 & 1 & 0

  \end{pmatrix} .
$$  

The adjacency matrix representation can be used to describe (un)directed/weighted/signed graph structures by changing the ones to real-valued weights or signs. We can efficiently check whether two specific nodes are connected using the adjacency matrix representation. However, real-world networks are usually sparse which means most elements in an adjacency matrix are zeros. The storage usage of an adjacency matrix is proportional to the square of the number of vertices, which is not space-efficient compared with the adjacency list representation. 

\subsection{Statistics on a Network.}
The aforementioned adjacency list and matrix can faithfully record the structure of a network. However, in many scenarios, we need to extract features from a network, e.g., by statistics. We classify the statistics on a network into node/edge-based statistics and subgraph-based statistics. 

\subsubsection{Node/Edge-based Statistics}
Note that node/edge-based statistics are not necessarily used to represent a node or an edge. For example, node degree can be used to represent a node, while average degree characterizes the entire network. To characterize and represent a network (or subgraph), we can calculate the size of a network (the number of nodes and edges), average degree, edge density (the ratio of the number of edges to the number of possible edges), etc. In fact, such statistics are widely used to describe the datasets.

\begin{figure}
\centering
\begin{subfigure}{}
  \centering
  \includegraphics[width=.4\linewidth]{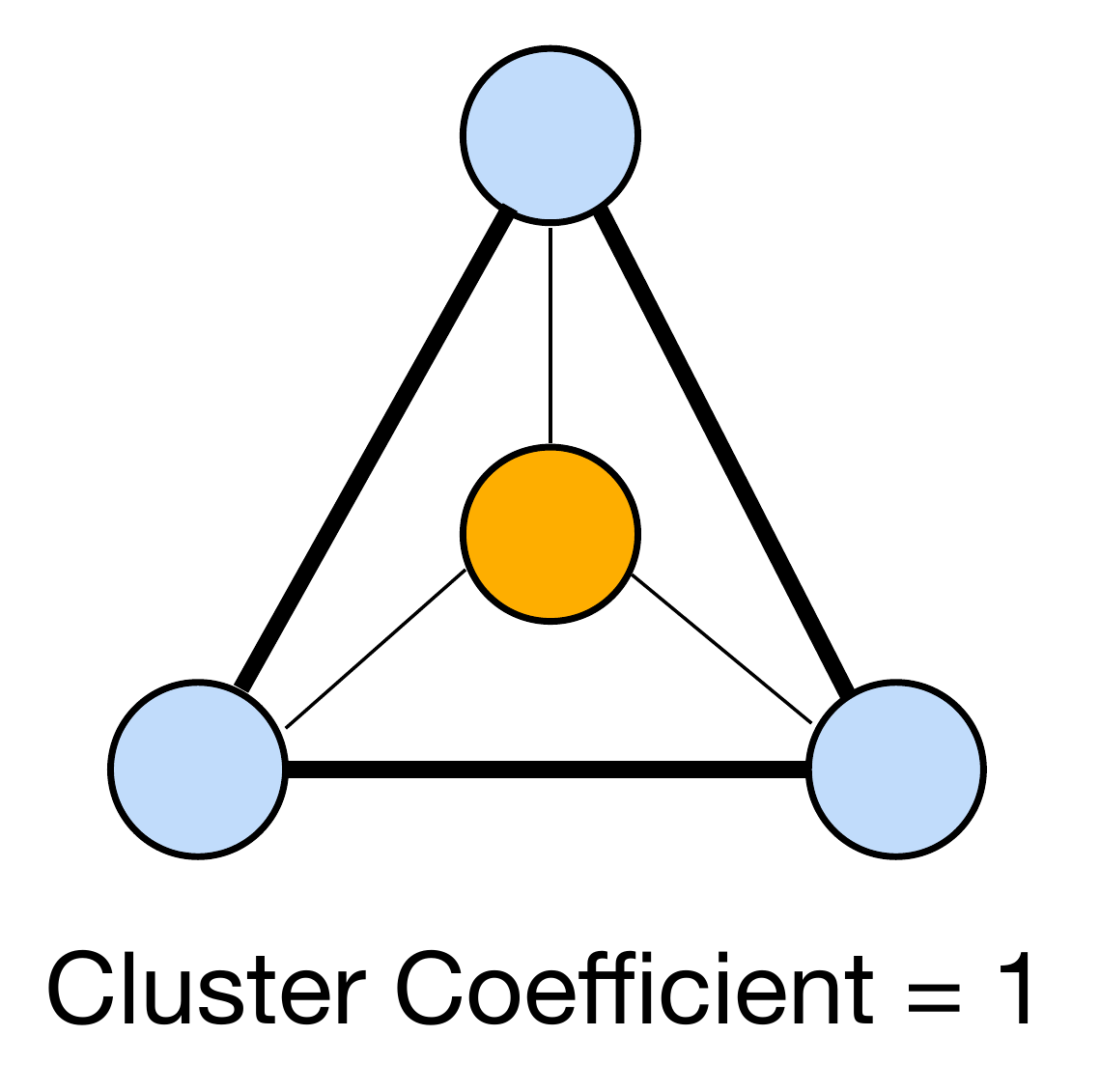}
  \label{fig:cc_1}
\end{subfigure}%
\begin{subfigure}{}
  \centering
  \includegraphics[width=.4\linewidth]{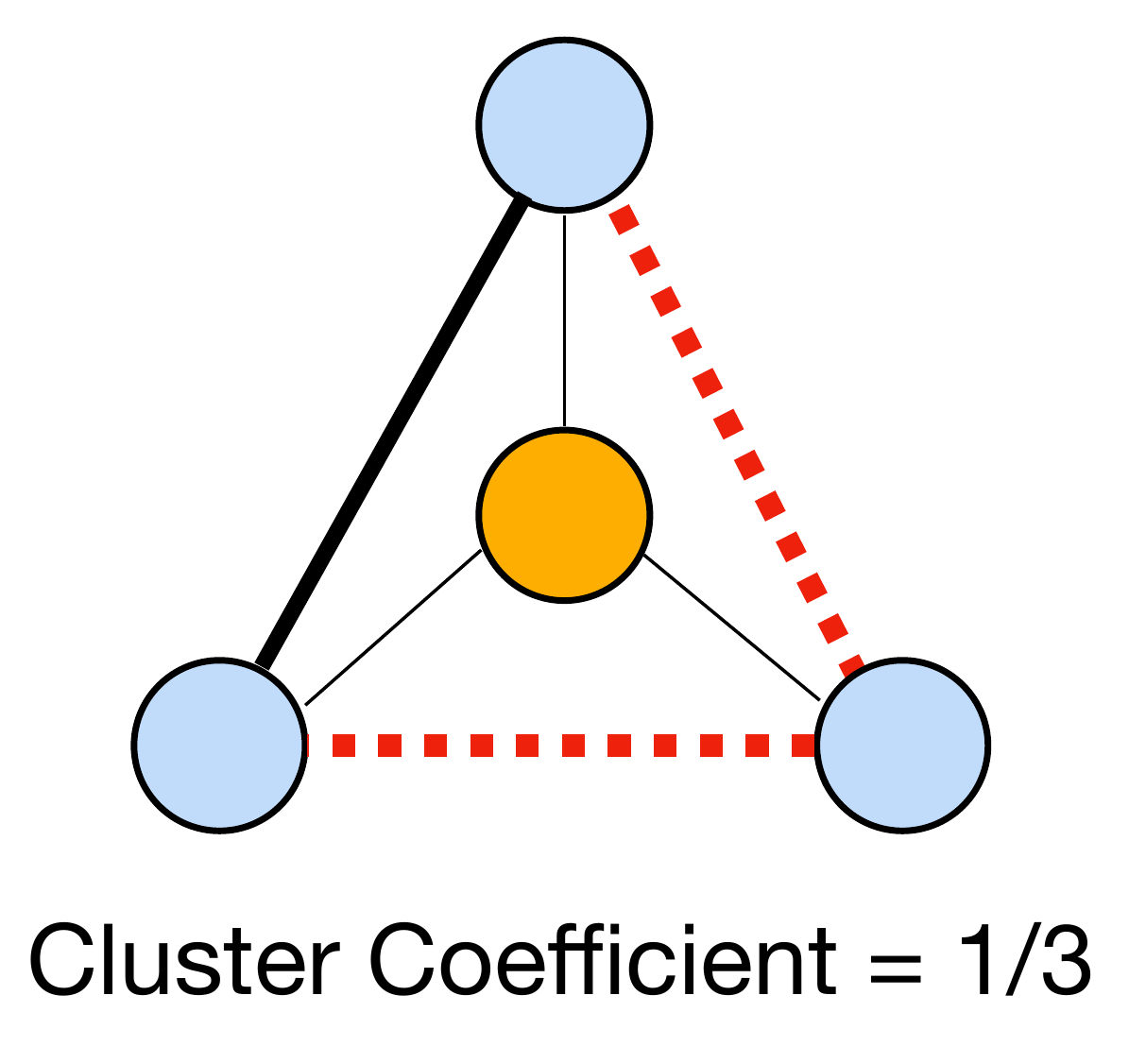}
  \label{fig:cc_13}
\end{subfigure}
\caption{An example of cluster coefficient.}
\label{fig:cc}
\end{figure}

In general, employing statistics to represent nodes is more common and useful in the studies of CSS because they usually need to model the behaviours or properties of individuals in a large (social) network. On one hand, the simplest statistics directly come from a node's behaviours or features, e.g. the number of a Facebook user (node)'s posts. On the other hand, the statistic-based representation can also come from a node's neighborhood structure. We will take local cluster coefficient as an illustrative example: As shown in Fig.~\ref{fig:cc}, the local clustering coefficient of a node identifies the local density and is defined by the proportion of the number of links between its neighbors divided by the number of links that could possibly exist between them. In addition, the statistics are also possible to be a mixture of node behaviours and network structure, e.g. the number of likes obtained from one's friends in an online social network.


\subsubsection{Subgraph-based Statistics}
Subgraph-based statistics can be further categorized into cluster-based and motif-based.

A cluster in a network contains a group of nodes with dense connections or similar characteristics. Clusters in a network can be either overlapped or disjoint. A cluster is also referred to as a community in many scenarios. The cluster assignment of a node can be used as its cluster-based representation as shown in Fig.~\ref{fig:reps_symbol_network}. Besides, cluster-based indices can be used to characterize the whole network as well. For example, modularity measures the strength that a network is divided into clusters by the fraction of the edges within the clusters minus the expected fraction if edges were randomly distributed. A larger modularity indicates dense connections within clusters and sparse connections between different clusters. 

On the other hand, motifs which are defined as recurrent and statistically significant subgraphs or patterns, are much smaller than communities, e.g., a triangle made up of 3 nodes or a square made up of 4 nodes. The frequencies of motifs are widely used as motif-based statistics. As shown in Fig.~\ref{fig:reps_symbol_network}, we can count the numbers of appearances of triangles and squares to represent the entire network. In addition, the global cluster coefficient, which is calculated as the proportion of the number of closed triplets (i.e., triangles) divided by the number of all triplets (either closed or not), can give an indication of the clustering in the whole network.

\begin{figure}[htb]
    \noindent
    \centering
    \includegraphics[width=0.9\columnwidth]{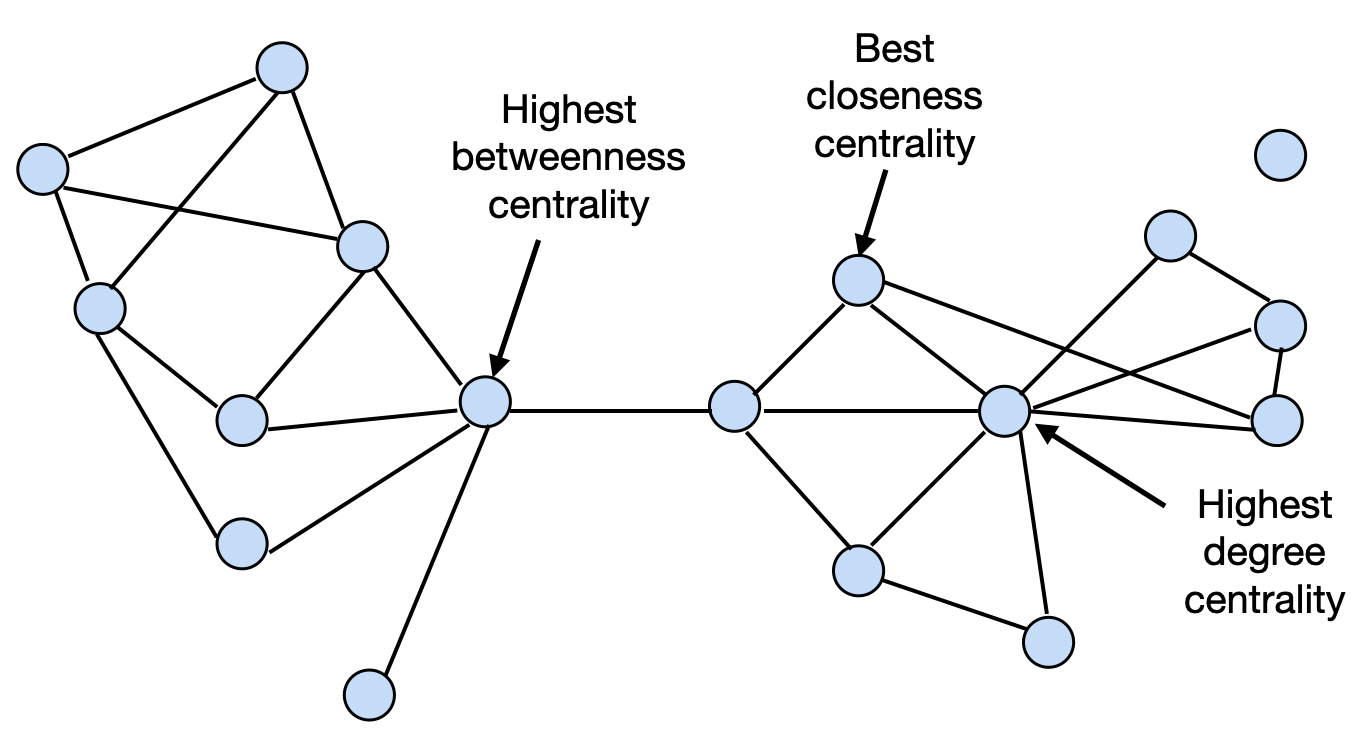}
    \caption{An example of centrality indicators.}
    \label{fig:net-centrality}
    \noindent
\end{figure}

\subsubsection{Centrality Indicators}
To characterize the properties of nodes in a network, there exist various of indicators ranging from simple statistics to designed indices. Among all such pre-defined or manually designed indicators, centrality indicators, which measure the importance of each node in a network, are the most widely used ones and thus we put them into a separate subsection. 

\textit{Node degree}, i.e., the number of edges connected to a node, is the simplest centrality indicator. Intuitively, a node with a larger degree will have a larger impact on the network. Besides, \textit{closeness centrality} of a node measures the average length of the shortest path between the node and all other nodes in the network. Hence, a node with smaller closeness centrality will be closer to all other nodes and thus be more central. \textit{Betweenness centrality} counts the number of times a node acts as a bridge along the shortest path between two other nodes. A node with larger betweenness centrality will probably control the information flow or communications in the network. Fig.~\ref{fig:net-centrality} shows the nodes with best degree/closeness/betweenness centrality. There are also many other centrality indicators such as eigenvector and PageRank, and readers are encouraged to learn more about them if interested\footnote{https://en.wikipedia.org/wiki/Centrality}.

\subsubsection{Specialized Modeling}
Real-world interaction systems are quite sophisticated and thus motivate many case-by-case representations of networks. Depending on how complicated a network representation is, we roughly divide them into designed index and probabilistic model.

Designed indices are usually a heuristic combination of multiple simple factors. For example, if we want to quantify how good a person works in a collaboration network as node representations, we can compute the weighted sum of his/her scores of error rate, decision time and peer evaluation. In detail, the score of decision time could be an exponentially time-decayed function. In contrast, probabilistic models are much more complicated. Besides the probabilistic modeling among a number of variables, differential equations are also widely used to characterize the dynamics in a network. In all, both designed index and probabilistic model are usually more complicated than previously mentioned simple statistics and highly specialized for a given problem. 




\begin{figure*}[htb]
    \noindent
    \centering
    \includegraphics[width=\textwidth]{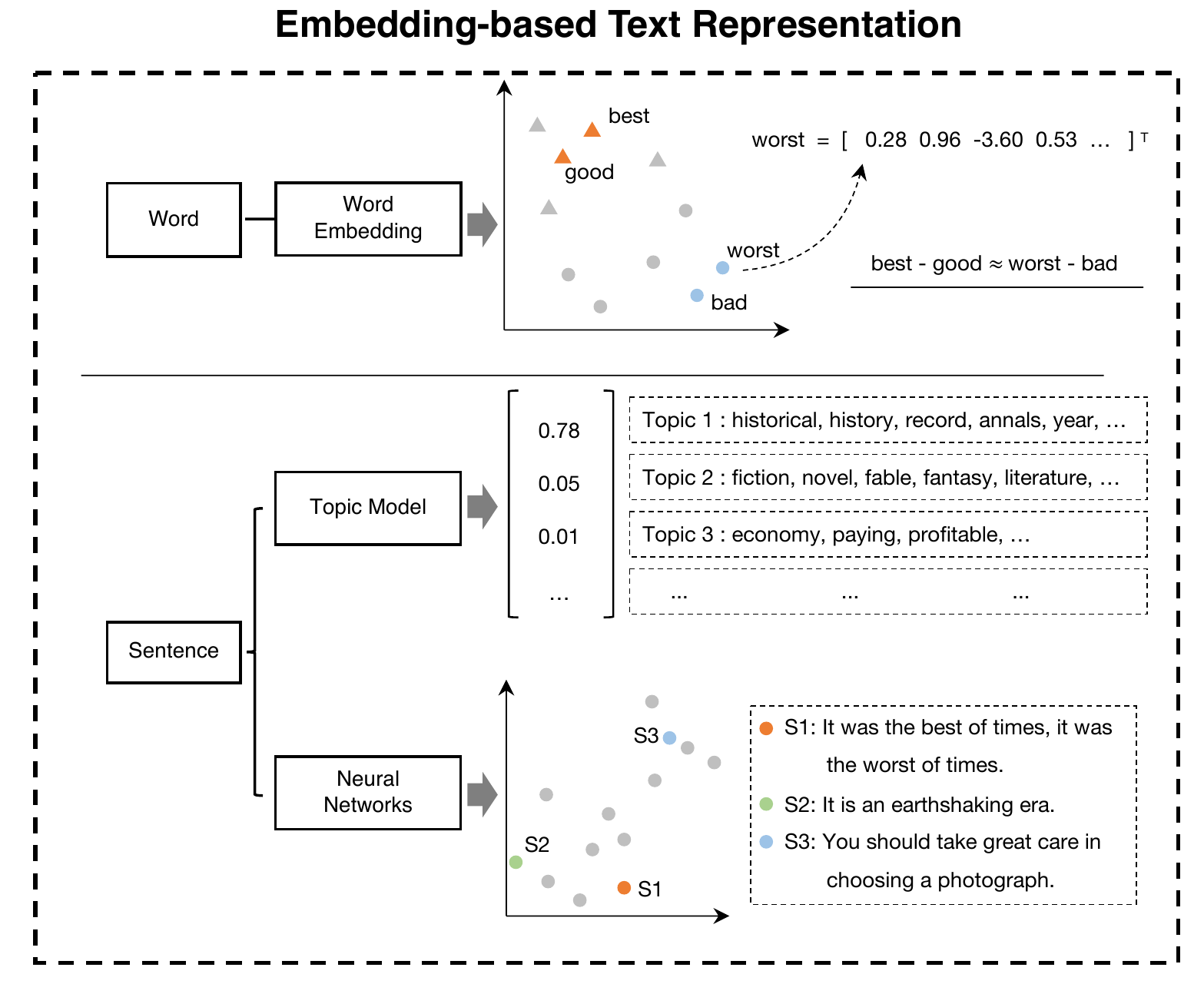}
    \caption{An illustration of embedding-based representation of text.}
    \label{fig:reps_embedding_text}
    \noindent
\end{figure*}

\section{Embedding-based Text Representation}
\label{sec:emb_text}
\noindent
Since text consists of multi-grained units as mentioned in section~\ref{sec:symbol_text}: from words to sentences, embedding-based text representation also follows the same composition principle. In this section, we will introduce the most widely used methods to learn the embedding-based representation of words and sentences, respectively. An illustrative demonstration is shown in Fig.~\ref{fig:reps_embedding_text}.

\subsection{Embedding-based Word Representation}
Approaches of learning embedding-based word representation aim to embed each word into a low-dimensional and dense vector and require that closer distance between two vectors in the space denotes more similar semantics between the corresponding words. The intuition behind these approaches is simple: words sharing similar contexts should have similar word embeddings. For instance, the word ``apple'' and ``banana'' will probably both appear in the context ``I like eating xxx'' or ``xxx trees'' from a large corpus and thus should have similar word vectors. Existing methods fall into two main groups, namely count-based models and prediction-based models~\cite{lenci2018distributional}. Next, we present each of them, respectively. 

\subsubsection{Count-based Models}
\begin{figure}[htb]
    \noindent
    \centering
    \includegraphics[width=0.9\columnwidth]{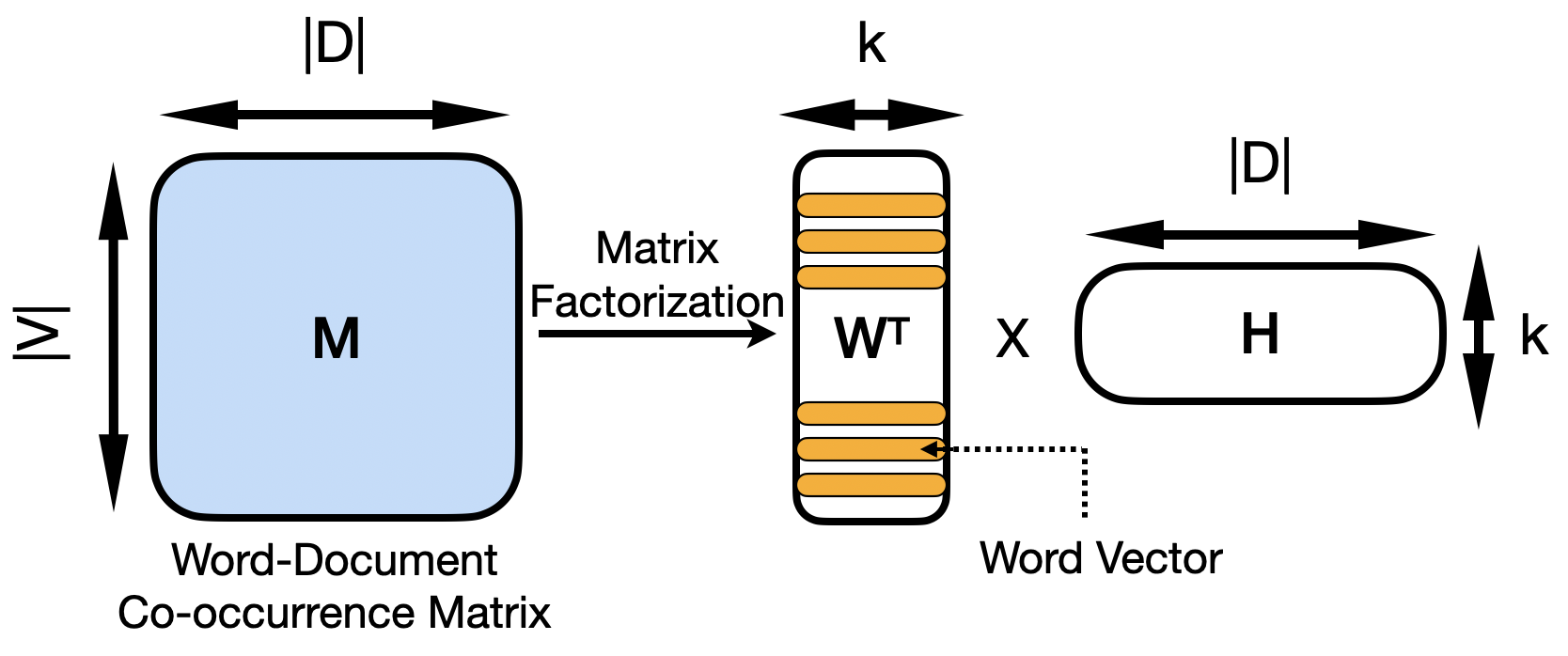}
    \caption{An example of count-based model LSA.}
    \label{fig:count_based}
    \noindent
\end{figure}
Count-based models establish distributional representations of words upon co-occurrence counting. A primary branch of these models works on transforming the co-occurrence matrix of words into a reduced space, with matrix factorization techniques such as singular value decomposition (e.g. Latent Semantic Analysis (LSA)~\cite{deerwester1990indexing}) or weighted least-squares regression (e.g. Global Vectors for Word Representation (GloVe)~\cite{pennington2014glove}). A brief example of LSA is shown in Fig.~\ref{fig:count_based}. Another branch of count-based models is Random Indexing (RI)~\cite{sahlgren2005introduction}, which learns distributional representation by assigning an initialized random vector to each word and then gradually updated according to the co-occurring contexts. It overcomes the difficulty of LSA by precluding expensive pre-processing of huge word-document matrices.

\subsubsection{Prediction-based Models}
Prediction-based models aim to create low-dimensional distributional representations through optimization of the probability that predicts a target word based on contexts or predicts the contexts of a target word. Word2vec~\cite{mikolov2013distributed} is one of the most popular toolkits of prediction-based models proposed by Google in 2013, which can efficiently learn word embeddings from a large corpus. It is equipped with two model variants: continuous bag-of-words (CBOW) and Skip-Gram. 

\begin{figure}[htb]
\centering
\includegraphics[width=0.6\columnwidth]{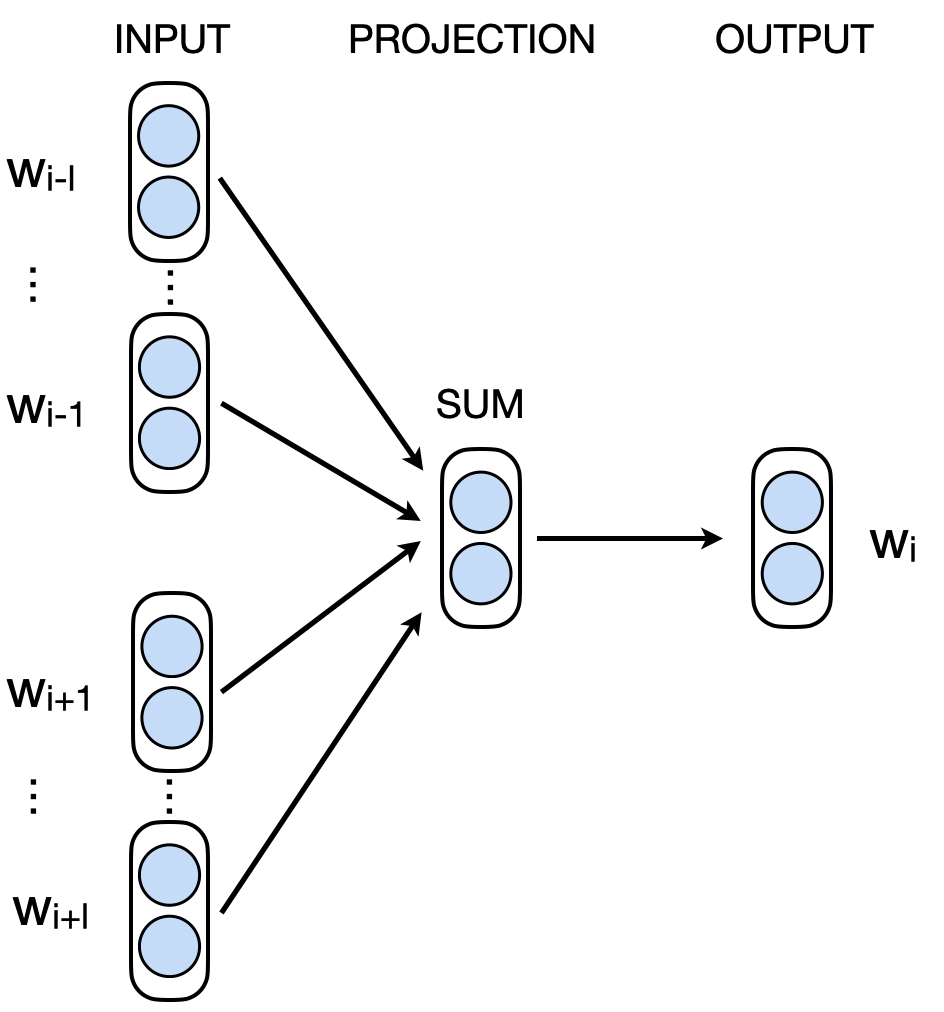}
\caption{The architecture of CBOW model.}
\label{fig:cbow}
\end{figure}

\textbf{CBOW} optimizes a training objective of predicting a target word given its context words. As shown in Fig.~\ref{fig:cbow}, CBOW predicts the center word given a window of context with the window size $l$. The window size $l$ is a hyper-parameter to be tuned.
Formally, CBOW predicts the probability of the $i$-th word $w_i$ in the corpus given its contexts of window size $l$ as:
\begin{equation}
  \begin{aligned}
\Pr(w_i|w_{i-l}\dots w_{i-1},w_{i+1}\dots w_{i+l})  = \\ \operatorname{softmax}\big( \mathbf{M}_c (\sum_{j:|j-i|\leq l, j \neq i} \mathbf{w}_j) \big),
\end{aligned}  
\end{equation}
where softmax is a normalization function that ensures the sum of the components of the output vector equals to $1$, $\mathbf{w}_i$ is the word vector of word $w_i$, $\mathbf{M}_c$ is the weight matrix in $\mathbb{R}^{|V|\times m}$, $V$ indicates the vocabulary, and $m$ is the dimension of word vectors. Then CBOW is optimized by maximizing the log likelihood:
\begin{equation}
\mathcal{L}_c=\sum_i \log \Pr(w_i|w_{i-l}\dots w_{i-1},w_{i+1}\dots w_{i+l}).
\end{equation}

\begin{figure}[htb]
\centering
\includegraphics[width=0.6\columnwidth]{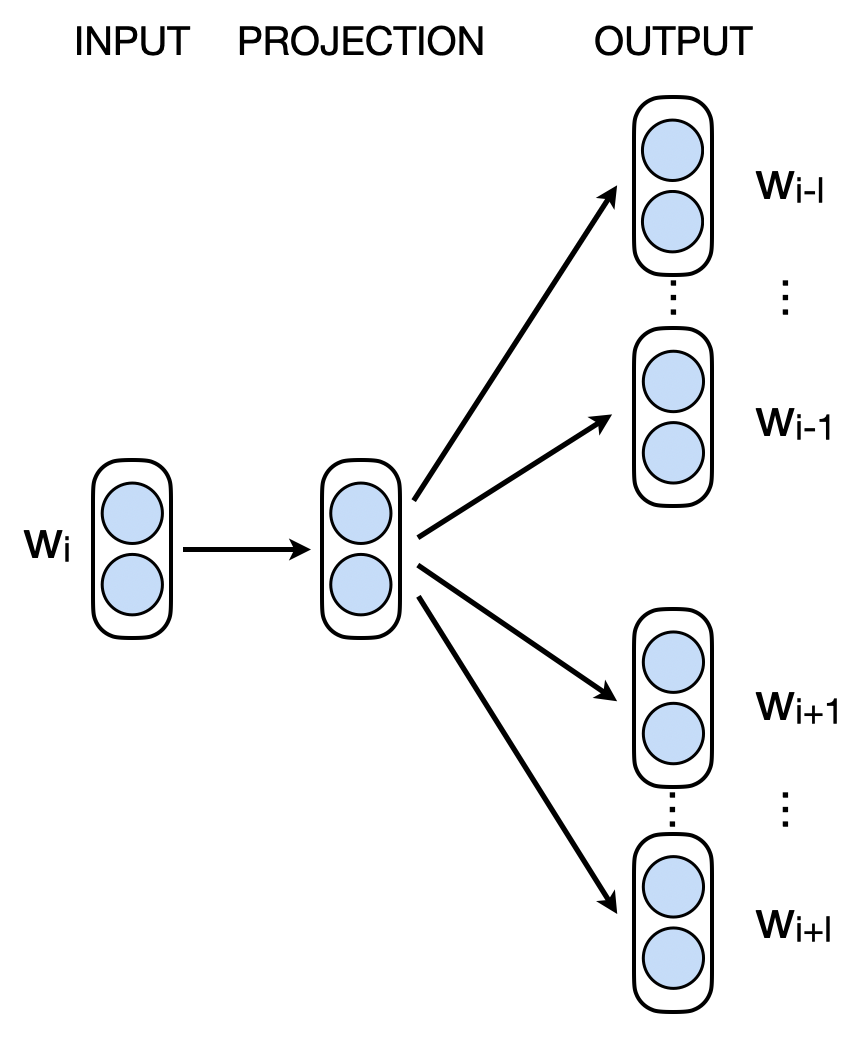}
\caption{The architecture of Skip-Gram model.}
\setlength{\abovecaptionskip}{0.cm}
\setlength{\belowcaptionskip}{-0.cm}
\label{fig:sg}
\end{figure}

\textbf{Skip-Gram} aims to predict the context words given a center one, as shown in Fig. \ref{fig:sg}. Formally, given a word $w_i$, Skip-Gram predicts each word $w_j$ $(|j-i|\leq l, j \neq i)$ in its context as:
\begin{equation}
\Pr(w_j|w_i) = \text{softmax}(\mathbf{M}_s\mathbf{w}_i),
\end{equation}
where $\mathbf{M}_s$ is the weight matrix. The optimization objective is defined as: 
\begin{equation}
\mathcal{L}_s=\sum_i \sum_{j:|j-i|\leq l, j \neq i} P(w_j|w_i).
\end{equation}

Word2vec further employs Hierarchical Softmax~\cite{morin2005hierarchical} and Negative Sampling~\cite{mnih2012fast} to speed up the computation process.

Though the algorithms differ, empirical results show that count-based models such as GloVe and prediction-based models such as CBOW perform comparably on semantic similarity and downstream tasks with certain system designs and optimized hyperparameters~\cite{levy2015improving}. Hence, we uniformly refer to them as word embedding-based representations.



\subsection{Embedding-based Sentence Representation}
Similar to word embedding, embedding-based sentence representation is also formed as a continuous and dense vector with rich semantic meanings. There are two main series of methods to learn the sentence representation: one is based on topic models, another is based on neural network models. Below we present each of them in detail.

\subsubsection{Topic Model-based Representation}

Topic models seek to represent a sentence (document) as a distribution of a series of topics, based on two assumptions: each document contains multiple topics; each topic contains multiple words. Here, we describe the most typical topic models including LSA, Latent Dirichlet Allocation (LDA)~\cite{blei2003latent}, Sturctural Topic Model (STM)~\cite{roberts2014structural}.  

\textbf{LSA} is one of the basic techniques for topic modeling, of which the core idea is to decompose the document-word matrix into independent document-topic matrices and topic-word matrices. Then each row vector in the document-topic matrix can be used to represent the corresponding document. However, the meaning of each dimension (i.e. topic) in the row vector is vague to us, though we can measure the similarity between two documents by calculating cosine similarity between two row vectors.

 

\begin{figure}[htb]
    \noindent
    \centering
    \includegraphics[width=0.9\columnwidth]{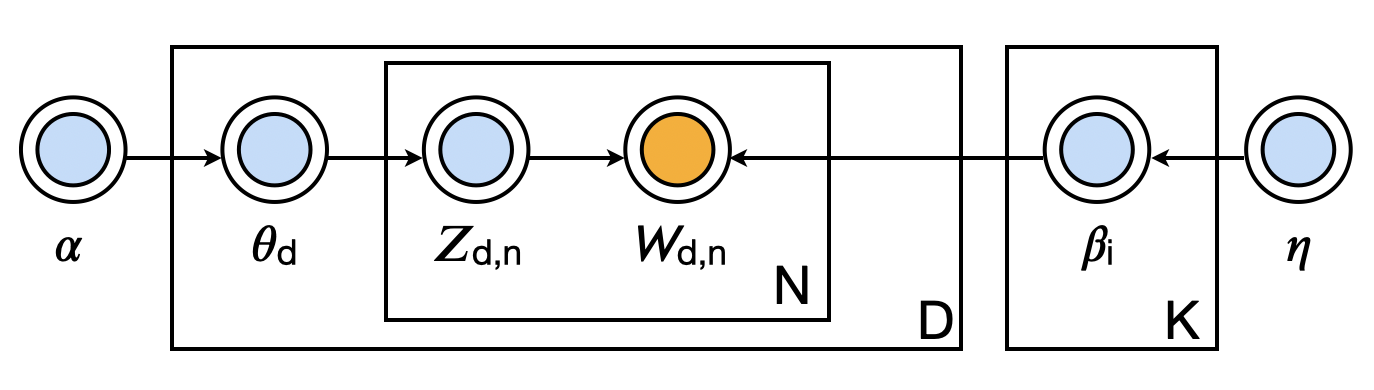}
    \caption{The architecture of graphical model for LDA.}
    \label{fig:LDA}
    \noindent
\end{figure}

\textbf{LDA} is the most widely used topic model, a member of the probabilistic graphical model. It introduces a probabilistic interpretation to the basic LSA through a generative model. Here we introduce the basic generative process of LDA, as shown in Fig.~\ref{fig:LDA}: 
    \begin{enumerate}
   \item For each document $d$, randomly choose a topic distribution $\theta_d$ over $K$ topics from the prior Dirichlet distribution with hyperparameters $\alpha$.
   \item For each word $w_{dn}$ in the document,
   \begin{itemize}
        \item randomly sample a topic $z_{d,n}$ from the topic distribution $\theta_d$;
        \item randomly choose a word distribution $\beta_{z_{d,n}}$ of topic $z_{d,n}$ over $N$ words, from another prior Dirichlet distribution with hyperparameters $\eta$;
        \item randomly sample the word $w_{d,n}$ from the word distribution $\phi_{d,n}$.
   \end{itemize}
 \end{enumerate}
Through this process, each document can be granted a representation (i.e. $\theta_d$) denoting the distribution over topics, with each topic assigned a probability distribution (i.e. $\beta_{z_{d,n}}$) over words. With the help of topic-word distribution, we can further capture the keywords of each topic and elucidate the meaning of each topic.
 
\textbf{STM} further extends LDA to account for meta-data of text, since documents usually entail time, geographic location, author, title, and other additional information. These can be formalized as covariates in the topic model so that each document can have its own prior distributions over topics and words depending on its covariates. This approach is widely used in CSS owing to the consideration of environmental variances of documents.

\subsubsection{Neural-based Representation}
Neural-based Representation is learned from neural network models, which are constructed based on a collection of connected artificial neurons inspired by the biological brain. These neurons are connected by edges with different weights which can be learned from the training process. The training process is operated by processing instances, each of which contains a given ``input'' and ``output''. Once training begins, neural network models will update their weighted associations to bridge the gap between inputs and outputs. At the end of the training process, the sentence representation will be refined automatically without manual design.  

Besides, the sentence representation based on neural network models can capture the complex internal structures of sentences owning to the flexible connections of neurons, such as sequential, hierarchical, and tree structures, which are essential for understanding sentences. Furthermore, neural network models allow us to imitate the cognitive mechanisms of the human brain, such as working memory~\cite{baddeley1992working} and attention mechanism~\cite{james1890principles}, to construct sentence representation.

In the following, we will introduce the most popular used neural network models for learning embedding-based sentence representation, including Convolutional Neural Network (CNN), Recurrent Neural Network (RNN), and Transformer. 

\begin{figure}[htb]
	\centering
	\includegraphics[width=0.6\columnwidth]{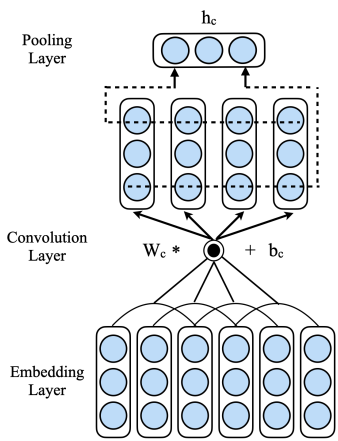}
	\caption{The architecture of CNN.} 
	\label{fig-cnn}
\end{figure}

\textbf{CNN} learns the sentence representation by two layers~\cite{kalchbrenner2014convolutional}: a convolution layer and a pooling layer, as shown in Fig.~\ref{fig-cnn}. The convolution layer extracts local features of the inputted sentence through multiple different filters. Formally, it behaves as a matrix multiplication between a convolution matrix and a sequence of word vectors in a sliding window centered on each word in the sentence. Afterwards, the pooling layer merges all local features to obtain a fixed-sized representation, with the max-pooling and mean-pooling layers most commonly used. These two layers can be represented as:
\begin{equation}
 \mathbf{h}_c= \operatorname{Pooling}[f(\mathbf{W}_c\mathbf{x}_i+\mathbf{b}_c)] , 
\end{equation}
where $\mathbf{W}_c$ denotes the convolution matrix, and $\operatorname{Pooling}$ indicates the pooling layer. $\mathbf{x}_i$ denotes the concatenation of word representations in the subsequence centered on $i$-th word. $f$ and $\mathbf{b}_c$ indicate a non-linear function and a bias vector in the convolution layer, respectively. $\mathbf{h}_c$ is the final sentence representation obtained from CNN model. 

To sum up, CNN adopts the convolutional layer so that it can focus on the sentence's local features and effectively reduce the parameters of the model. Besides, the utilization of the pooling layer endows the sentence representation with translational invariance to features, making it more robust to positions of local features.
	
\begin{figure}[htb]
	\centering
	\includegraphics[width=0.5\columnwidth]{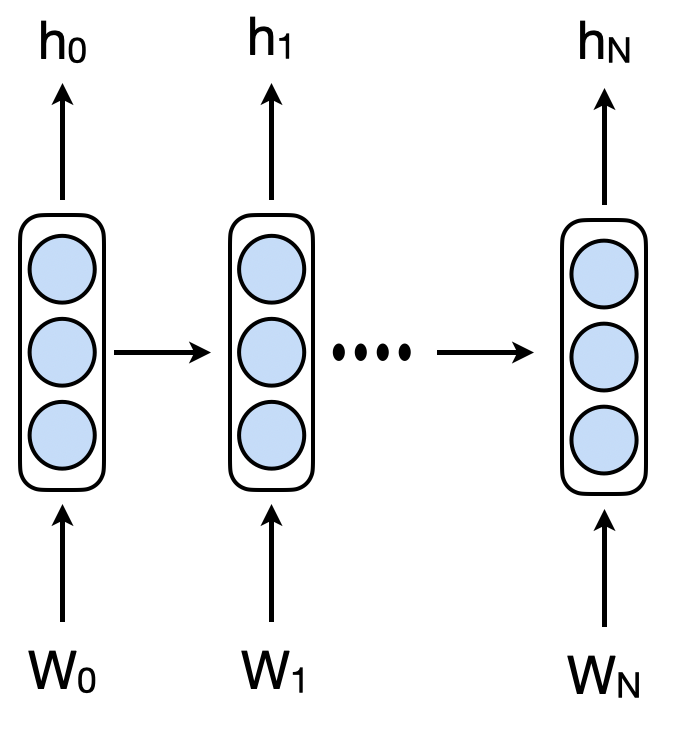}
	\caption{The architecture of RNN.} 
	\label{fig_rnn}
\end{figure}

\textbf{RNN} models the sequential structure of sentence through continuously accumulating previous information of sentence~\cite{mikolov2010recurrent}, namely hidden states. Formally, as shown in Fig.~\ref{fig_rnn}, in each time step $t$, the hidden state $\mathbf{h}_t$ is dependent on the previous hidden state $\mathbf{h}_{t-1}$ and the present word representation $\mathbf{w}_t$. It can be represented as:
\begin{equation}
    \mathbf{h}_t = f(\mathbf{W}_{r1}   \mathbf{h}_{t-1} + \mathbf{W}_{r2}   \mathbf{w}_{t} + \mathbf{b}_r), 
\end{equation}
where $\mathbf{W}_{r1}$ and $\mathbf{W}_{r2}$ are weighted matrices, and $\mathbf{b}_r$ is bias vector. The representation of the sentence can be defined as the final hidden state $\mathbf{h}_N$, with N denoted as the length of the sentence. Several extended versions of RNN model have been proposed and applied to sentence modeling, such as Gated Recurrent Unit (GRU)~\cite{cho2014learning} and Long Short-Term Memory Network (LSTM)~\cite{hochreiter1997long}, with an extra gating mechanism.

Owing to the portrait of sequential structure in text, the representation learned by RNN is more sensitive to the word and phrase order in a sentence, which is crucial for semantic caption.

\begin{figure}[htb]
	\centering
	\includegraphics[width=0.7\columnwidth]{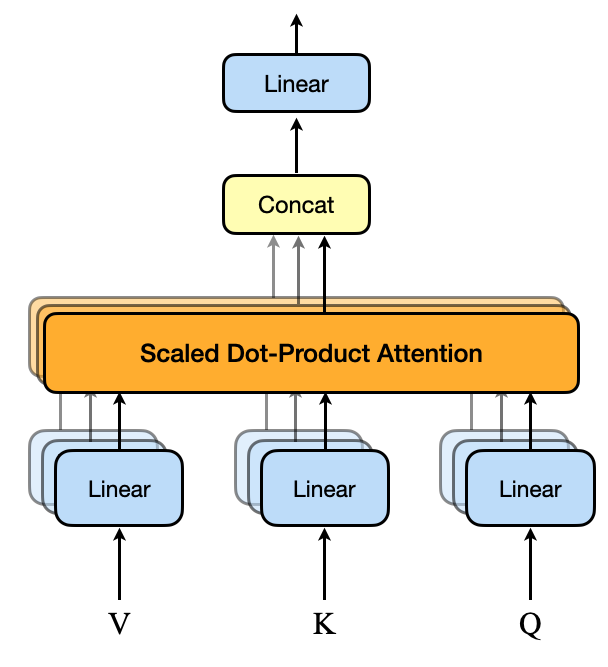}
	\caption{Multi-head attention mechanism of Transformer. $V$, $K$, and $Q$ denote value, key, and query in the attention mechanism, respectively.} 
	\label{fig_mh}
\end{figure}

\textbf{Transformer} is a deep neural network proposed by ~\cite{vaswani2017attention}, which alleviates two issues of RNN model: one is the long-distance dependence problem which means previous information will be depleted for a long sentence, another is the incapability of parallel training due to the sequential modeling. Instead of the sequential dependence, Transformer proposes a multi-head self-attention mechanism to directly connect the hidden state in each time step, as shown in Fig.~\ref{fig_mh}, which can store the information of a sentence in all positions equally and be trained parallel. Meanwhile, the multi-head mechanism can also attend to information from different vector sub-spaces.
Based on this architecture, a series of pre-trained language models have been developed, with Bidirectional Encoder Representations from Transformers (BERT)~\cite{devlin2018bert} and Generative Pre-Training (GPT)~\cite{radford2018improving} as the most representative models. They have achieved state-of-the-art performance on numerous Natural Language Processing (NLP) tasks. 


\begin{figure*}[htb]
    \noindent
    \centering
    \includegraphics[width=1.0\textwidth]{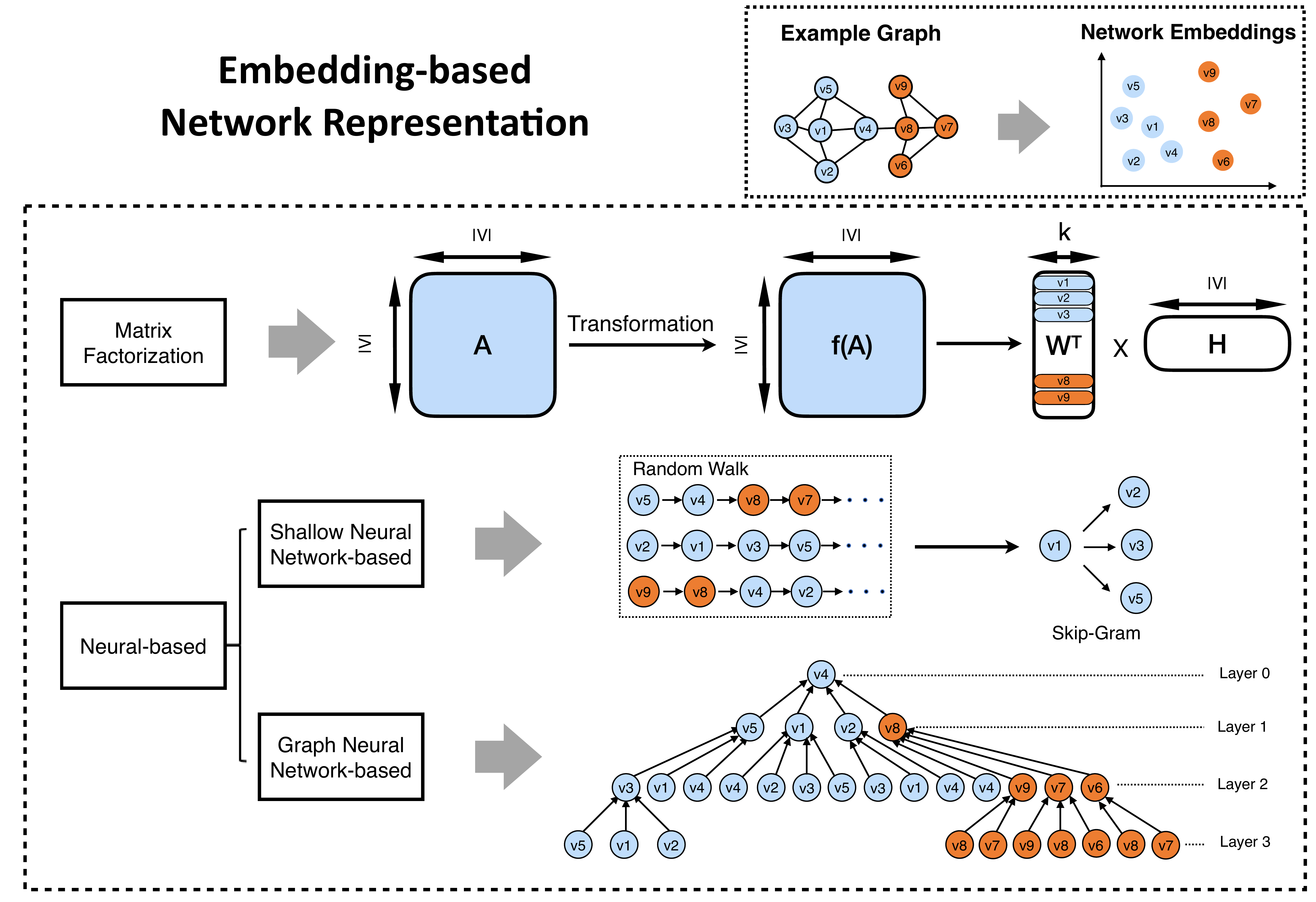}
    \caption{An illustration of embedding-based network representation.}
    \label{fig:reps_embed_network}
    \noindent
\end{figure*}


\section{Embedding-based Network Representation}
\label{sec:emb_network}
\noindent
Network embedding has attracted much attention in deep learning and data mining areas since DeepWalk~\cite{perozzi2014deepwalk} was proposed in 2014. Before that, matrix factorization-based methods were widely adopted to project nodes in a network into real-valued vectors. In this section, we classify embedding-based network representation methods into matrix factorization-based and neural-based ones.

\subsection{Matrix Factorization-based Methods} 
\noindent
Matrix factorization-based methods usually set up an optimization objective, which can be reformalized in matrix form, and then solve the optimization by eigenvector decomposition. We will introduce Laplacian Eigenmap~\cite{Belkin2001LaplacianEA} as a representative of these methods.


Given graph $G=(V,E)$ where $V$ is the vertex set and $E$ is the edge set, Laplacian Eigenmap~\cite{Belkin2001LaplacianEA} aimed to minimize the sum of the distances of all connected nodes, where the distance between two nodes is measured by Euclidean distance of their embeddings:
\begin{equation}
\sum_{(v_i,v_j)\in E} \Vert \mathbf{v}_i-\mathbf{v}_j\Vert^2,
\end{equation}
where $\mathbf{v}_i$ is the embedding of vertex $v_i$.

Assume that $R$ is a $|V|$-by-$d$ matrix where the $i$-th row of $R$ is the $d$-dimensional embedding $\mathbf{v}_i$ of node $v_i$. Laplacian Eigenmap added a constraint to avoid the trivial all-zero solution:
\begin{equation}
R^TDR=I_d,
\end{equation}
where $D$ is the $|V|$-by-$|V|$ degree matrix with $D_{ii}$ is the degree of node $v_i$ and $I_d$ is the $d$-by-$d$ identity matrix. Then the optimal solution of $R$ is proved to be the eigenvectors with $d$ smallest nonzero eigenvalues of Laplacian matrix $L$, \textit{i.e.,} the difference of diagonal matrix $D$ and adjacency matrix $A$.

During the last decade, gradient descent techniques are also used to solve the optimization problem in matrix factorization instead of eigenvector decomposition, especially when the close-form solution does not exist. Gradient descent techniques make it easier to train matrix factorization-based methods and help this line of work gets popular. By the way, the topic model introduced in sentence embedding methods can also be viewed as a general factorization process of the document-word cooccurrence matrix.
\subsection{Neural-based Methods}
Neural-based methods can take advantage of neural networks as well as deep learning techniques to build their optimization objectives. Their model could be deep or non-linear, and thus more flexible than the matrix factorization-based ones. Therefore, neural-based methods have become the mainstream for learning network embeddings in recent years. We further categorize relevant methods into shallow neural network-based and graph neural network-based ones.
\subsubsection{Shallow Neural Network-based}
\noindent
Now we will first introduce three popular unsupervised network embedding algorithms, i.e., DeepWalk, node2vec and LINE. Then we will briefly illustrate the idea of graph neural networks, a powerful neural architecture to encode structural information and feasible for supervised or semi-supervised end-to-end training.

\textbf{DeepWalk.} Inspired by the great success of word2vec~\cite{mikolov2013distributed}, as shown in Table~\ref{tab:analogy}, DeepWalk~\cite{perozzi2014deepwalk} makes an analogy between word/sentence and node/random walk, and adopts word2vec algorithm for learning node embeddings. The intuition behind is that node frequency in short random walks and word frequency in documents both follow power law.  

\begin{table}[htb]
\centering
\begin{tabular}{c|c|c|c}
  \hline
  Method & Object & Input & Output \\
  \hline
  word2vec & word & sentence & word embedding \\
  DeepWalk & node & random walk & node embedding\\
  \hline
\end{tabular}
\caption{The analogy between word2vec and DeepWalk.}
\label{tab:analogy}
\end{table}

Formally, a random walk $(v_1,v_2,\dots, v_{i})$ is a node sequence started from node $v_1$ and each node $v_k$ is randomly selected from the neighbors of node $v_{k-1}$. Random walks have been used in many network analysis tasks, such as similarity measurement~\cite{fouss2007random} and community detection~\cite{andersen2006local}. Therefore, the structural information can be encoded into sampled random walks. 

Then DeepWalk treats sampled random walks as sentences from a text corpus, and employs Skip-Gram and hierarchical softmax model for learning node embeddings. The overall objective function can be obtained by summing up every node in every sampled random walk.

By preserving structural information in learned node embeddings, DeepWalk outperforms traditional symbol-based representations such as adjacency matrix on both efficiency and effectiveness by alleviating the computation and sparsity issues. Besides, compared with the adjacency matrix, random walks can better characterize the network structure by capturing the similarity between the nodes that are not directly connected. Thus we can achieve better performance on downstream tasks with more structural information provided. 

\textbf{Node2vec.} Note that DeepWalk generates random walks by choosing the next node from a uniform distribution. Node2vec~\cite{grovernode2vec} further generalizes DeepWalk with Breadth-First Search (BFS) and Depth-First Search (DFS) on random walks. Specifically, node2vec proposes a neighborhood sampling strategy for generating random walks and can smoothly interpolate between BFS (microscopic local neighborhoods) and DFS (macroscopic community information).

Formally, given a random walk arriving at node $v$ through edge $(t,v)$, node2vec defines the unnormalized transition probability of edge $(v,x)$ for next walk step as $\pi_{vx}=\alpha_{pq}(t,x)$, where

\begin{equation}
\alpha_{pq}(t,x) = \left\{
             \begin{array}{lcl}
             {\frac{1}{p}} &\text{if} &d_{tx}=0 \\
             {1} &\text{if} &d_{tx}=1 \\
             {\frac{1}{q}} &\text{if} &d_{tx}=2 \\
             \end{array}
             \right.
\end{equation}
and $d_{tx}$ denotes the shortest path distance between node $t$ and $x$. $p$ and $q$ are controlling hyper-parameters: a small $p$ will increase the probability of revisiting and restrict the random walk in a local neighborhood while a small $q$ will encourage the random walk to move to distant nodes. The operations of node2vec after the generation of random walks are the same as DeepWalk.

\textbf{LINE.} LINE~\cite{tang2015line} parameterizes first-order and second-order proximities between vertices for learning network embeddings. The first-order proximity denotes directly connected nodes and second-order proximity represents nodes sharing common neighbors.

Formally, LINE models the first-order proximity between node $v_i$ and $v_j$ as the probability 
\begin{equation}
p_1(v_i,v_j)=\frac{1}{1+\exp(-\mathbf{v}_i\cdot \mathbf{v}_j)},
\end{equation}
where $\mathbf{v}_i$ is the embedding of vertex $v_i$.

The target probability is defined as the weighted average $\hat{p}_1(v_i,v_j)=w_{ij}/\sum_{(v_i,v_j)\in E}w_{ij}$ where $w_{ij}$ is the edge weight. The optimization objective is to minimize the distance between parameterized probability $p_1$ and target probability $\hat{p}_1$:
\begin{equation}
\mathcal{L}_1=D_{\text{KL}}(\hat{p}_1\left|\right|p_1),
\label{eq:line1}
\end{equation}
where $D_{\text{KL}}(\cdot\left|\right|\cdot)$ is the KL-divergence between two probability distributions.

For modeling the second-order proximity, the probability that node $v_j$ appears in $v_i$'s context (\textit{i.e.,} $v_j$ is a neighbor of $v_i$) is parameterized as:
\begin{equation}
p_2(v_j|v_i)=\frac{\exp(\mathbf{c}_k\cdot \mathbf{v}_i)}{\sum_{k=1}^{|V|}\exp(\mathbf{c}_k\cdot \mathbf{v}_i)},
\end{equation}
where $\mathbf{c}_j$ is the context embedding of node $v_j$. Given two nodes sharing many common neighbors, their embeddings will have large inner products with the context embeddings of common neighbors. Therefore, their embeddings will be similar and thus can capture the second-order proximity.

Similar to Eq.~(\ref{eq:line1}), the target probability is defined as $\hat{p}_2(v_j|v_i)=w_{ij}/\sum_k w_{ik}$ and the optimization objective is to minimize
\begin{equation}
\mathcal{L}_2=\sum_i \sum_k w_{ik} D_{\text{KL}}(\hat{p_2}(\cdot,v_i)\left|\right|p_2(\cdot,v_i)).
\end{equation}

The first-order and second-order proximity embeddings are learned independently. After the training phase, we can concatenate them as node embeddings.

\subsubsection{Graph Neural Network-based Methods}
Graph Neural Network (GNN) can be seen as a special kind of convolutional neural network that operates on graphs. There are three common points between GNN and CNN: local connection, shared weights, and multi-layer architectures. Each sliding window in a CNN becomes the enumeration of every node's neighborhood in a GNN, i.e., a node and all its neighbors. Therefore, in each layer of GNN, every node will update its embedding by aggregating the embeddings of its neighbors as well as itself in the previous layer. Weight matrices and non-linear functions are also employed in the update process. Taking one of the most widely used GNN architecture, graph convolutional neural network (GCN)~\cite{Kipf2017SemiSupervisedCW}, as an example, the update rule in the $t$-th layer of GCN can be formalized as
\begin{equation}
    \mathbf{H}^{(t)} = f({D}^{-\frac{1}{2}}{A}{D}^{-\frac{1}{2}}\mathbf{H}^{(t-1)}\mathbf{W}^{(t)}),
\end{equation}
where matrix $\mathbf{H}^{(t)}$ indicates the embeddings of all the nodes in a network, $D$ is the degree matrix and $A$ is the adjacency matrix with self-loops, and $\mathbf{W}^{(t)}$ is the trainable weight matrix in the $t$-th layer. The output embeddings can be directly fed into classifiers for an end-to-end training process.

\section{Applications in Computational Social Science}
\label{sec:app}
\noindent
Computational social science has received widespread attention after decades of development. As a typical inter-disciplinary area, it is involved in multifarious disciplines, including not only five primary sub-disciplines of traditional social science, namely sociology, anthropology, psychology, politics, and economics but also other disciplines of humanities, such as linguistics, communication, and geography. Hence, we choose three of the most cited and prestigious multidisciplinary academic journals: Nature\footnote{We only choose the articles in the social science subject of Nature as candidate pool to ensure the relevance.}, Science\footnote{Articles in the main journal and the sub-journal Science Advances are considered to ensure representativeness and relevance as well.} and PNAS\footnote{Papers from its social science category are examined, of which the link is \url{https://www.pnas.org/category/social-sciences}.} to investigate the applications of symbol-based and embedding-based representations in CSS in recent ten years (2011-2020). Specifically, we first sort published papers in these journals by the number of citations each year,\footnote{The number of citations is crawled from Bing search engine.} since we believe the number of citations is an important indicator of the influence and representativeness of an article. Afterwards, top-cited papers utilizing one or more types of symbol-based or embedding-based representations in CSS each year, are selected for our survey. Note that the number of citations of papers published in years closer to now makes less sense, so we list all relevant papers in 2019 and 2020.

Since CSS is a highly intertwined discipline between social science and computer science, we further examine the number of applications using these two representations in computer science. We select $3$ top conferences closely related to CSS in computer science, namely ACL\footnote{The Association for Computational Linguistics.}, WWW\footnote{The International World Wide Web Conference} and KDD\footnote{The International Conference on Knowledge Discovery and Data Mining }, involving the research areas of natural language processing, data mining, and network analysis. We follow the similar settings in the above three journals and choose top-cited papers each year between $2011$ and $2020$ for text and network, respectively. 



Considering the comprehensiveness of the audience and the diversity of the topics, we highlight the representative applications in three journals in the main text. An overview of all the applications is presented in the appendix. 

In this section, we first formalize the main tasks utilizing text and network data in CSS, respectively. Afterwards, we group the applications following their task formalizations, and present how existing studies utilize symbol-based and embedding-based representations to serve these tasks, in order from symbols to embeddings, and texts to networks. At last, we further summarize and compare the advantages and disadvantages between these two kinds of representations according to their applications.

\subsection{Task Formalization}
\begin{figure*}[htb]
	\centering
	\includegraphics[width=0.95\linewidth]{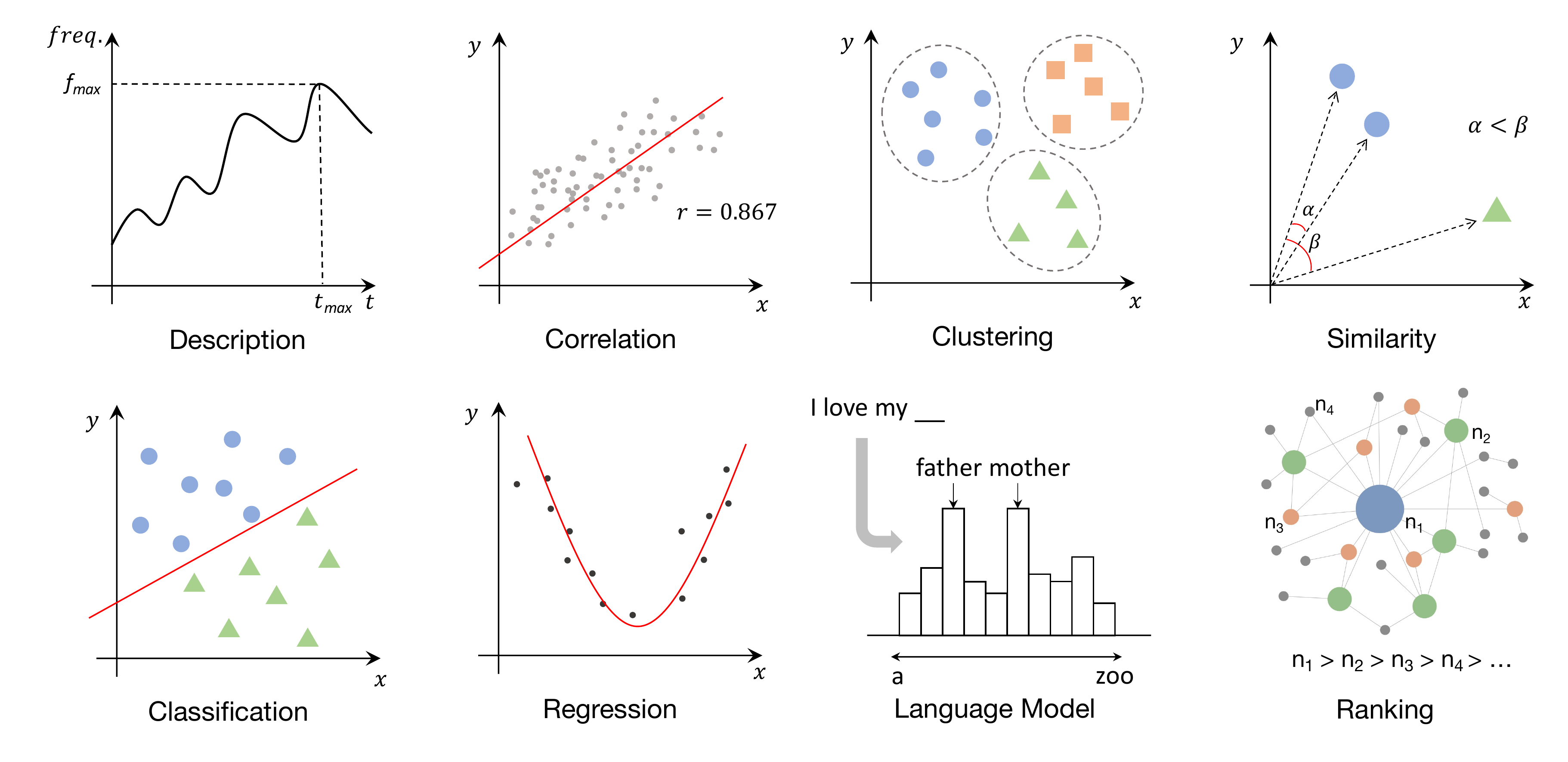}
	\caption{A simple illustration of eight prototypical tasks.}
	\label{fig:tasks}
\end{figure*}

Though the explosive growth in research topics, applications employing text and network data in CSS can be summarized mainly in eight prototypical tasks, i.e. description, correlation, similarity, clustering, classification, regression, language model and ranking. A simple illustration of these formalized tasks is shown in Fig.\ref{fig:tasks}. In the followings, we will give explicit definitions of them, respectively. 

\textbf{Description} denotes quantitative depiction of characteristics of data, including frequency, distribution, etc. Distinguished from inferential statistics, it is a direct summarization of the observed data, while inferential statistics aims to infer the properties of a larger population based on the analysis of observed data. 

\textbf{Relation} aims to measure the relationship between two variables, with correlation and causality as the most typical relationships. If under certain circumstances, one variable changes, another variable also moves, then these two variables are correlated. Causality can be regarded as a kind of continuous and stable correlation, regardless of whether other variables exist and how they change. 

\textbf{Similarity} aims to measure if two objects have similar characteristics, such as semantics, sentiment, or styles. From the technical perspective, this task is the basis for many other tasks such as clustering and classification. 

\textbf{Clustering} is to group a set of objects so that objects in the same group (i.e. cluster) are more similar than objects in other groups. Note that it automatically explores the features of different categories existing in data, without requirement of the specific definition of each category.

\textbf{Classification} focuses on classifying each object into one or multiple specific categories in line with its properties. Different from clustering, these categories are manually defined in advance. 

\textbf{Regression} is similar to the classification task, with the difference existing in that regression focuses on predicting a continuous target for each object, rather than a discrete category.

\textbf{Language model} is a unique task for text analysis, which calculates the probability distribution over sequences of words. It is generally implemented by calculating the conditional probability of a word given its context. Taking the word sequence \{`I', `love', `my', `mother'\} in Fig.~\ref{fig:tasks} as an example, it behaves as:
    \begin{equation}
    \small
     \begin{aligned}
    &P(\text{`I'},\text{`love'},\text{`my'},\text{`mother'}) = \\ &P(\text{`I'}) \times 
    P(\text{`love'}|\text{`I'})  \dots \times  P(\text{`mother'}|\text{`I'},\text{`love'},\text{`my'}),
     \end{aligned}
    \end{equation}
    where $P(\text{`mother'}|\text{`I'},\text{`love'},\text{`my'})$ denotes the conditional probability of predicting word `mother' known the previous context subsequence \{`I', `love', `my'\}.
    
\textbf{Ranking} is a task mainly for network analysis, aiming to find out the most important or influential nodes in a network. In other words, we need to score the nodes and rank them for our purpose.

Based on the above task formalizations in CSS, we can gain an overview of the scenarios in which symbol-based and embedding-based representations can be applied, to further consider what type of tasks they are expert in.

\begin{sidewaystable}
\centering
\resizebox{0.9\textwidth}{!}
{
\begin{tabular}{c|c|l|lllllll}
\hline
\multicolumn{3}{c|}{\multirow{2}{*}{Representations}}                                              & \multicolumn{7}{c}{Tasks}                                                                                                                                                                                                                         \\ \cline{4-10} 
\multicolumn{3}{c|}{}                                                                              & \multicolumn{1}{c|}{Description} & \multicolumn{1}{c|}{Relation} & \multicolumn{1}{c|}{Similarity} & \multicolumn{1}{c|}{Clustering} & \multicolumn{1}{c|}{Classification} & \multicolumn{1}{l|}{Regression} & \multicolumn{1}{c}{Language Model} \\ \hline
\multirow{5}{*}{Symbol}    & \multirow{3}{*}{Word}                          & Frequency-based      
& \multicolumn{1}{l|}{
\begin{tabular}[c]{@{}l@{}}
~\cite{yang2013ontogeny}, ~\cite{michel2011quantitative}, ~\cite{bruch2018aspirational}\\
~\cite{lupia2020does}, ~\cite{sheshadri2019public}, ~\cite{golder2011diurnal}\\
\end{tabular}
}           
& \multicolumn{1}{l|}{
\begin{tabular}[c]{@{}l@{}}
~\cite{alanyali2013quantifying}, ~\cite{sheshadri2019public}
\end{tabular}
}          & \multicolumn{1}{l|}{}          & \multicolumn{1}{l|}{}              & \multicolumn{1}{l|}{}          & \multicolumn{1}{l|}{}              &                             \\ \cline{3-10} 
                           &                                                & Feature-based        
& \multicolumn{1}{l|}{
~\cite{dodds2015human}
}       
& \multicolumn{1}{l|}{
~\cite{dodds2015human}
}          & \multicolumn{1}{l|}{}          & \multicolumn{1}{l|}{}              & \multicolumn{1}{l|}{}          & \multicolumn{1}{l|}{~\cite{huth2016natural}}              &
\\ \cline{3-10} 
                           &                                                & Network-based        
& \multicolumn{1}{l|}{}           & \multicolumn{1}{l|}{}          & \multicolumn{1}{l|}{
\begin{tabular}[c]{@{}l@{}}
~\cite{stella2018bots}, ~\cite{ramiro2018algorithms}, ~\cite{jackson2019emotion}
\end{tabular}
}          
& \multicolumn{1}{l|}{
\begin{tabular}[c]{@{}l@{}}
~\cite{jackson2019emotion}, ~\cite{rule2015lexical}
\end{tabular}
}              & \multicolumn{1}{l|}{
~\cite{bovet2018validation}
}          & \multicolumn{1}{l|}{}              &                             \\ \cline{2-10} 
                           & \multirow{2}{*}{Sentence}                      & Frequency-based      
& \multicolumn{1}{l|}{
\begin{tabular}[c]{@{}l@{}}
~\cite{citron2015patterns}, ~\cite{eichstaedt2018facebook}
\end{tabular}
}           
& \multicolumn{1}{l|}{~\cite{eichstaedt2018facebook}}          
& \multicolumn{1}{l|}{}          & \multicolumn{1}{l|}{}        
& \multicolumn{1}{l|}{
\begin{tabular}[c]{@{}l@{}}
~\cite{eichstaedt2018facebook}, ~\cite{green2020elusive}, ~\cite{bakshy2015exposure}
\end{tabular}
}          & \multicolumn{1}{l|}{}              & ~\cite{piantadosi2011word}                            \\ \cline{3-10} 
                           &                                                & Feature-based       
& \multicolumn{1}{l|}{
\begin{tabular}[c]{@{}l@{}}
~\cite{futrell2015large}, ~\cite{jordan2019examining}, ~\cite{frank2013happiness}
\end{tabular}
}           
& \multicolumn{1}{l|}{~\cite{kryvasheyeu2016rapid}}          & \multicolumn{1}{l|}{
\begin{tabular}[c]{@{}l@{}}
~\cite{klingenstein2014civilizing}, ~\cite{catalini2015incidence}, ~\cite{boyd2020narrative}\\
~\cite{hughes2012quantitative}
\end{tabular}
}          & \multicolumn{1}{l|}{}              
& \multicolumn{1}{l|}{
\begin{tabular}[c]{@{}l@{}}
~\cite{kramer2014experimental}, ~\cite{brady2017emotion}, ~\cite{jones2017distress}\\
~\cite{kryvasheyeu2016rapid}, ~\cite{stella2018bots}, ~\cite{catalini2015incidence}\\
~\cite{del2016echo}, ~\cite{bovet2018validation}, ~\cite{alizadeh2020content}\\
\end{tabular}
}          
& \multicolumn{1}{l|}{~\cite{lupia2020does}}              &                             \\ \hline
\multirow{3}{*}{Embedding} & Word                                           & Word Embedding-based 
& \multicolumn{1}{l|}{}           & \multicolumn{1}{l|}{}         & \multicolumn{1}{l|}{
\begin{tabular}[c]{@{}l@{}}
~\cite{garg2018word}, ~\cite{caliskan2017semantics}, ~\cite{sivak2019parents}\\
\end{tabular}
}          
& \multicolumn{1}{l|}{}              & \multicolumn{1}{l|}{}          & \multicolumn{1}{l|}{}              &                             \\ \cline{2-10} 
                           & \multicolumn{1}{l|}{\multirow{2}{*}{Sentence}} & Topic Model-based   
& \multicolumn{1}{l|}{
\begin{tabular}[c]{@{}l@{}}
~\cite{gerow2018measuring}, ~\cite{lupia2020does}
\end{tabular}
}           & \multicolumn{1}{l|}{~\cite{bokanyi2016race}}      
& \multicolumn{1}{l|}{
\begin{tabular}[c]{@{}l@{}}
~\cite{bokanyi2016race}, ~\cite{Farrell2016Network}, ~\cite{roy2015predicting}\\
\end{tabular}
}          
& \multicolumn{1}{l|}{
\begin{tabular}[c]{@{}l@{}}
~\cite{2014Quantifying}, ~\cite{farrell2016corporate}
\end{tabular}
}              
& \multicolumn{1}{l|}{
\begin{tabular}[c]{@{}l@{}}
~\cite{jaidka2020estimating}, ~\cite{eichstaedt2018facebook}
\end{tabular}
}          & \multicolumn{1}{l|}{}              &                             \\ \cline{3-10} 
                           & \multicolumn{1}{l|}{}                          & Neural-based 
& \multicolumn{1}{l|}{}           & \multicolumn{1}{l|}{}          & \multicolumn{1}{l|}{
\begin{tabular}[c]{@{}l@{}}
~\cite{sheshadri2019public}, 
\end{tabular}
}          & \multicolumn{1}{l|}{}              
& \multicolumn{1}{l|}{
\begin{tabular}[c]{@{}l@{}}
~\cite{fetaya2020restoration}, ~\cite{hahn2020universals}, ~\cite{mooijman2018moralization}
\end{tabular}
}          
& \multicolumn{1}{l|}{}              & 
~\cite{hahn2020universals}                            \\ \hline
\end{tabular}
}
\caption{Applications of symbol-based and embedding-based text representations.}
\label{tab:text_apps}

\end{sidewaystable}

\begin{sidewaystable}
\centering
\resizebox{0.9\textwidth}{!}
{
\begin{tabular}{c|c|l|lllllll}
\hline
\multicolumn{3}{c|}{\multirow{2}{*}{Representations}}                                                  & \multicolumn{7}{c}{Tasks}                                                                                                                                                                                  \\ \cline{4-10} 
\multicolumn{3}{c|}{}                                                                                  & \multicolumn{1}{c|}{Description} & \multicolumn{1}{c|}{Relation} &\multicolumn{1}{c|}{Similarity} & \multicolumn{1}{c|}{Clustering} & \multicolumn{1}{c|}{Classification} & \multicolumn{1}{c|}{Regression} & \multicolumn{1}{c}{Ranking} \\ \hline
\multirow{6}{*}{Symbol}    & \multirow{4}{*}{Node}     & Node \& Edge-based Statistics    
& \multicolumn{1}{l|}{~\cite{grinberg2019fake},~\cite{li2014comparative}}  
& \multicolumn{1}{l|}{
\begin{tabular}[c]{@{}l@{}}
~\cite{grinberg2019fake},~\cite{turetsky2020psychological},~\cite{apicella2012social}\\
~\cite{li2014comparative},~\cite{boardman2012social}\\
~\cite{clauset2015systematic},~\cite{trujillo2018document}, ~\cite{wesolowski2012quantifying}\\
~\cite{jo2014spatial},~\cite{parkinson2018similar},~\cite{wu2019large}
\end{tabular}
} 
& \multicolumn{1}{l|}{~\cite{park2018strength}, ~\cite{parkinson2018similar}, ~\cite{boardman2012social}}          
& \multicolumn{1}{l|}{}          
& \multicolumn{1}{l|}{~\cite{garcia2017leaking}}              
& \multicolumn{1}{l|}{

\begin{tabular}[c]{@{}l@{}}

\end{tabular}
}          
& \multicolumn{1}{l}{}       
                             
\\ \cline{3-10} 
                          &                           
                               
& Centrality-based                              
& \multicolumn{1}{l|}{}  
& \multicolumn{1}{l|}{~\cite{turetsky2020psychological}}  
& \multicolumn{1}{l|}{}          
& \multicolumn{1}{l|}{}          
& \multicolumn{1}{l|}{}              
& \multicolumn{1}{l|}{}          
&
\multicolumn{1}{l}{
\begin{tabular}[c]{@{}l@{}}
~\cite{schich2014network}, ~\cite{manrique2016women}, ~\cite{reino2017networks} \\
~\cite{hart2017effects}, ~\cite{fraiberger2018quantifying}\\
~\cite{bruch2018aspirational}, ~\cite{pei2014searching}, ~\cite{waniek2018hiding}
\end{tabular}
} 
\\ \cline{3-10} 
                          &                           
                        & Designed index                        
& \multicolumn{1}{l|}{~\cite{dankulov2015dynamics}}  
& \multicolumn{1}{l|}{
\begin{tabular}[c]{@{}l@{}}
~\cite{asikainen2020cumulative},~\cite{charoenwong2020social},~\cite{aral2012identifying}\\~\cite{eom2014generalized},~\cite{wu2019large}
\end{tabular}
}  
& \multicolumn{1}{l|}{~\cite{asikainen2020cumulative}}      
& \multicolumn{1}{l|}{}          
& \multicolumn{1}{l|}{}              
& \multicolumn{1}{l|}{~\cite{ganin2017resilience}}          
& \multicolumn{1}{l}{~\cite{pei2014searching}}    
\\ \cline{3-10} 
                          &                           
                          & Probabilistic model                           
& \multicolumn{1}{l|}{~\cite{dankulov2015dynamics}}  
& \multicolumn{1}{l|}{}  
& \multicolumn{1}{l|}{}          
& \multicolumn{1}{l|}{}          
& \multicolumn{1}{l|}{~\cite{massucci2016inferring}}              
& \multicolumn{1}{l|}{~\cite{teng2016collective}}          
& \multicolumn{1}{l}{~\cite{medo2016identification},~\cite{teng2016collective}}       
\\ \cline{2-10} 
                          & \multirow{2}{*}{Subgraph} 
& Motif-based statistics/coefficients/index     
& \multicolumn{1}{l|}{}  
& \multicolumn{1}{l|}{~\cite{asikainen2020cumulative},~\cite{kovanen2013temporal}}  
& \multicolumn{1}{l|}{
~\cite{asikainen2020cumulative}, ~\cite{kovanen2013temporal}
}          
& \multicolumn{1}{l|}{}          
& \multicolumn{1}{l|}{}              
& \multicolumn{1}{l|}{}          
& \multicolumn{1}{l}{}       
\\ \cline{3-10} 
                          &                           & Cluster-based statistics/coefficients/index 
& \multicolumn{1}{l|}{}  
& \multicolumn{1}{l|}{~\cite{trujillo2018document}}  
& \multicolumn{1}{l|}{~\cite{jackson2019emotion}}          
& \multicolumn{1}{l|}{~\cite{expert2011uncovering}}          
& \multicolumn{1}{l|}{}              
& \multicolumn{1}{l|}{}          
& \multicolumn{1}{l}{~\cite{waniek2018hiding}}       
\\ \hline
\multirow{2}{*}{Embedding} & \multirow{2}{*}{Node \& Subgraph}     
& Matrix Factorization                          
& \multicolumn{1}{l|}{}  
& \multicolumn{1}{l|}{}  
& \multicolumn{1}{l|}{~\cite{yang2011like}}          
& \multicolumn{1}{l|}{~\cite{sachan2012using}}          
& \multicolumn{1}{l|}{~\cite{kosinski2013private}, ~\cite{yang2011like}}              
& \multicolumn{1}{l|}{~\cite{kosinski2013private}}          
& \multicolumn{1}{l}{~\cite{wang2012magnet}}                                    
\\ \cline{3-10} 
                          &                           
                          & Neural-based                                  
& \multicolumn{1}{l|}{}  
& \multicolumn{1}{l|}{}  
& \multicolumn{1}{l|}{~\cite{yang2019revisiting}}          
& \multicolumn{1}{l|}{}          
& \multicolumn{1}{l|}{
\begin{tabular}[c]{@{}l@{}}
~\cite{zhang2017regions}, ~\cite{fan2019graph}, ~\cite{yang2019revisiting}\\
~\cite{wu2019dual}
\end{tabular}
}              
& \multicolumn{1}{l|}{~\cite{zhang2017regions}}          
& \multicolumn{1}{l}{}     
\\ \hline
\end{tabular}
}
\caption{Applications of symbol-based and embedding-based network representations.}
\label{tab:network_apps}
\end{sidewaystable}

\subsection{Applications of Symbol-based Text Representation }
\label{sec:symbol_app}
As a traditional way, symbol-based representation has been widely applied in CSS over the past decade, when analyzing text data. In this section, we sort out the applications according to the type (i.e. word or sentence) of representations they employ and prototypical tasks they are formalized into. Top half of the Table~\ref{tab:text_apps} lists the sorted applications using symbol-based text representations. 

\subsubsection{Applications of Symbol-based Word Representation}
Symbol-based word representation is mainly applied to the task forms of description, relation, similarity, clustering, with few in classification and regression. 

\textbf{Description.} Symbol-based word representation is extensively used for description, especially frequency-based word representation, owing to its intuitiveness and interpretability. 
    
Researchers usually define specific words as representatives of an abstract concept such as culture, linguistic grammar, and sentiment, and demonstrate the development and variations of the concept by observing their frequency changes across time and space. ~\citep{michel2011quantitative} tracks the words expressing time such as ``1880'' and ``1973'' in millions of digitized books from $1800$ to $2000$, and find that people forget past faster as time goes by, with ``1973'' declined to half its peak three times faster than ``1880''. Similarly, it also finds that we absorb the technology faster than before with words of the invention widespread more rapidly. ~\cite{yang2013ontogeny} counts the frequency of two determiners ``a'' and ``an'' paired with nouns in the data where young children learning American English, respectively, and calculates the empirical probabilities of nouns co-occurring with these two determiners. Compared with expected probabilities, it discovers that young children's language is equipped with a productive grammar rather than memorization of caregivers' speech. ~\cite{lupia2020does} calculates the most distinguished words co-occurring with the ``National Science Foundation'' or ``NSF'' between the Republicans and Democrats according to the count of words, and further finds their different concerns for NSF. ~\cite{bruch2018aspirational, sheshadri2019public, golder2011diurnal} all extract sentiment words from the corpus, and regard the frequency of them as the indication of individuals' mood or the framing polarity of news. 
    
    
Outside of frequency-based word representation, feature-based representation is also applied to the task of description, although it generally requires a large amount of manual effort. ~\cite{dodds2015human} manually labeled happiness value of $10,000$ most common words in $10$ languages, and derive that a universal positive bias exists in natural language through observing the distributions of happiness scores across different languages. 
    
\textbf{Relation.} Symbol-based word representation is also frequently utilized to investigate the relationship between two variables, including the correlation and causality. ~\cite{alanyali2013quantifying} counts the frequency of a company's mentions in the news, and discovers a positive correlation between the frequency and its daily transaction volume. ~\cite{dodds2015human} examines how the happiness scores of words vary across $10$ languages and find a strong correlation between any two languages. Apart from the relationship of correlation, ~\cite{sheshadri2019public} investigates the causality of negative polarity of news framing and public approval, as well as legislation, where the polarity is represented by the frequency of negative sentiment words.

\textbf{Similarity.} Symbol-based word representation can be applied in the task of calculating the similarity between objects, where the main method is based on network-based word representation. Researchers can use network analysis methods to calculate the similarity between two networks' structures or two nodes in a network.  ~\cite{stella2018bots} builds two networks according to hashtag co-occurrences of two polarized groups on Twitter, and calculates the consistency of common nodes in two networks to reveal semantic similarity of two polarized groups. Different from using consistency to calculate similarity, ~\cite{ramiro2018algorithms} defines semantic similarity of a word's two senses based on their conceptual proximity in the network, constructed following the taxonomic hierarchy structure of a word form-sense dictionary. It demonstrates that a word extends its senses mainly through a nearest-neighbor chain based on the above similarity. ~\cite{jackson2019emotion} constructs colexification networks of $24$ emotion concepts across $2474$ spoken languages, and further uses adjusted Rand indices (ARIs) to quantify the similarity of two networks' structures. Depending on the similarity calculation, it reveals the significant difference of emotion semantics across different language families.
    
\textbf{Clustering.} Since the task of clustering is principally based on the calculation of similarity, it also commonly applies network-based word representations. We can implement clustering of words by network analysis algorithms such as community detection. ~\cite{rule2015lexical} constructs a semantic network of word co-occurrences in the annual State of the Union address (SoU) corpus, and identifies discursive categories in political discourse through the community detection algorithm. ~\cite{jackson2019emotion} clusters the emotion colexification networks using the community detection algorithm as well.


\textbf{Classification \& Regression.} Symbol-based word representation has also been used in predictive tasks, mainly classification and regression. As for frequency and feature-based representations, they usually denote as clues for objectives in a specific class or with a particular value. ~\cite{huth2016natural} represents each word in narrative stories as a vector comprising of the numbers of co-occurrences with a set of $985$ common English words. Then it adopts regularized linear regression to predict the blood-oxygen-level-dependent (BOLD) responses for each subject when subjects listen to the narratives. In this manner, it reveals the semantic map across the cerebral cortex of humans. As for network-based representation, it allows for label or value propagation through the connections of nodes in the network, with the label indexing a singular category.  To classify a series of hashtags into two classes: ``pro-Clinton'' or ``pro-Trump'', ~\cite{bovet2018validation} constructs a network according to hashtag co-occurrences on Twitter, with several labeled hashtags as initial seed nodes. Afterwards, it spreads labels to other hashtags in the light of connections among nodes with different labels. After several iterations, it obtains a stable label for each node in the network, namely, endows each hashtag with a fitting class. 

\subsubsection{Applications of Symbol-based Sentence Representation}
Symbol-based sentence representation is widely used in prototype tasks of prediction, mainly classification, while also introduced into the tasks of similarity, description, and relation. Besides, it also can be employed in the exclusive task for text, namely language model.

\textbf{Classification.}
Symbol-based sentence representation is well received in classifying text through defined inputs. It occurs two main branches in our investigation, namely attitude classification, and content classification, which we will describe below.

As for attitude classification, it covers the classifications of sentiment, emotion, and stance, etc expressed from the text. In this branch, frequency-based and feature-based representations are often utilized jointly, while both can be used separately. Concerning frequency-based representation alone, ~\citep{eichstaedt2018facebook} leverages the unigrams and bigrams to represent posts on Facebook to predict posting users' depression status. ~\cite{green2020elusive} uses the same frequency-based representations to classify the partisanship of tweets’ authors and further examine the polarization in elite communication about the COVID-19 pandemic. With regard to feature-based representations, ~\cite{kramer2014experimental,brady2017emotion,jones2017distress,kryvasheyeu2016rapid} all adopt dictionary-based features to study the sentiment or emotion of posts on social media, with LIWC most well-known. Besides dictionary-based features, ~\cite{stella2018bots} incorporates lexical features such as emoticons and acronyms, and syntactic features such as polarity shifts due to connectors to decide the sentiment of tweets. ~\cite{catalini2015incidence} considers lexical features such as part-of-speech to represent each citation of a paper (i.e. sentences that contain the reference to another paper), and assigns the citations to two types of interest: objective and negative. When it comes to jointly use these two kinds of representations, ~\cite{del2016echo} integrates n-grams and TF-IDF, etc. (frequency-based), with emoticons, negations, and sentiment words from a predefined dictionary, etc. (feature-based) to classify the emotion of posts on Facebook. ~\cite{bovet2018validation} extracts BOW (frequency-based), hashtags, and emoticons, etc. (feature-based) to represent tweets, and investigates users' stance for Clinton and Trump in the context of $2016$ US Presidential Election.

As for content classification, it aims to classify the text according to its substantive things such as topic and meaning. ~\cite{bakshy2015exposure} applies frequency-based representations (i.e. unigrams, bigrams, trigrams) to classify news into ``hard'' (e.g. national, politic, or world affairs) or ``soft'' (e.g. sports, entertainment, or travel) content. ~\cite{alizadeh2020content} combines feature-based representations with frequency-based representations to distinguish influence operations from organic activity in social media, which contains URL and words in LIWC dictionary appeared in a tweet, in addition to unigrams and bigrams.

\textbf{Similarity.} The similarity of two sentences or documents is usually regarded as the agreement degree of their extracted predefined features, when applying symbol-based sentence representation. Researchers normally adopt features from an existing dictionary or design features from a customized vocabulary. For example, \cite{klingenstein2014civilizing} represents each trail as the probability distribution over synonym sets in Roget’s Thesaurus, and then calculates the divergence between violent and nonviolent trials using Kullback-Leibler (KL) divergence. It shows that trials for violent and nonviolent offenses become progressively distinct through analysis of $150$ year of oral testimony in the English criminal justice system. Besides, \cite{hughes2012quantitative,boyd2020narrative} studies the stylistic similarity of literature by representing each literary work as the distribution over a list of defined content-free words, while \cite{boyd2020narrative} focuses on the structural similarity but through the representation of distribution over LIWC dictionary. Except above dictionary-based sentence representation, frequency-based sentence representation can also be employed, though the dimension of the representation could be relatively large. \cite{citron2015patterns} represents each scientific article based on 7-grams occurred in the article, and investigate the text reuse in scientific corpus through calculating overlapping 7-grams between any two articles.

\textbf{Description.} 
Symbol-based sentence representation can also be used to describe the data. It generally relies on the statistics of some artificially defined features to disclose some phenomena, different from symbol-based word representation relying on frequency-based manner mostly. \cite{jordan2019examining} defines two scores of psychological processes: analytic thinking and clout in language of political leaders and cultural institutions, based on the statistics of function words in LIWC dictionary appeared in their text. It is derived from that people's thinking and attention patterns are reflected in their use of function words. Similarly, ~\cite{frank2013happiness} defines the happiness score of a tweet depending on the usage of words in the labMT dictionary. ~\cite{futrell2015large} leverages the syntactic features through calculating the dependency lengths of sentences across $37$ languages, and unearths that dependency length minimization is a universal property of languages. 

\textbf{Relation.}
Because of the intuitive and interpretable nature of the symbol-based sentence representation, it can be used with confidence to detect relationships between internal variables of sentences or with other external variables. For instance, \cite{eichstaedt2018facebook} examines the use of words from LIWC in tweets, and observes the association of these features with users' depression status who post them.

\textbf{Language model.} To alleviate the data sparsity problem caused by the exponentially many sequences, language model generally refers to the N-gram language model in a symbol-based manner. It is assumed that the probability of the word occurred after the context history can be approximated by the probability of the word occurred after the preceding $N-1$ words, namely independent of words before these $N$ words. Therefore, the calculation of language model depends on the frequency of N-gram occurring together in the corpus, which is widely used to measure the creativity and information presented in a sentence. For example, ~\cite{piantadosi2011word} quantities the information provided from a word by calculating the N-gram language model across 10 languages, and reveals that information content predicts word length better than frequency.


\subsection{Applications of Symbol-based Network Representation}
Symbol-based network representation is still the mainstream used in CSS applications. The top half of Table~\ref{tab:network_apps} lists the applications using symbol-based network representations. We split them into the representations of node and subgraph.

\subsubsection{Applications of Node-based Representation}
Symbol-based node representations except network centrality are mainly applied to the task of description/relation, where qualitative/quantitative connections between data characteristics and a specific phenomenon or property are discussed. In contrast, centrality-based representations naturally fit the ranking task.

\textbf{Description.} Some work explored node-based statistics or designed indices to describe the patterns of network data. ~\cite{grinberg2019fake} studied fake news on Twitter during the 2016 US presidential election, with the help of the co-exposure network, where nodes are news websites and edges are shared-audience relationship. They employed a number of node-based simple statistics and designed indices, mostly percentage ratios, to draw their conclusions, e.g., only 1\% of individuals accounted for 80\% of fake news source exposures. ~\cite{li2014comparative} employed simple node-based statistics such as degree distributions to describe the patterns of large mobile phone calling networks. ~\cite{dankulov2015dynamics} computed and visualize node-based temporal indices (e.g., the distributions of the interactivity time for users and tags) including complex probabilistic ones, to describe the dynamic patterns of users and tags in a Questions \& Answers system.

\textbf{Relation.} Studying the correlation between two factors or variables is the most popular task in network analysis. Regression analysis and correlation coefficients are the most used mathematical tools for quantifying the relations.

For regression analysis, ~\cite{grinberg2019fake} studied fake news on Twitter during the 2016 US presidential election. They employed a regression model to show the relation between two variables, e.g., the sharing of content from fake news sources (as a binary variable) was positively associated with tweeting about politics. ~\cite{turetsky2020psychological} studied students' peer social network and represent each student's network positions by a number of centrality-based indicators (degree, betweenness, closeness, etc) and simple statistics. They built multiple regression models between the indicators and whether a student is perturbed by a psychological intervention. As a result, they found the intervention has positive social effects. ~\cite{apicella2012social} characterized the social network of the Hadza hunter-gatherers in Tanzania, which may reveal the behaviours of early humans. They used regression analysis to evaluate the relationship between personal characteristics (sex, age, height, etc.) and degree (campmate ties and gift ties). For example, they found that taller people are more socially active and  attractive. ~\cite{boardman2012social} discussed about how genetic factors (i.e., genotypes) can be predicted based on the genotype of his/her friends as well as the environment context. They also used regression analysis to detect the relationship between node-based factors and genotypes. ~\cite{charoenwong2020social} employed regression analysis to understand the relation between social connections and the compliance with mobility restrictions under COVID-19 pandemic. ~\cite{wu2019large} focused on the citation network, and employed regression analysis to reveal the relation between the team size and a number of statistics/designed indices. For instance, disruption percentile measures whether a team search more deeply into the past, which could be disruptive to science and may succeed in the future. They concluded that ``large teams develop and small teams disrupt''.

For correlation coefficients, ~\cite{li2014comparative} computed the Spearman/Pearson correlation coefficients between two calling networks' node degree and edge weight distributions to analyze their sharing patterns. ~\cite{clauset2015systematic} studied the inequality and hierarchy in faculty hiring networks of universities. They first construct a network of institutions, where each directed edge represents a faculty member at one institution who received his/her doctorate from another. Then a prestige score for each institution is computed by node-based statistics. They showed that institutional prestige correlates well with the U.S. News \& World Report rankings, and concluded that institutional prestige leads to increased faculty production and better faculty placement. ~\cite{eom2014generalized} validated the generalized friendship paradox that your friends have on average more friends than you have in complex networks. In specific, they designed several indices as node characteristics and analyzed the degree-characteristic correlation. 

For others, ~\cite{wesolowski2012quantifying} analyzed travel networks of people and parasites between settlements and regions based on mobile phone data. They identified the relation between human travel and parasite movement mainly by visualization and simple statistics.

\textbf{Similarity.} For the similarity task, node-based representations are usually used for analyzing the strength of links or how likely a link will appear between two nodes. ~\cite{park2018strength} measured the similarity of two nodes (i.e., the strength of the edge between them) as the frequency of bidirected mentions and the total bidirected call volume in seconds. They concluded that long-range edges are nearly as strong as those within a small circle of friends. ~\cite{parkinson2018similar} studied the social network of first-year graduate students, and scanned subjects’ brains during the viewing of naturalistic movies. Through the statistics of the significance test, they showed that similar neural responses can help predict the friendship. ~\cite{boardman2012social} employed descriptive statistics to represent genetic and social factors, and demonstrated the genetic homophily (persons with the same genotype tend to be friends) by significance test. ~\cite{asikainen2020cumulative} studied the tendency of similar people to be connected to each other by choice homophily (measured by node-based designed index) and the strength of triadic closure (measured by motif-based designed index).


\textbf{Classification.} To classify a node, node-based representations are usually built by integrating the information of neighbors. ~\cite{garcia2017leaking} predicted the hidden profiles (i.e., sexual orientation and relationship status) of nonusers given the profiles of disclosing users. They formalized the problem as binary classification, and simply averaged the profiles of a nonuser's friends as the node representation for prediction. ~\cite{massucci2016inferring} proposed to infer the propagation paths of perturbations in a network (e.g., the spread of epidemics). They also treated the problem as binary classification and used a probabilistic model to estimate the probability of each unobserved node being perturbed given its neighbors.

\textbf{Regression.} Though regression analysis is widely used in CSS for detecting the correlations, only a few work targets on the regression problem. ~\cite{ganin2017resilience} studied the efficiency and resilience of transportation networks, where intersections are mapped to nodes and road segments between the intersections are mapped to links. They designed node-based indices to model the commuter flows, and constructed a regression model to estimate travel delays in 20 different urban areas, with another 20 areas for calibration. ~\cite{teng2016collective} first built a probabilistic information spreading model to characterize the behaviours of nodes and estimate the collective influence of multiple spreaders. Then they will identify the most influential spreaders that maximize the influence.

\textbf{Ranking.} Centrality indicators perfectly suit the ranking task, where most work aims at finding the most important or influential nodes in a network. ~\cite{pei2014searching} utilized various network centrality coefficients to detect the most influential information spreaders in online social networks. 
To study the cultural history and discover cultural centers, ~\cite{schich2014network} constructed a directed network of cities in Europe and North America based on migration, where the endpoints of each edge in the network represent the birth and death locations of a notable individual. Then they used PageRank centrality to identify the most influential cities. ~\cite{manrique2016women} investigated the online network of ISIS (Islamic State) members. With the help of centrality indicators, they found that although men dominate numerically, women emerge with superior network connectivity that can benefit the underlying system’s robustness and survival. ~\cite{hart2017effects} built a similarity network of 200 Iroquoian village sites dating from A.D. 1350 to 1600, and concluded the importance of a specific location in population dispersal. ~\cite{fraiberger2018quantifying} investigated the exhibition history of half a million artists, constructing the coexhibition network that captures the movement of art between institutions. Centrality is further employed to capture institutional prestige and help understanding the career trajectory of individual artists. ~\cite{bruch2018aspirational} identified the most desirable users in an online dating network by PageRank centrality. Then they conducted analysis on users' strategies, e.g., both men and women pursue partners who are on average about 25\% more desirable than themselves. ~\cite{medo2016identification} developed a probabilistic model to find out the discovers who are repeatedly and persistently among the first to collect the items that later become hugely popular. They also showed that traditional centrality indicators fail in this scenario.

\subsubsection{Applications of Subgraph-based Representation}
Subgraph-based representation can be further divided into motif-based and cluster-based ones. Generally, subgraph-based representation is less popular than node-based representation in terms of both paper number and task coverage.


\textbf{Relation.} Both cluster-based and motif-based representations are utilized in relation analysis. ~\cite{trujillo2018document} focused on document co-citation analysis and used $\chi^2$ test to validate whether subject communities are related to co-citation communities. In other words, $\chi^2$ test measures the correlation between two community assignments. ~\cite{kovanen2013temporal} studied the tendency of similar individuals to participate in communications. Besides similarity analysis, they also investigate how different representations correlate, e.g., edge weights and motif counts.

\textbf{Similarity.} Motif-based representations are still used for the similarity analysis between two nodes, while cluster-based ones are used for characterizing more high-level similarities, such as the similarity of two networks. ~\cite{kovanen2013temporal} focused on the tendency of similar individuals to participate in communications by calculating the ratio score of temporal motifs (e.g., repeated call, returned call, chains, etc.). ~\cite{asikainen2020cumulative} studied the tendency of similar people to be connected to each other by choice homophily (measured by node-based designed index) and the strength of triadic closure (measured by motif-based designed index). To understand the universality and diversity in how humans understand and experience emotion, ~\cite{jackson2019emotion} built a network of emotion concepts (e.g. ``angry'' and ``fear'') for each of 2,474 spoken languages, where two concepts are connected if their meanings appear in the same word. Then they used adjusted Rand indices (ARIs), which measures the alignment of two cluster assignments, to quantify the similarity of two networks (i.e. languages).

\textbf{Clustering.} Cluster-based coefficients naturally fits the need of clustering. But most work only used the clustering of a network as their intermediate products. Therefore, they did not develop their own clustering algorithms, but directly employed traditional community detection methods instead. Thus we only present one example work here. Modularity is defined as the number of edges within given clusters minus the expected number in a network with edges placed at random, and can characterize to what extent a network can be divided into clusters. ~\cite{expert2011uncovering} specialized the modularity to spatial networks (e.g., road networks and location-based social networks), in order to discover space-independent communities.



\textbf{Ranking.} Most work studied the ranking of nodes in a network, and thus cluster-based representations are rarely used. ~\cite{waniek2018hiding} studied an interesting problem: can individuals or groups actively manage their connections to evade social network analysis tools? Here each node's importance is measured by centrality, and each community's concealment is measured by a manually designed cluster-based index. They showed that simple heuristic strategies are effective to hide from the above measurements.

\subsection{Applications of Embedding-based Text Representation}
\label{sec:emb_text}
With the rapid development of natural language processing and deep learning, embedding-based text representation receives increasing attention from social scientists and computational scientists. In the following subsections, we will describe the applications employing embedding-based word and sentence representations and present them according to their formalized tasks. The bottom half of the Table~\ref{tab:text_apps} lists the sorted applications using embedding-based text representations. 

\subsubsection{Applications of Embedding-based Word Representation}
Owing to the excellent performance in capturing the semantic relation, embedding-based word representation is popularly introduced into the task of similarity. 

\textbf{Similarity.} 
Different from symbol-based word representation, the semantic similarity of words is reflected by the distance of word embeddings in the vector space, not based on symbol matching, so it can be used to measure the similarity of the abstract concepts. For example, ~\cite{garg2018word,caliskan2017semantics} both compute the average distance between word embeddings of words denoting genders and a series of words indicating occupations, and view the difference between men and women as the indicator of occupational stereotypes. They compare the occupational bias reflected in the embeddings with occupation participation rates and stereotypes investigated in the traditional survey, and identify a strong association between them. Besides, it can also be utilized to expand words outside of our knowledge with similar semantics. ~\cite{sivak2019parents} uses word embeddings to detect the similar words of ``son'' and ``daughter'', and investigates public mentions of them on social media. It finds that both men and women mention sons more frequently than daughters in their posts, which reveals that gender inequality may start early in life.



\subsubsection{Applications of Embedding-based Sentence Representation}
Sentence representations obtained from topic models and neural network models are quite different in learning mechanisms and applied tasks, though both representations are based on embeddings. Therefore, we will present the applications of these two types of sentence representations separately.

For topic models, although each of its dimensions is still unintelligible, we can infer the meaning of each dimension of the representation by its probability distribution over the word list and further artificially define it as a specific topic of the text. 
Therefore, it has been used in various tasks of similarity, clustering, classification, description, and relation. 

\textbf{Similarity.}
Topic models represent a sentence as a distribution over a series of topics, so researchers can measure the similarity of text in semantic topics. ~\cite{Farrell2016Network} investigates the similarity of contrarian organizations' text and text from media and politics in the climate change counter-movement by LSA model, and finds growth in the semantic similarity between them from 1993 to 2013. ~\cite{bokanyi2016race} also applies LSA to study the language use patterns of counties in the USA, and mines the similarity between these counties in the semantic space. To measure the linguistic distinctiveness of the context where the child produces a word, ~\cite{roy2015predicting} utilizes LDA model to extract the topic distribution for each first appeared word, and uses KL-divergence to compare it with the background topic distribution.

\textbf{Clustering.}
Based on the topic model-based representations, we can cluster these sentences based on topics. ~\cite{2014Quantifying} clusters Wikipedia into $100$ different semantic topics by LDA model, and quantifies the search volume of these topics in Google search engine before stock market moves.
~\cite{farrell2016corporate} uses STM to obtain representations for written and verbal texts produced by individuals and organizations participating in climate change counter-movement, and clusters them into $30$ topics such as ``CO2 is Good'' and ``Energy Production''. Based on the clustering results, it reveals that corporate funding influences the written and disseminated texts of these organizations.

\textbf{Classification.}
Topic model-based representation is often operated as one of the features for the classification task, since it can supply the semantic information of text. For instance, ~\cite{jaidka2020estimating} leverages the representation learned from LDA model to predict the subjective well-being from Twitter. Besides,  ~\cite{eichstaedt2018facebook} also uses LDA to represent posts on Facebook and predict the depression of users. 

\textbf{Description.}
Since each dimension's meaning of the topic model-based representation can be inferred to some extent, it can facilitate the semantic description of the text. 
~\cite{lupia2020does} applies STM to model the statements mentioned NSF in the Congressional Record to find the distinctive topics of Democrats and Republicans, e.g. Democrats care about technology and education more than Republicans. 
~\cite{gerow2018measuring} defines the discursive influence of scholarly articles the extent to which they shape the future discourse and uses the topic model to describe the influence. In other words, it estimates an article's influence as the divergence between topic distributions learned with and without this article. 

\textbf{Relation.}
Sentence representations learned from topic models have also been exploited to assess the relationship between different variables. ~\cite{bokanyi2016race} studies the relation of regional patterns in language use and socioeconomic and cultural status of counties in the USA, such as ethnicity and tourism, where the regional pattern in language use is represented through the LSA model.
 
For neural network models, due to their powerful ability to fit data and capture deep semantics, neural-based representations have been gradually introduced into classification and similarity tasks in CSS. Besides, neural-based representations are also skilled at the task of language model. Below we will introduce each of them respectively.

\textbf{Classification.} 
Neural-based sentence representation is mainly applied to the classification of abstract concepts or objects of which the feature definition needs hard human efforts. As for abstract concepts, ~\cite{mooijman2018moralization} uses the LSTM neural network to automatically predict moral values involved in Twitter posts and suggests an association between moralization and protest violence. The complexity and ambiguity of human languages are also predicted by the LSTM neural network when investigating the languages' efficiency~\cite{hahn2020universals}.
As for objects with hard feature definition, ~\cite{fetaya2020restoration} also applies the LSTM neural network to predict the missing Babylonian text, of which the restorations require extensive expert knowledge of each genre and a large corpus of texts. 


\textbf{Similarity.}
The similarity task can be tackled by measuring the distance or similarity of neural-based sentence representations in embedding space. ~\cite{sheshadri2019public} utilizes the Paragraph Vector model to obtain the representations of news articles first and uses cosine similarity to measure the similarity between these articles in hyperconcentrated news periods. It further demonstrates that high similarity between articles Granger causes (G-causes)~\cite{granger1969investigating} public attention changes and legislation. 

\textbf{Language model.}
Owing to the strength in fitting text, neural-based representations behave excellently in the language model task. LSTM neural networks are applied to construct the language model, which is a general and solid indication of language's surprisal and complexity~\cite{hahn2020universals}.

\subsection{Applications of Embedding-based Network Representation}
Lower half of Table~\ref{tab:network_apps} lists the applications using embedding-based network representations. Since embedding-based methods are still undergoing the emergence period in CSS, especially in the analysis of network data, only a few works on Nature, Science and PNAS adopted embedding-based representations. Hence we also add a couple of works from WWW in this subsection.




\textbf{Similarity.} Both two work on similarity analysis studied the user-user friendship in a social network. ~\cite{yang2011like} applied matrix factorization to the social network including user-user friendship network and bipartite user-item interaction network for learning user and item embeddings, which were employed for friend and item recommendations. ~\cite{yang2019revisiting} characterized a location-based social network containing both user mobility data and the corresponding social network as a hypergraph where a friendship is represented by an edge between two user nodes and a check-in is represented by a hyperedge among four nodes (a user, an activity type, a timestamp and a POI). Network embedding methods are then employed for both friendship and location predictions.

\textbf{Clustering.} ~\cite{sachan2012using} employed topic model to build the relationship of user, community, and topic. Then they solved the optimization and computed the community distribution of each user in the social network. Note that topic models can be seen as a special kind of matrix factorization.

\textbf{Classification \& Regression.} Embedding-based methods are widely used for the classification and regression tasks in computer science area. Besides the above mentioned methods~\cite{yang2011like,yang2019revisiting}, ~\cite{kosinski2013private} applied singular value decomposition to the user-like matrix for learning user embeddings, which were further utilized for predicting private traits. ~\cite{zhang2017regions} constructed a heterogeneous network with three types of nodes, i.e. location, time and text, from geo-tagged social media (GTSM) data. Then they jointly encoded all spatial, temporal, and textual units into the same embedding space to capture the correlations for modeling people’s activities in the urban space. More recently, graph neural network-based methods~\cite{fan2019graph,wu2019dual} were also proposed for social recommendation, where a user friendship network and a user-item interaction network are given as input to predict future user-item interactions.

\textbf{Ranking.} For the ranking task, ~\cite{wang2012magnet} aimed at discovering magnet communities, which are communities that attract significantly more people’s interests. In detail, ~\cite{wang2012magnet} learned cluster-based representations via matrix-based optimization, and ranked given communities in a domain based on their attractiveness to people among the communities of that domain.

\section{From Symbols to Embeddings}
\begin{figure}[htb]
	\centering
	\includegraphics[width=0.98\linewidth]{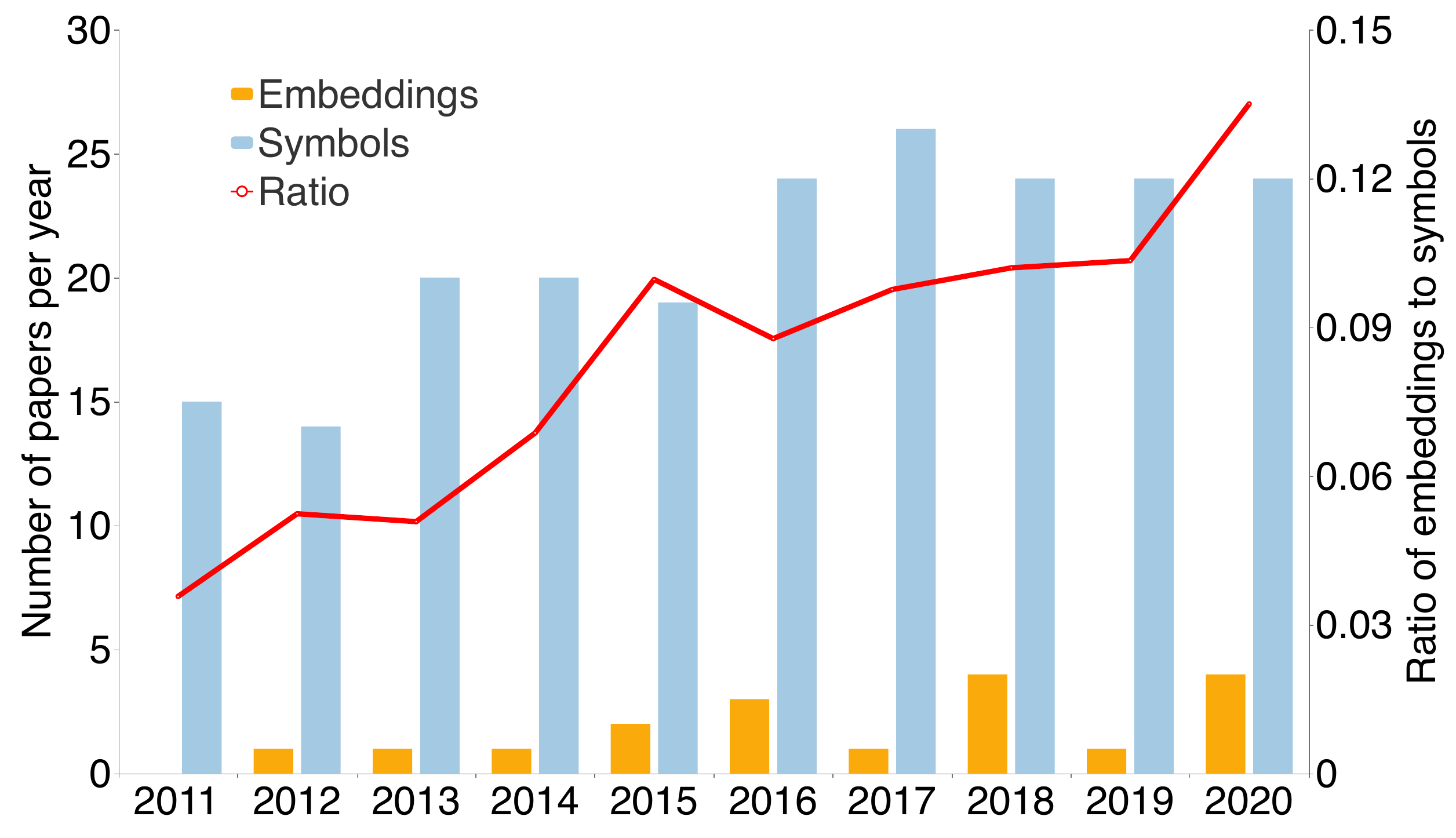}
	\caption{Number of papers applying symbol-based representation or embedding-based representation and the ratio between them over the last decade in Nature, Science, and PNAS. The line is smoothed by taking a 3-year average. Detailed settings where we selected these papers are shown in Section~\ref{sec:app}. } 
	\label{fig:papernum_nsp}
\end{figure}

\begin{figure}[htb]
	\centering
	\includegraphics[width=0.98\linewidth]{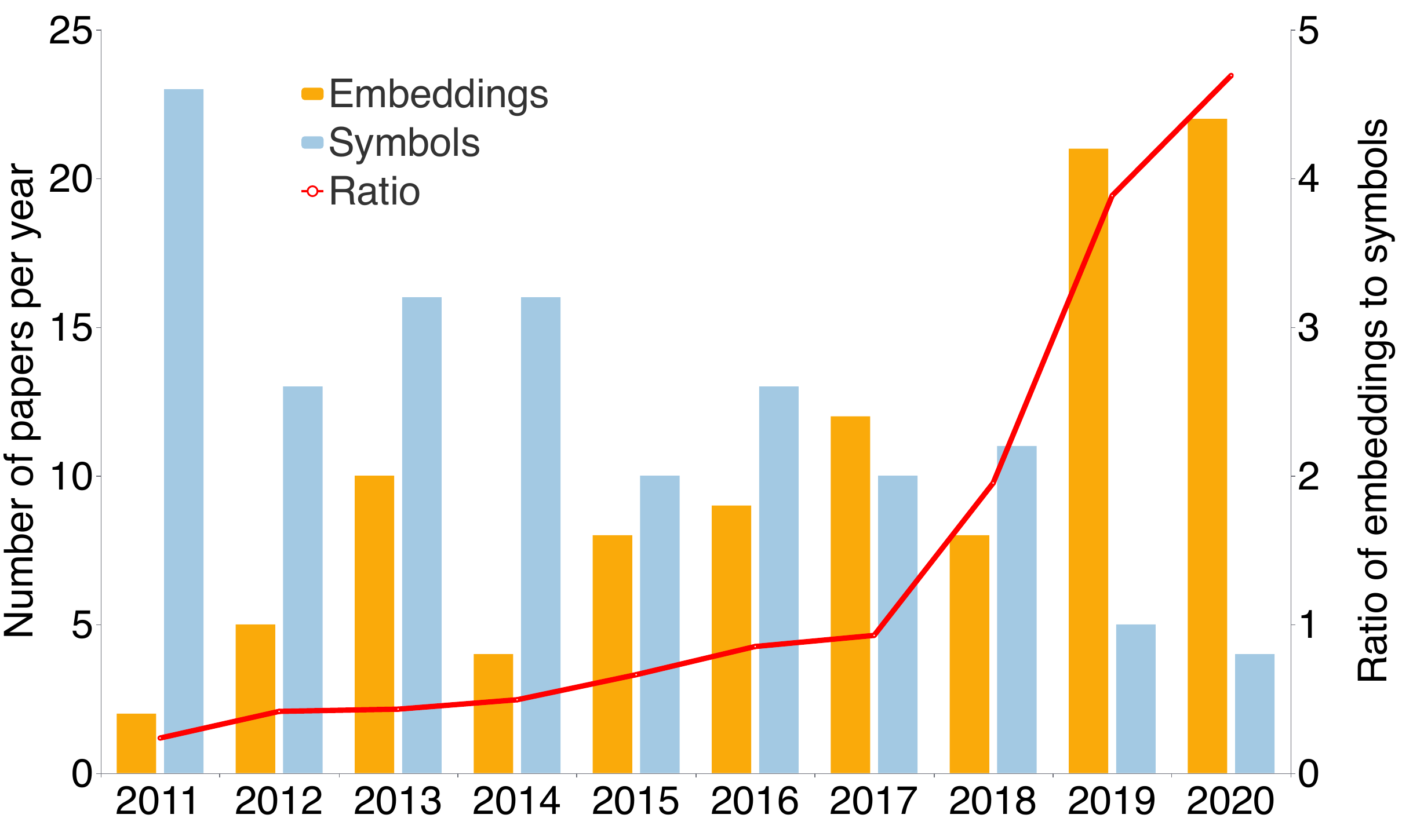}
	\caption{Number of papers applying symbol-based representation or embedding-based representation and the ratio between them over the last decade in ACL, WWW, and KDD. The line is smoothed by taking a 3-year average. Detailed settings where we selected these papers are shown in Section~\ref{sec:app}.} 
	\label{fig:papernum_awk}
\end{figure}

\noindent
\label{sec:etos}
Based on the introduction of applications in the previous section, we can observe that both symbol-based and embedding-based representations have been considerably adopted in CSS. To investigate their coverage definitely, we count the number of works utilizing one or both of the two representations each year, as shown in Fig.~\ref{fig:papernum_nsp}. By comparisons, we can find that the proportion of articles using embedding-based representations is gradually increasing over the last decade in Nature, Science, and PNAS. This indicates that more and more works in CSS have considered and benefited from the embedding-based representations. We also make the same statistics in conferences of ACL, WWW, and KDD. Fig.~\ref{fig:papernum_awk} shows the comparison between the numbers of applications using symbol-based and embedding-based representations in these three conferences. From the figure, we can find that the number of articles using embedding-based representations has significantly exceeded those using symbol-based representations. However, compared with Fig.~\ref{fig:papernum_nsp}, there is a large gap between the volume of embedding-based representations in computer science conferences and the three multidisciplinary journals. This prompts us to deepen and amplify the interdisciplinary integration between social science and computer science, despite the slight shift in their research concerns.

\textbf{To sum up, embedding-based representations have emerged and performed an increasingly critical role in CSS over the last decade.}

We further discuss the underlying reasons for this trend and summarize the expert areas of both representations. Based on their internal mechanisms and existing applications, we conclude three key points as follows.


\textbf{Symbol-based representations excel at the tasks of description and relation}, due to their explicitness and interpretability. Each value in the symbol-based representation denotes certain and human-readable meaning, so we can use it directly to observe the distribution of data, as well as to extract relations between objects. For example, as we introduced in Section~\ref{sec:symbol_app}, frequency-based word representations are applied to observe cultural changes and capture the relationship between the number of mentions in news and the stock trading volume of a company. While topic model-based representations and some neural-based representations are equipped with practical meanings to some degree~\cite{zhang2018visual,belinkov2019analysis}, they are still fuzzy and less compelling for researchers in social science. 

\textbf{Embedding-based representations perform better in the tasks of prediction (e.g. classification and regression) and similarity}, owing to the powerful ability of neural networks to fit the data and to extract deep semantics. On the one hand, neural networks achieve efficient input-output mapping functions through the connections of large-scale neurons. On the other hand, it realizes the extraction of deep semantics and abstract concepts by the constructions of multi-layer networks. Existing researches have demonstrated that the deep layer captures the more abstract features relative to the shallow one~\cite{zeiler2014visualizing}. As presented in Section~\ref{sec:emb_text}, abstract concepts such as social biases and moralizations are all well measured by embedding-based representations. Although we mentioned that symbol-based representations can stand for abstract concepts through some defined symbols, such representations are still partial and shallow, and hard to capture their full picture.


\textbf{Embedding-based representations require fewer human efforts}. Symbol-based representations usually require a large amount of expert knowledge to define the features of research objects, which is labor-intensive. Besides, for some abstract concepts or objects without well-founded features, their performances will be limited. Different from them, embedding-based representations are automatically extracted from data, with few human interventions and even complements for human knowledge. For example, as introduced in the application section, we can use neural networks to automatically restore the missing Babylonian text, which is challenging even for experts. In addition, embedding-based representations are qualified to portray the complexity and ambiguity of the language without manual definition.

\section{Discussions on Future Directions}
\label{sec:future}
\noindent
Although the tendency from symbols to embeddings has emerged in the past ten years, there are still many challenges and open issues to be explored. Going forward, we list some essential and potential future directions involved with data representations in CSS.


\textbf{Pre-trained language models.} In recent years, pre-trained language models have received considerable attention and achieved great success in processing textual data~\cite{devlin2018bert,liu2019roberta}. The models learn rich semantic information from massive textual data such as encyclopedias and books, with merely fine-tuned in downstream tasks to achieve efficient embedding-based representations. Therefore, for CSS, we can obtain more generalized and robust textual representations with the aid of pre-trained language models. The representations can not only be used to analyze social phenomena from text more extensively and accurately, but also reduce the manual annotations for those tasks requiring enormous labeled data, compared to representations learned from traditional neural network models.

\textbf{Graph neural networks.} Through the message passing mechanism, graph neural networks~\cite{zhou2020graph} can effectively model both the network topology and node/edge features (e.g., text information) simultaneously, thus providing a unified framework to take advantage of information from heterogeneous sources. Many scenarios in CSS need to deal with a social network as well as individual characteristics. Therefore, graph neural network techniques have great application potentialities for CSS studies, which can learn representations integrating the information of both text and network. In fact, various applications in computer science, such as natural language processing~\cite{wu2021deep} and recommendation systems~\cite{ying2018graph}, have already adopted graph neural networks for modeling.

\textbf{Design as prediction and similarity.}
Embedding-based representations are well-known for rich and deep semantics, while symbol-based representations are usually preserved in partial and shallow semantics. Meanwhile, embedding-based representations are skilled at the task of prediction and similarity. Therefore, to take full advantage of the strong semantics in embeddings, researchers in CSS are encouraged to design the research problem as a prediction or similarity task whenever possible. For example, we can design the problem of social bias as a similarity measurement between the embeddings of gender words and neutral words~\cite{caliskan2017semantics,garg2018word}. In addition, the complexity of human language can be designed as a predictive task, which views the predicted probability of a word or sentence using language model as the indicator~\cite{hahn2020universals}.
 


\textbf{Interpretability.} Admittedly, a drawback of embedding-based methods is the lack of interpretability. This problem would harm the application for decision-critical systems related to ethics, safety, or privacy. Though the interpretability of embedding models, especially neural network models, has not been fully addressed yet, researchers in the computer science area have made some efforts towards better explainability of neural-based models~\cite{arrieta2020explainable}. Therefore, taking advantage of both embedding-based models and explainability analysis methods for effective and (partially) explainable predictions would be an intriguing direction.

\noindent
\section{Conclusion}
\label{sec:conclusion}
\noindent
As an emerging and promising inter-disciplinary field, computational social science has attracted considerable research interests over recent years.  Two main types of data, namely text and network data, are widely used in studies of CSS. In this survey, we first summarize the data representation into symbol-based and embedding-based representations and further introduce typical methods when constructing these representations. Afterwards, we conduct a comprehensive review on the applications of these two classes of representations based on more than $400$ top-cited literature from $6$ classic journals and conferences. According to the statistics of these applications, a tendency that embedding-based representations of text and network in CSS are emerging and growing is discovered, which we further discuss the reason contributed to. Finally, we suggest four challenges and open issues in CSS, which are essential and potential directions to be explored.

\zihao{5-}
\noindent
\renewcommand\refname{\zihao{5}\textbf{References}}

\bibliographystyle{plain}
\bibliography{JSC}

\appendix

\section{Appendix}

With the explosive growth in research topics in CSS, we divide the topics of applications we investigated into $9$ domains, which is inspired from the $5$ primary sub-disciplines in traditional social science, namely sociology, anthropology, psychology, politics, economics,  and other fields of humanities including linguistics, communication, geography, and environment. Distribution of these applications are listed in Tab.~\ref{tab:domain-distribution-table}. Note that each work can exist in multiple domains if relevant simultaneously. We also list all relevant papers on the GitHub link: \url{https://github.com/thunlp/CSSReview}.

In the following, we will divide these applications into different domains, and introduce them from text to data according to the data type used, and further present them from symbol-based representation to embedding-based representation according to the representation type used.
\subsection{Text}

\subsubsection{Symbol-based representation}

Symbol-based representation of text is mostly used in the fields of sociology, followed by linguistics, psychology, geography, politics, and communication respectively, with few utilized in economics, environment, and anthropology. 

In the domain of \textbf{Sociology},  a series of works utilize the symbol-based representation of text to analyze and detect misinformation and misbehaviour in online world, such as rumor~\cite{boididou2014challenges,zhao2015enquiring,kumar2016disinformation,popat2017truth,yang2012automatic}, fake news~\cite{bhatt2018combining} or image~\cite{gupta2013faking}, low quality wikipedia~\cite{flekova2014makes}, hate speech~\cite{badjatiya2017deep,zannettou2018gab}, abusive language and behaviour~\cite{nobata2016abusive,chatzakou2017measuring}, social bots~\cite{davis2016botornot,stella2018bots}, sockpuppets~\cite{kumar2017army}, cybercriminal activity~\cite{portnoff2017tools}, influence operations~\cite{alizadeh2020content}, text reuse in scientific papers~\cite{citron2015patterns}, where they usually manipulate N-gram, BOW, TF-IDF features, as well as linguistics features such as length, URL, hashtag in the text, accompanied with syntactic features such as part-of-speech tagging and dependency relations. Extra lexicon such as LIWC, is also widely used to extract keywords in the above analysis and detection works. In addition, the privacy issue is a hot topic where researchers use the symbol-based text features to prevent privacy disclosure across multiple online sites~\cite{goga2013exploiting,jain2013seek}. These text features also benefit the search for informative posts~\cite{imran2013practical,imran2014aidr} and assessment of damage~\cite{cresci2015linguistically} in a disaster, while they can also contribute to the social power relation prediction~\cite{bramsen2011extracting}. Different from the above studies, a list of works aims to explore the behaviour law of human in online social media based on the symbol representations of words, for instance, hashtag adoption~\cite{yang2012we} and collective attention on Twitter~\cite{lehmann2012dynamical}, pursuit in online dating markets~\cite{bruch2018aspirational}, user feedback in application store~\cite{fu2013people}.

\begin{figure}[t!]
	\centering
	\includegraphics[width=0.98\linewidth]{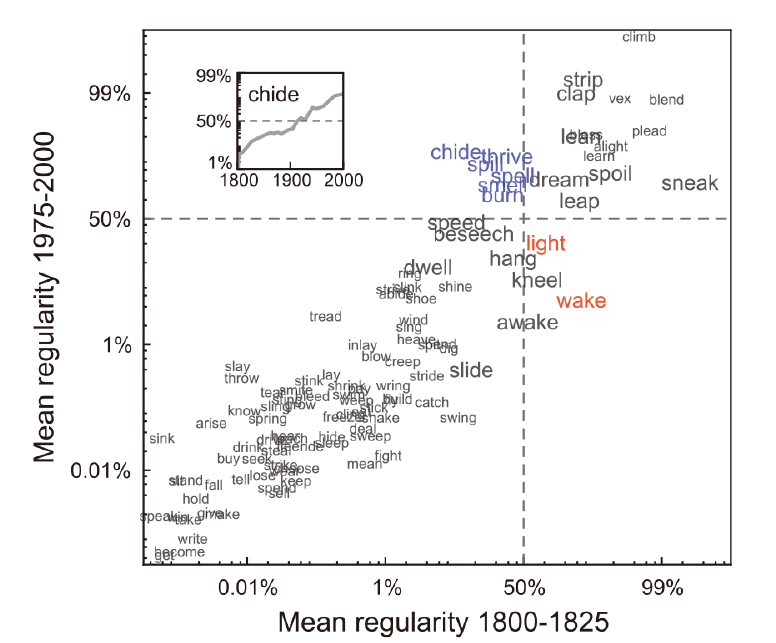}
	\caption{Scatterplot of irregular verbs in \cite{michel2011quantitative}, which locate in the position with its regularity in the early 19th century as \textit{x} coordinate and regularity in the late 20th century as \textit{y} coordinate.}
	\label{fig_lings-exap}
\end{figure}

In the domain of \textbf{Linguistics}, a set of works focus on the study of linguistic phenomenon and trend. Some researchers count the frequencies of linguistic features, such as N-gram and emoticon, to investigate linguistic phenomenon including the evolution of grammar~\cite{michel2011quantitative}, and the correlation with socio-economic variables~\cite{hovy2015user}. Taking \cite{michel2011quantitative} as an example, as shown in Fig.~\ref{fig_lings-exap}, it counts the regular forms (added ``-ed") and irregular forms (conjugated extraordinarily) of verbs from 1800 to 2000, such as ``strived" and ``strove" of ``strive".  Through the quantitative analysis, it finds the linguistic fact that irregulars generally yield to regulars, with $16\%$ of irregulars changed into regularity of more than $10\%$. Besides the linguistic trend, N-gram language model is used to approximate the information content of each word~\cite{piantadosi2011word} or distinctiveness of language~\cite{danescu-niculescu-mizil-etal-2013-computational,danescu2012you}.
Another set of works concentrate on the text analysis of various genres, such as debate and narrative. \cite{boyd2020narrative} count the function and cognitive words across each text with LIWC, to analyze the structures of narratives in different types. \cite{jordan2019examining} also use the LIWC to measure analytic thinking and clout in leaders' debates and speeches and find a general decline in analytic thinking and a rise in confidence. In addition, designed linguistic lexicons, accompanied with semantic and syntactic features, are also popularly adopted in language quality detection, such as the detection of politeness~\cite{danescu-niculescu-mizil-etal-2013-computational}, popularity~\cite{tan-etal-2014-effect}, and biased statements~\cite{hube2018detecting}.

In the domain of \textbf{Psychology}, dictionary-driven text representations are widely utilized, with LIWC and Language Assessment by Mechanical Turk (LabMT)~\cite{dodds2015human} as mostly popular dictionaries. ~\cite{kramer2014experimental} uses LIWC to define the emotion of posts and finds the emotional contagion through social networks. ~\cite{frank2013happiness} employs the LabMT to measure the happiness expressed in language and discovers that happiness increases with distance from people's average location. Moral Foundation Dictionary is also incorporated to assist the prediction of moral values involved in Twitter posts~\cite{mooijman2018moralization}. Besides emotion and happiness, dictionary-driven representations are also extensively used to detect depression in social media~\cite{eom2014generalized,chen2018mood}. Despite the wide adoption of dictionary-driven representations, ~\cite{jaidka2020estimating} makes a comparison between unsupervised dictionary-driven and supervised data-driven methods, and verifies that the latter is more robust for well-being estimation from social media data. Therefore, outside of the dictionary-driven representations, linguistic features such as N-gram, BOW, are also applied to represent text in psychology, combined with the supervised machine learning method. For instance, ~\cite{chang2020don} uses them to distinguish a person's intention and others’ perception of the same utterance, while ~\cite{kern2019social} take them as signals for personality prediction.

In the domain of \textbf{Geography}, a set of studies focus on geo-location inference, in which case the location where a textual message is generated is discovered ~\cite{ikawa2012location,ryoo2014inferring,wing2011simple}, and route navigation, with the aim to provide a more promising route according to sentiments detected from geo-tagged documents in social media ~\cite{kim2014socroutes}. As for geo-location inference, ~\cite{ikawa2012location} proposed a method to learn associations between a location and its pertinent keywords extracted from historical messages, while \cite{ryoo2014inferring} extracted the spatial correlation between texts and GPS locations from tweets with GPS-tags. Besides, ~\cite{wing2011simple} adopted simple supervised approaches on the textual content of documents as well as a geodesic grid, so as to acquire the discrete representation of the earth's surface. With regard to route navigation, \cite{wing2011simple} presented a system to recommend routes based on sentiments exposed from Twitter tweets towards places, by combining eight existing sentiment analysis tools, including LIWC, Happiness Index, SentiWordNet, SASA, PANAS-t, Emoticons, SenticNet, and SentiStrength.



In the domain of \textbf{Politics}, most studies focus on investigating political activities and analyzing ideology applying symbol-based text representations. As regards political activities, \cite{alizadeh2020content} utilizes series of defined features such as N-gram, URL, and LIWC, to predict social media influence operations, while \cite{lupia2020does} extracts the most distinguished words and sentiment words from statements in the Congressional Record, to explore the Congressional concern about National Science Foundation. Besides,  \cite{jordan2019examining} uses LIWC lexicon to analyze the style of political leader's language and further discusses the long-evolving political trends. With regarding to ideology analysis,  \cite{preoctiuc2017beyond} and \cite{bovet2018validation} use similar symbol-based features as N-gram, URL, emoticon, to predict political ideology and opinion toward presidential candidates of Twitter users, while \cite{burfoot2011collective} aims to predict sentiments in Congressional floor-debate transcripts with unigram features.

In the domain of \textbf{Communication}, symbol-based text representation is employed to mining the content in communication. ~\cite{jenders2013analyzing} takes hashtags, mentions and sentiments as symbol features to predict viral tweets. ~\cite{sheshadri2019public} utilizes N-gram features to analyze the news framing and explore its  public and legislative impact, while ~\cite{green2020elusive} adopts similar features to represent tweets sent by political elites and further analyze the polarization in elite communication on the COVID-19 pandemic.  ~\cite{arous2020opencrowd} calculates TF-IDF scores of words as tweet features to help the detection of social influencers in communication.

In the domain of \textbf{Economics}, researchers were greatly interested in revealing economical phenomenon based on the relationship between financial news and the stock market, using text-based correlational analyses ~\cite{alanyali2013quantifying} and the combination of several basic linguistic features ~\cite{xie2013semantic}. Specifically, ~\cite{alanyali2013quantifying} adopted correlational analyses, according to daily number of mentions in the Financial Times for each company of interest, for the purpose of quantifying the relationship between decisions made in stock market and situation in financial news.
~\cite{xie2013semantic} utilized scores for words in the Dictionary of Affect in Language (DAL) ~\cite{agarwal2011sentiment} via part-of-speech, along with bag-of-words in order to predict change in stock price according to financial news. 

In the domain of \textbf{Environment}, the main research interest lies in analyzing social media text data generated before, during and after the occurrences of natural disasters, such as earthquake, hurricane and etc. In ~\cite{kryvasheyeu2016rapid}, LIWC ~\cite{pennebaker2001linguistic} and SentiStrength ~\cite{thelwall2010sentiment} were adopted for analyzing the sentiments embedded in social media texts, posted before, during and after Hurricane Sandy, in order to investigate if the sentiment signal indicated the damage inflicted by the hurricane. Besides, ~\cite{ghosh2018class} proposed modified TF-IDF based approaches to better classify disaster related social media tweets so that the rescue and relief operations can be better launched when natural disasters occur.

In the domain of \textbf{Anthropology}, studies related to cultural evolution served as the major interests of researchers. Specifically, two kinds of cultural shifts were studied, namely the cultural changes accompanying the monopolization of violence by the state ~\cite{klingenstein2014civilizing} and the cultural universality and diversity in music~\cite{mehr2019universality}. In particular, ~\cite{klingenstein2014civilizing} applied a bag-of-words model as a symbol-based representation of texts to coarsely categorize the words that occur in jury trials into several predefined classes and further analyze the extent to which the patterns of talking in a criminal trail varied from violent to nonviolent offenses and how these differences evolved over time. \cite{mehr2019universality} conducted a systematic analysis regarding the features of worldwide vocal music, where four kinds of representations were derived for each song. Using machine classifiers, they managed to observe the universality and variability in musical behaviour, reflecting cultural evolution in forms of music.




\subsubsection{Embedding-based Representation}
\noindent

Embedding-based representation of text mostly benefits the sociology, then geography, politics, psychology, environment, economics, and linguistics successively, with few adopted in communication and anthropology, as shown in Tab.\ref{tab:domain-distribution-table}. 

\begin{figure}[t!]
	\centering
	\includegraphics[width=0.98\linewidth]{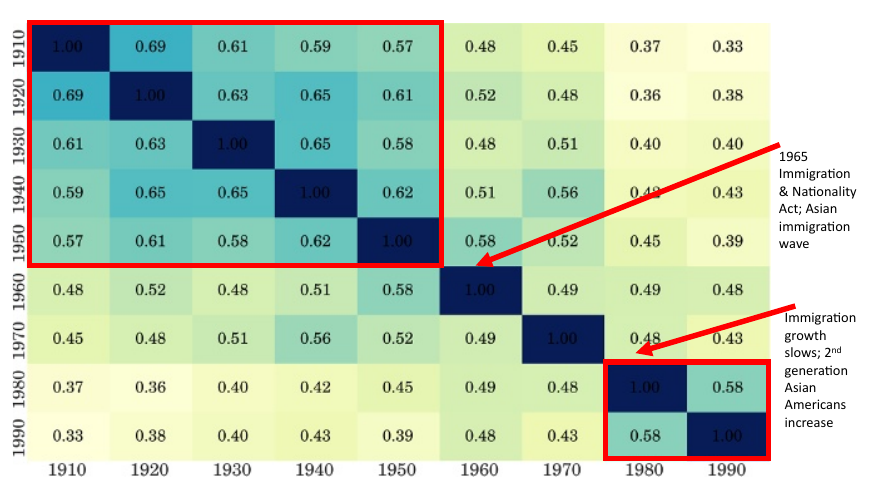}
	\caption{Pearson correlation of Asian bias scores between each decade in the 20th century~\cite{sivak2019parents}.} 
	\label{fig_bias-exap}
\end{figure}

In the domain of \textbf{Sociology}, embedding-based text representation is mostly adopted in content mining, misinformation, and misbehavior detection, as well as human trait prediction. As for content mining, the topic model is widely used. \cite{singer2017we} adopts it to extract the topic in Wikipedia and is eager to understand why we read Wikipedia. \cite{fu2013people} and \cite{sachan2012using} uses it to analyze content users writing and discovers users' preferences and interests, while \cite{weerasinghe2020pod} uses it to mining underlying topics of comments by pods, aiming to increase the popularity of user content effectively. Further, \cite{gerow2018measuring} builds a dynamic topic model to measure how the content shapes future scholarship, namely its discursive influence of a paper across scholarship. \cite{zhang2017regions} and \cite{wang2012historical} incorporate extra data outside the text such as region and time, to uncover the spatial and temporal topics. As for misinformation and misbehavior detection, word embedding methods and deep neural networks are widely used in this direction. Word embedding methods such as Skip-Gram and GLOVE, are commonly used in social bias detection~\cite{caliskan2017semantics,garg2018word,sivak2019parents}, e.g. gender bias and ethnic bias. For example, ~\cite{sivak2019parents} computes the average distance between word embeddings of last names in various groups and a series of adjectives, and views the difference in distance between the common group and Asian group as the score for Asian bias. As shown in Fig.~\ref{fig_bias-exap}, there are two-phase shifts in Asian bias with each correlated with the increase of Asian immigration into the United States in the 1960s and the appearance of the second-generation Asian-American in the 1980s, respectively. Besides bias detection, rumor and fake news detection are also hot topics employing embedding-based representation. They usually use RNN models as basic frameworks to encode the text representation~\cite{ma2018detect,bhatt2018combining}, with VAE~\cite{khattar2019mvae}, GAN~\cite{ma2019detect}, and Bayesian model~\cite{zhang2019reply} further improving the performance. Detection of other misinformation and misbehavior such as toxicity triggers\cite{almerekhi2020these}, abuse language~\cite{nobata2016abusive} and hate speech detection~\cite{badjatiya2017deep}, apply the deep neural networks to obtain the text representation as well. As for the human trait prediction, researchers endeavor to use LSTM to predict human age and gender~\cite{wang2019demographic}, as well as activity~\cite{wilson-mihalcea-2019-predicting}, while ~\cite{pan2019twitter} learns the representation of bios of each user with GCN to predict the user's occupation. 

In the domain of \textbf{Geography}, embedding representation of texts was applied in multiple application scenarios using geo-tagged social media data, including geo-location estimation ~\cite{ahmed2013hierarchical, miura2017unifying}, geographical topical analysis ~\cite{yuan2013and, hong2012discovering}, urban dynamics discovery ~\cite{zhang2017regions}, human mobility modelling ~\cite{zhang2016gmove} and local event detection ~\cite{zhang2017triovecevent}. Specifically, topic models were employed for obtaining location-specific topics for tweets ~\cite{ahmed2013hierarchical} and discovering language characteristics along with common topics exposed in geo-tagged Twitter streams ~\cite{hong2012discovering}. In addition, multi-modal signals, in the form of spatial, temporal and texts, were utilized for different research purposes. For instance, ~\cite{zhang2017regions} proposed a novel cross-modal representation learning method to embed all spatial, temporal and textual units into the same vector space in order to uncover urban dynamics. ~\cite{yuan2013and} discovered spatio-temporal topics for Twitter users by using a probabilistic generative model for user behavior modelling from the geographic and temporal perspectives. ~\cite{zhang2017triovecevent} presented a method to leverage multi-modal embeddings for the purpose of accurately detecting local events. Furthermore, ~\cite{miura2017unifying} adopted a complex neural network, which was able to unify the representations learned from text, metadata and user network, and an attention mechanism to better infer geo-locations of tweets.

In the domain of \textbf{Politics}, existing researches can be divided into three classes using embedding-based text representations: political ideology detection, political relation extraction, political technique analysis. Regarding political ideology detection, topic model, word embedding, and neural networks all have been adopted. For instance, ~\cite{farrell2016corporate} uses STM to discover ideological polarization around climate change, and further examine the influence of corporate funding on it. ~\cite{preoctiuc2017beyond} uses Word2vec to assist political ideology prediction. Hierarchical LSTM and FastText are also applied to detect political perspective~\cite{li2019encoding} and stance~\cite{stefanov2020predicting}. As to political relation extraction, the topic model is mainly used, to look at the relationship between Republican legislators~\cite{nguyen2015topic}  or extract events between political actors from news  corpora~\cite{DBLP:conf/acl/OConnorSS13}. For political technique analysis, topic models and neural networks are also used to catch a glimpse of processes of framing~\cite{DBLP:conf/acl/TsurCL15}, propaganda techniques~\cite{da-san-martino-etal-2020-prta} and political ads~\cite{Silva2020Facebook}.

In the domain of \textbf{Psychology}, most studies concentrate on mental health identification with embedding-based text representations. For instance, the topic model is used to mine topics from statuses and predict depression of patients~\cite{eichstaedt2018facebook}. RNN and CNN models are employed to represent typing data when using mobile phone and users' posts in Reddit, for mood detection~\cite{cao2017deepmood} and suicide risk assessment~\cite{gaur2019knowledge}, respectively. Besides the mental health, embedding-based text representations are also used in other psychological sphere, such as moralization~\cite{mooijman2018moralization}, intention~\cite{yu2020identifying}, and happiness~\cite{frank2013happiness}.

In the domain of \textbf{Environment}, issues related to climate change and air quality prediction attract the most attention from scholars. In ~\cite{Farrell2016Network}, Latent Semantic Analysis (LSA) ~\cite{wild2015lsa} was adopted to examine the impact of different climate-contrarian organizations' ideas, regarding climate change counter-movement, on news media and bureaucratic politics. With respect to air quality, ~\cite{DBLP:conf/acl/JiangSWY19} deployed a deep learning model,  based on convolutional neural network and overtweet-pooling, on social media data to enhance air quality prediction.




In the domain of \textbf{Economics}, the majority of works utilized embedding-based representations of text to investigate
issues related to stock market ~\cite{2014Quantifying, DBLP:conf/www/YangNSD20, DBLP:conf/acl/CohenX18, nguyen2015topic, xie2013semantic}, while others concentrated on electronic commerce (e-commerce) ~\cite{liu2016repeat} and socio-economic indicators ~\cite{chakraborty2016predicting}. In particular, topic modelling techniques, such as LDA, were employed in a set of studies, where ~\cite{liu2016repeat} managed to predict loyal buyers for e-commerce and ~\cite{2014Quantifying} quantified the semantics of search behavior of Internet users and identified topics of interest before stock market moves. In addition, scholars were also interested in combining textual contents and other types of signals for stock market related research. For instance, ~\cite{nguyen2015topic} incorporated sentiment signals from social media into topic models to better predict stock price movement; \cite{DBLP:conf/www/YangNSD20} proposed a novel model architecture based on Transformer ~\cite{vaswani2017attention} to harness the textual and audio information for predicting future stock price volatility; ~\cite{DBLP:conf/acl/CohenX18} designed a deep generative model for stock movement prediction by jointly exploiting text and price signals.

In the domain of \textbf{Linguistics}, embedding-based representation is mostly used in two fashions as graphical model, especially topic model, and word embedding. As for graphical model, it is utilized to capture the latent information behind the text, such as linguistic topics. ~\cite{hong2012discovering}  uses topic model to discover geographical patterns in language use, while ~\cite{bokanyi2016race} further relate the patterns to demographics. ~\cite{roy2015predicting} also adopts it to capture the topic distributions of context, where children accumulate interactions and learn words. ~\cite{doyle2016robust} proposes a graphical model to model the linguistic alignment in Twitter interactions, which is an important measure of accommodation. As for word embedding, it is usually used to detect the linguistic change across corpora and time, because of the ability to capture semantics. ~\cite{kulkarni2015statistically} proposes an approach to detect the linguistic change in the meaning and usage of words by Skip-Gram, while ~\cite{gonen2020simple} designs a more simple, interpretable and stable method with Skip-Gram as well.

In the domain of \textbf{Communication}, topic model is most observed to obtain text representation. ~\cite{DBLP:conf/acl/TsurCL15} applies it to analyze the statements from Congress, which attempts to gain insights about about agenda setting. ~\cite{2011Topic} and ~\cite{Farrell2016Network} both aim to find topical aspects of actors and identify most influential actors in a network. Besides, ~\cite{sheshadri2019public} employs paragraph vector to estimate similarity between two news documents, devoted to the impact of news framing.

In the domain of \textbf{Anthropology}, ~\cite{fetaya2020restoration} concentrated on the reconstruction of a lost ancient heritage, with the help of embedding-based representation of text. In particular, ~\cite{fetaya2020restoration} employed recurrent neural networks to reconstruct the damaged and missing ancient Akkadian texts from Achaemenid period Babylonia.

%


\subsection{Network}
\subsubsection{Symbol-based representation}
In the domain of \textbf{Sociology}, most studies were conducted on the social network. For online social networks, the research data usually came from popular websites or communication applications, such as Twitter~\cite{romero2011differences,meeder2011we,wu2011says}, Facebook~\cite{lewis2012social,schmidt2017anatomy,eckles2016estimating}, Yahoo~\cite{kayes2015social} and Wechat~\cite{qiu2016lifecycle}. For offline social or friendship networks, the studied scenarios are quite diverse, such as the dating network~\cite{lewis2013limits}, the social network structure of potential male raiders~\cite{glowacki2016formation}, problem-solving networks~\cite{braha2020patterns} where people worked by groups and collaborated with each other, and even the social network of cooperative bird species~\cite{dakin2020reciprocity}.

In terms of research problem and methodology, we summarize the following four patterns of these literature:

The first category is to study whether a phenomenon exists in the network. For example, ~\cite{lewis2013limits} studied cross-racial communication to detect the existence of racial prejudice. ~\cite{gupte2011finding} aimed to find the social hierarchy and stratification among humans in social networks. This line of work usually employed simple statistics or proposed indices involving related features or factors for their methods.

The second category is to find out the structural patterns leading to a specific property. For example, ~\cite{glowacki2016formation} tried to find out how the formation of social network structure will lead to a potential male raider. This line of work usually analyzed the patterns of subgraphs (e.g. the frequency of specific subgraphs~\cite{braha2020patterns}) for modeling the correlations.

The third category is to identify the most important nodes in a network. For example, ~\cite{teng2016collective} aimed at identifying the most influential spreaders that maximize information flow. This line of work usually employed various of network centrality coefficients (e.g. the degree of a node) as the measurements.

The fourth category is to predict the future behaviours of users in a network. Common scenarios include recommendation system~\cite{liu2013soco} and information diffusion (e.g. the spread of rumors or misinformation~\cite{del2016spreading}). This line of work need to model the temporal dynamics and user preferences for predicting future behaviours. The detailed models are quite personalized and differ from each other.



In the domain of \textbf{Anthropology}, most works employed similar symbol-based network representations for analysis, such as PageRank score or betweenness coefficient for extracting the most important nodes. Therefore, it would be more interesting to see what kind of networks they built to solve their problems.

The first kind is location networks. To study cultural history and discover cultural centers, ~\cite{schich2014network} constructed a directed network of cities in Europe and North America based on migration, where the endpoints of each edge in the network represent the birth and death locations of a notable individual. ~\cite{hart2017effects} built a similarity network of 200 Iroquoian village sites dating from A.D. 1350 to 1600, and concluded the importance of a specific location in population dispersal.  ~\cite{lulewicz2019social} also constructed a network of sites from the southern Appalachian region between ca. AD 800 and 1650, to study the variation of Mississippian sociopolitics.

The second kind is social or friendship network. ~\cite{fowler2011correlated} studied the correlation between genotypes and friendship networks, and identified a positively correlated (homophily) one and a negatively correlated (heterophily) one from all six available genotypes. ~\cite{boardman2012social} also discussed genotypes and friendship networks, but with more consideration of environment context.  ~\cite{apicella2012social} characterized the social network of the Hadza hunter-gatherers in Tanzania, which may reveal the behaviours of early humans. ~\cite{roberts2016wild} used the social bonds between wild chimpanzees to inspire the study of human evolution.

There are also other kinds of networks. For instance, ~\cite{hilger2017intelligence} constructed a brain network to understand human intelligence, where nodes correspond to regions in a grey matter and edges represent high positive correlations of signals between nodes.

In the domain of \textbf{Linguistics}, only a few work utilized network structure for their study. They studied the networks of concepts, words or languages, and usually used simple statistics or cluster coefficients for analysis. 

To understand the universality and diversity in how humans understand and experience emotion, ~\cite{jackson2019emotion} built a network of emotion concepts (e.g. ``angry'' and ``fear'') for each of 2,474 spoken languages, where two concepts are connected if their meanings appear in the same word. ~\cite{youn2016universal} explored a more general problem, i.e., the universal structure of human lexical semantics. To be more specific, ~\cite{youn2016universal} built a weighted network of concepts using cross-linguistic dictionaries: sometimes a single “polysemous” word from one language can express multiple concepts that another language represents using distinct words. The frequency of such polysemies between two concepts can be seen as a measure of their semantic similarity. ~\cite{sizemore2018knowledge} focused on the sparsity (i.e., knowledge gaps) of semantic feature networks of humans, where words correspond to nodes and are connected by shared features, to understand the process of language learning. ~\cite{ronen2014links} constructed a co-spoken language network to figure out the influence of different languages.

In the domain of \textbf{Psychology}, the most widely used network types are the brain and social networks.

For brain networks, ~\cite{taruffi2017effects} found that compared with happy music, sad music was linked to greater centrality of the nodes of the Default Mode Network (i.e., a set of brain regions typically active during rest periods). ~\cite{schmalzle2017brain} studied the functional connectivity of brain regions under social inclusion or exclusion.

For social networks, ~\cite{turetsky2020psychological} showed that psychological interventions can strengthen the connections of peer social network in terms of node degree, closeness, betweenness, etc. ~\cite{morelli2017empathy} studied how psychological traits correlate to centrality in social networks, and concluded that people high in well-being were central to the ``fun'' networks, while people high in empathy were central to the ``trust'' networks. ~\cite{kim2019social} showed that occupying a bridging position in a social network may alleviate the impact of depressive symptoms among older men, whereas the opposite holds true for older women. ~\cite{ito2020influence} studied how networks
of general trust will affect the willingness to communicate in English for Japanese people, via the analysis of centrality indices.

In the domain of \textbf{Geography}, the wide adoption of mobile devices significantly benefits the collections of location-based data, and thus facilitates relevant researches in this area. There are two main types of networks discussed in this work: transportation network and location-based social network. 

Transportation networks include road/street networks and travel/mobility networks. For example, ~\cite{bao2017planning} utilized bike trajectory data on road networks to develop bike lane construction plans. ~\cite{ganin2017resilience} studied the efficiency and resilience of transportation networks, where intersections are mapped to nodes and road segments between the intersections are mapped to links. ~\cite{barrington2020global} analyzed a time series of street network and discussed road building in new and expanding cities for urban development. Taking the London rail network as an example, ~\cite{yadav2020resilience} found that topological attributes designed for maximizing efficiency in urban transport networks will make the network more vulnerable under intense flood disasters. On the other hand, ~\cite{wesolowski2012quantifying} analyzed travel networks of people and parasites between settlements and regions based on mobile phone data. ~\cite{bonaccorsi2020economic} studied the effect of lockdown restrictions on the economic conditions of individuals and local governments based on the Italian mobility network. ~\cite{santi2014quantifying,vazifeh2018addressing,liu2016rebalancing} focused on vehicle-shareability networks for better taxi or bike-sharing services. ~\cite{riascos2020networks} also analyzed taxi trip data and built a directed weighted origin-destination network for the study of long-range mobility. There are also some work ~\cite{karduni2016protocol}~\cite{kujala2018collection} proposed for data collections of transportation networks. 

A location-based social network can be either online or offline. For online networks, ~\cite{wang2011human} found that two individuals’ movements strongly correlates with their proximity in the social network. ~\cite{cho2011friendship} tried to understand the basic laws of human motion and dynamics based on location-based social networks and cell phone location data. ~\cite{li2012towards} focused on profiling users' home locations in the context of a social network. ~\cite{yang2012socio} proposed Socio-Spatial Group Query (SSGQ) to select nearby attendees with close social relation based on users' social networks on Facebook, as well as their spatial locations from Facebook Checkin records. For offline networks, ~\cite{sun2013understanding} studied a time-resolved in-vehicle social encounter network on public buses in a city. ~\cite{sekara2016fundamental} explored the dynamic social network of about 1,000 individuals and their interactions measured via Bluetooth, telecommunication networks or online social media, etc.

In addition, other relevant work studied infrastructure networks of urban microgrids~\cite{halu2016data} or the community detection problem on general spatially-embedded networks~\cite{expert2011uncovering}. 

In the domain of \textbf{Economics}, the most discussed networks are financial institution networks and trade networks.

For financial institution networks, ~\cite{battiston2016price} studied the multi-layer networks of financial institutions connected by contracts and common assets, and showed that the complexity of financial networks may increase the social cost of financial crises. ~\cite{bardoscia2017pathways} also studied the network of financial institutions, and discussed how the instability of model ecosystems is relevant to the dynamical processes on complex networks.

For trade networks, ~\cite{porfirio2018economic} studied the structural changes in the global agricultural trade network under greenhouse gas emissions. It is worth noting that ~\cite{porfirio2018economic} employed matrix factorization, a technique widely used in network embedding learning, for their modeling. However, they only utilized the singular values and discarded the vectors in singular value decomposition. Thus, we still classify this work as a symbol-based one. ~\cite{ren2020bridging} characterized the international trading system with a multi-layer network with each layer representing the transnational trading relations of a product. They studied a nation's economic growth by analyzing node degrees and product rankings over time.

For others, ~\cite{anderson2017skill} built a network of skills based on their relationship in the market, and showed that workers with diverse skills can earn higher wages than those with more specialized skills. ~\cite{bonaccorsi2020economic} studied the effect of lockdown restrictions on economic conditions of individuals and local governments based on the Italian mobility network.
 
In the domain of \textbf{Politics}, most researches were conducted on social media or online social networks and discussed ideology or elections.

For the ideology topic, ~\cite{farrell2016corporate} constructed an organization network and concluded that organizations with corporate funding were more likely to write and spread texts that lead to ideological polarization on the climate change issue. By analyzing the retweeting behaviours in online social networks, ~\cite{brady2017emotion}  found that the expression of moral emotion is key for the spread of moral and political ideas or ideology. ~\cite{bail2018exposure} extracted a following network of 4,176 opinion leaders on Twitter, and used the first component of the adjacency matrix to create liberal/conservative ideology scores. ~\cite{pado2019sides} built a bipartite network of actors and claims to understand the structures of political debates. ~\cite{burfoot2011collective} analyzed the sentiment of U.S. Congressional floor-debate transcripts with the help of document networks where one speaker cites another was annotated. To predict the frames used in political discourse, ~\cite{DBLP:conf/acl/JohnsonJG17} assumed that politicians with shared ideologies are likely to frame issues similarly and retweet and/or follow each other on Twitter network.  ~\cite{rule2015lexical} studied the network of correlated words in textual corpora that span a long time, and identified that terms, concepts, and language use changes in American political consciousness since World War I in 1917.

For the election topic, ~\cite{bovet2019influence} studied the dynamics and influence of fake news on Twitter during the 2016 US presidential election by analyzing the retweet networks formed by the top 100 news spreaders of different media categories. ~\cite{grinberg2019fake} studied the same problem with the help of a co-exposure network, where nodes are news websites and edges are shared-audience relationships. ~\cite{bovet2018validation} inferred the opinion of Twitter users in the context of the 2016 US Presidential Election based on both social network and hashtag co-occurrence network. ~\cite{volkova2014inferring} also inferred user's political preferences between Democrat and Republican based on the Twitter social graph. 

In the domain of \textbf{Environment}, the network types used in different work are quite diverse. However, simple statistics and network centrality coefficients are still the most popular techniques for network analysis.

~\cite{farrell2016corporate} constructed an organization network and concluded that organizations with corporate funding were more likely to write and spread texts that lead to ideological polarization on the climate change issue. ~\cite{Farrell2016Network} built a bipartite graph of the climate contrarian network with 4,556 individuals and 164 contrarian organizations, in order to uncover the institutional and corporate structure of the climate change counter-movement. ~\cite{barnes2016social} studied the information-sharing networks among tuna fishers to reveal how these social networks will affect the incidental catch of sharks, a global environmental issue.  ~\cite{reino2017networks} studied the global trade network of wild-caught birds with network centrality to analyze the bird invasion risk of different regions.  ~\cite{camara2019indigenous} proposed indigenous knowledge networks to describe the wisdom of indigenous people on plant species and the services they provide.  ~\cite{zheng2013u,hsieh2015inferring} utilized air quality monitoring data, human mobility, road network structures, and other information to suggest the best locations of new monitoring stations.



In the domain of \textbf{Communication}, most works discussed the phenomenon of information diffusion (e.g. the spread of ideas, opinions or products) in online or offline social networks.

For example, ~\cite{guille2012predictive} employed feature engineering and a simple probabilistic model to characterize the temporal dynamics of information diffusion in social networks. ~\cite{gomez2016explosive} also studied the spread of social phenomena such as behaviours, ideas, or products in the contact network of individuals. ~\cite{pei2014searching,zhang2019identifying} utilized various network centrality coefficients to detect the most influential information spreaders in online social networks. ~\cite{brady2017emotion} analyzed the retweeting behaviours in online social networks and found that the expression of moral emotion is key for the spread of moral and political ideas. ~\cite{gao2014quantifying} compared the contact networks of users under emergency events and non-emergency events, in order to figure out how human communications will affect the propagation of situational awareness. 

Some of these work especially focused on the spread of rumor, misinformation, and fake news, ~\cite{quattrociocchi2014opinion} studied opinion dynamics on the network containing the interactions between gossipers, the influence network between gossiper and media, and the leader-follower relationship between media. ~\cite{bovet2019influence} studied the dynamics and influence of fake news on Twitter during the 2016 US presidential election by analyzing the retweet networks formed by the top 100 news spreaders of different media categories. ~\cite{shao2018spread} studied the spread of low-credibility content in a retweet network, and concluded that social bots played an important role in spreading articles from low-credibility sources. In contrast, ~\cite{vosoughi2018spread} also studied the spread of false news on Twitter networks, and found that robots accelerated the spread of true and false news at the same rate, indicating that false news spreads faster because of human.

Besides, there is some work crossed with other domains. To find out whether restricting mobility or spreading disease prevention information better helps the control of diseases, ~\cite{lima2015disease} modeled human mobility and communications by an interconnected multiplex structure where each node represents the population in a geographic area and extended the model with a social network where relevant disease prevention information spreads. ~\cite{luo2017inferring} measured individuals' location and influence in the social network from mobile and residential communication data, and found that an individual’s location is highly correlated with personal economic status. ~\cite{gomez2019clustering} studied how the distributions of knowledge and ability within a network of collective problem solvers contribute to the performance of the entire group. ~\cite{Farrell2016Network} built a bipartite graph of the climate contrarian network with 4,556 individuals and 164 contrarian organizations, in order to uncover the institutional and corporate structure of the climate change counter-movement.

\subsubsection{Embedding-based Representation}
In the domain of \textbf{Sociology}, most studies were conducted on the social network. For online social networks, the research data usually came from popular websites or communication applications, such as Twitter~\cite{romero2011differences,meeder2011we,wu2011says}, Facebook~\cite{lewis2012social,schmidt2017anatomy,eckles2016estimating}, Yahoo~\cite{kayes2015social} and Wechat~\cite{qiu2016lifecycle}. For offline social or friendship networks, the studied scenarios are quite diverse, such as the dating network~\cite{lewis2013limits}, the social network structure of potential male raiders~\cite{glowacki2016formation}, problem-solving networks~\cite{braha2020patterns} where people worked by groups and collaborated with each other, and even the social network of cooperative bird species~\cite{dakin2020reciprocity}.

In the domain of \textbf{Sociology}, matrix factorization and topic models were widely used for learning user embeddings in a user-user network or user-item interaction network in the early 2010s. Recently, deep learning models such as graph neural networks are becoming the mainstream to encode structural information.

Many studies focused on the completion task, such as inferring the missing attributes or recommending potential friends/items. ~\cite{kosinski2013private} applied singular value decomposition to the user-like matrix for learning user embeddings, which were further utilized for predicting private traits. ~\cite{pan2019twitter} utilized graph convolutional network to embed the users in a user network for occupation prediction. ~\cite{yang2011like} applied matrix factorization to the social network including user-user friendship network and bipartite user-item interaction network for learning user and item embeddings, which were employed for friend and item recommendations. ~\cite{fan2019graph,wu2019dual} used graph neural networks for social recommendation. Besides graph neural networks, ~\cite{tang2020knowing} also employed LSTM for temporal modeling.

Other works can be formalized as sequential prediction, binary classification, and clustering, respectively. ~\cite{li2017deepcas} sampled diffusion sequences of users from the diffusion network, and applied RNN to encode the sequences and predict future users that would be influenced. ~\cite{qiu2018deepinf} employed graph neural networks to encode the ego network of each user, and then used the embeddings to classify whether the user will be influenced during information diffusion. ~\cite{lu-li-2020-gcan} used graph attention network on user network for fake news detection. 
~\cite{zhong2020integrating} applied graph convolutional network on reply relationship network for controversy detection. ~\cite{sachan2012using} employed a topic model to compute the community distribution of each user in a social network.

In the domain of \textbf{Geography}, embedding-based representations of road networks are most widely used.

For example, ~\cite{wang2020traffic} employed deep learning models including recurrent neural networks, attention mechanism, and graph neural networks to project each road into embedding-based representations and characterize the dynamics of traffic flow in road networks. ~\cite{deng2016latent} learned embeddings for each road in a road network via non-negative matrix factorization. The embeddings can encode both topological and temporal properties for traffic prediction. ~\cite{li2018multi} focused on travel time estimation in a road network and employed a multi-task learning framework to encode links and spatial-temporal factors. ~\cite{sun2018exploring} developed a spatial-temporal latent factor model to identify the latent travel patterns and demands of urban region visitors. ~\cite{pan2019urban} used graph attention network to encode road networks, and RNN to further embed the temporal sequence of traffics for urban traffic prediction. 

Besides, social media and social networks related to geo-locations are also explored. ~\cite{zhang2017regions} constructed a heterogeneous network with three types of nodes, i.e. location, time, and text, from geo-tagged social media (GTSM) data. Then they jointly encoded all spatial, temporal, and textual units into the same embedding space to capture the correlations for modeling people’s activities in the urban space. ~\cite{miura2017unifying} applied attention mechanism to user mention network extracted from Twitter to enhance the performance of geolocation prediction. ~\cite{yang2019revisiting} characterized a location-based social network containing both user mobility data and the corresponding social network as a hypergraph where a friendship is represented by an edge between two user nodes and check-in is represented by a hyperedge among four nodes (a user, an activity type, a timestamp, and a POI). Network embedding methods are then employed for both friendship and location predictions.

For others, ~\cite{yuan2012discovering} constructed a region transition network by connecting origin and destination regions of human mobility, and used the topic model to learn functional topic distributions for each region. ~\cite{wang2018you} developed a driving state transition graph to characterize time-varying driving behaviour sequence, where nodes denote driving states (e.g. acceleration, turning right, etc.), and the weights of edges can be the frequency of state changes or the duration of state changes between two driving states. Then they employed a deep autoencoder to transform graphs into low-dimensional vectors and utilized RNN to incorporate temporal patterns.

In the domain of \textbf{Economics}, a recent work~\cite{zhong2020financial} employed an attributed heterogeneous information network to characterize the behaviours and relationships between users, merchants and devices. Then they used a fully neural-based model to model the representations of users for default probability prediction.

In the domain of \textbf{Politics}, ~\cite{stefanov2020predicting} applied node2vec~\cite{grovernode2vec} to a User-to-Hashtag graph and a User-to-Mention graph to learn user embeddings, which can help better predict the stance and political leaning of media. ~\cite{li2019encoding} employed Graph Convolutional Network (GCN)~\cite{Kipf2017SemiSupervisedCW} to embed the social information graph as well as text features, for identifying the political perspective of news media.

In the domain of \textbf{Environment}, ~\cite{shang2014inferring} estimated traffic conditions in a road network by filling the missing entries in an affinity matrix, where time slot embeddings, road embeddings, and feature embeddings are learned by matrix factorization.

In the domain of \textbf{Communications}, ~\cite{2011Topic} aimed to find the most influential users in a network on a specific topic and how the influential users connect with each other. They characterized each user with topic distributions learned by a topic model, which can be seen as a non-negative embedding for each user.

\begin{table*}[]
\centering
\resizebox{0.8\textwidth}{!}{
\begin{tabular}{l|l|l|l|l}
\hline
\centering

                       & \multicolumn{2}{c|}{Symbol} & \multicolumn{2}{c}{Embedding}                                        \\ \hline
                       & \multicolumn{1}{c|}{Text} & \multicolumn{1}{c|}{Network} &  \multicolumn{1}{c|}{Text} & \multicolumn{1}{c}{Network} \\ \hline
                       
\textbf{Sociology}     
& 
\begin{tabular}[c]{@{}l@{}}
~\cite{bruch2018aspirational}, 
~\cite{alizadeh2020content},
~\cite{catalini2015incidence},
~\cite{jones2017distress}\\
~\cite{stella2018bots},
~\cite{lakkaraju2011attention}, ~\cite{lehmann2012dynamical},
~\cite{yang2012we}\\
~\cite{gupta2013faking}, ~\cite{imran2013practical},
~\cite{goga2013exploiting}, ~\cite{jain2013seek}\\
~\cite{imran2014aidr}, ~\cite{boididou2014challenges},
~\cite{flekova2014makes}, ~\cite{zhao2015enquiring}\\ ~\cite{cresci2015linguistically},
~\cite{nobata2016abusive}, ~\cite{davis2016botornot},
~\cite{popat2017truth}\\
~\cite{kumar2017army}, ~\cite{chatzakou2017measuring},
~\cite{portnoff2017tools}, 
~\cite{zannettou2018gab}\\
~\cite{bhatt2018combining},
~\cite{volkova2018misleading}, ~\cite{tesfay2018read},
~\cite{almerekhi2020these}\\ ~\cite{parimi2011predicting},
~\cite{yang2012automatic}, ~\cite{fu2013people},
~\cite{ramakrishnan2014beating}\\ 
~\cite{chen2014non}, ~\cite{li2014social}, ~\cite{rayana2015collective},
~\cite{goga2015reliability}\\ 
~\cite{madaio2016firebird},
~\cite{frank2013happiness}, ~\cite{del2016echo},
~\cite{bovet2018validation}\\ ~\cite{shi2016detecting},
~\cite{bergsma2013using}, ~\cite{kozareva2013multilingual},
~\cite{rosenthal2011age}\\
~\cite{park2011contrasting},
~\cite{bakshy2015exposure},
~\cite{castillo2011information}, ~\cite{lin2011smoothing}\\
~\cite{potthast-etal-2018-stylometric}, ~\cite{ma2017detect},
~\cite{bramsen2011extracting}, ~\cite{DBLP:conf/kdd/HassanALT17}\\
~\cite{dong2014inferring}, ~\cite{hooi2016fraudar}, ~\cite{brady2017emotion}, ~\cite{hughes2012quantitative}\\

\end{tabular}

& 
\begin{tabular}[c]{@{}l@{}}
~\cite{romero2011differences}, ~\cite{wu2011says}, 
~\cite{budak2011limiting}, ~\cite{ratkiewicz2011truthy}\\
~\cite{gupte2011finding},
~\cite{meeder2011we}, 
~\cite{jamali2011modeling}, ~\cite{ghosh2012understanding}\\ 
~\cite{yang2012analyzing}, ~\cite{yang2012we},
~\cite{ver2012information}, ~\cite{guille2012predictive}\\
~\cite{beutel2013copycatch}, ~\cite{goga2013exploiting},
~\cite{liu2013soco},
~\cite{jain2013seek}\\
~\cite{lou2013mining}, ~\cite{cheng2014can},
~\cite{myers2014information}, ~\cite{buntain2014identifying}\\
~\cite{flekova2014makes},
~\cite{pavalanathan2015identity},
~\cite{kayes2015social},
~\cite{singer2015hyptrails}\\
~\cite{davis2016botornot},
~\cite{yin2016communication},
~\cite{qiu2016lifecycle},
~\cite{su2016effect}\\
~\cite{romero2016social},
~\cite{garcia2016discouraging},
~\cite{li2016world},
~\cite{kumar2017army}\\
~\cite{zannettou2018gab},
~\cite{resende2019mis},
~\cite{clauset2015systematic},
~\cite{friedkin2016network}\\
~\cite{manrique2016women},
~\cite{massucci2016inferring},
~\cite{mukherjee2017nearly},
~\cite{garcia2017leaking}\\
~\cite{trujillo2018document},
~\cite{bruch2018aspirational},
~\cite{park2018strength}\\
~\cite{grinberg2019fake},
~\cite{asikainen2020cumulative},
~\cite{charoenwong2020social}\\
~\cite{jin2011likeminer},
~\cite{parimi2011predicting},
~\cite{papadimitriou2011display},
~\cite{li2011social}\\
~\cite{matsubara2012rise},
~\cite{chen2014non},
~\cite{li2014social},
~\cite{rayana2015collective}\\
~\cite{goga2015reliability},
~\cite{rand2011dynamic},
~\cite{facchetti2011computing},
~\cite{fowler2011correlated}\\
~\cite{roca2011emergence},
~\cite{lewis2012social},
~\cite{boardman2012social},
~\cite{dall2012collaboration}\\
~\cite{mills2013transformation},
~\cite{jiang2013calling},
~\cite{kovanen2013temporal},
~\cite{lewis2013limits}\\
~\cite{rutherford2013limits},
~\cite{saramaki2014persistence},
~\cite{rzhetsky2015choosing},
~\cite{petersen2015quantifying}\\
~\cite{del2016spreading},
~\cite{paluck2016changing},
~\cite{sekara2016fundamental},
~\cite{barnes2016social}\\
~\cite{glowacki2016formation},
~\cite{coman2016mnemonic},
~\cite{brady2017emotion},
~\cite{schmidt2017anatomy}\\
~\cite{morelli2017empathy},
~\cite{han2017emergence},
~\cite{bail2018exposure},
~\cite{stella2018bots}\\
~\cite{guilbeault2018social},
~\cite{tamarit2018cognitive},
~\cite{stadtfeld2019integration},
~\cite{yang2019network}\\
~\cite{sardo2019quantification},
~\cite{almaatouq2020adaptive},
~\cite{dakin2020reciprocity},
~\cite{rovira2013predicting}\\
~\cite{li2013coevolving},
~\cite{eom2014generalized},
~\cite{li2014comparative},
~\cite{jo2014spatial}\\
~\cite{dankulov2015dynamics},
~\cite{wang2016suppressing},
~\cite{gomez2016explosive}\\
~\cite{teng2016collective},
~\cite{medo2016identification},
~\cite{aral2017exercise},
~\cite{del2017modeling}\\
~\cite{parkinson2017spontaneous},
~\cite{luo2017inferring},
~\cite{battiston2017layered},
~\cite{shen2017coevolution}\\
~\cite{wu2017impact},
~\cite{shao2018spread},
~\cite{parkinson2018similar},
~\cite{waniek2018hiding}\\
~\cite{bovet2018validation},
~\cite{altenburger2018monophily},
~\cite{iacopini2019simplicial},
~\cite{johnson2019hidden}\\
~\cite{lee2019homophily},
~\cite{gomez2019clustering},
~\cite{schuchard2019bots},
~\cite{zhu2020cooperation}\\
~\cite{david2020herding},
~\cite{ito2020influence},
~\cite{pomeroy2020dynamics}, 
~\cite{pickard2011time}\\
~\cite{aral2012identifying}, ~\cite{vosoughi2018spread},
~\cite{young2011dynamics}, ~\cite{contractor2014integrating}\\ ~\cite{eckles2016estimating},
~\cite{becker2017network},
~\cite{sayles2017social},
~\cite{hu2018local}\\
~\cite{ma2018scientific}, ~\cite{wei2013scientists},
~\cite{ehlert2020human},
~\cite{gargiulo2014driving}\\
~\cite{braha2020patterns}, ~\cite{wang2011human},
~\cite{ma2017detect}, ~\cite{calais2011bias}\\
~\cite{dong2014inferring}, ~\cite{hooi2016fraudar},
~\cite{youyou2015computer}, ~\cite{schlosser2020covid}\\ ~\cite{expert2011uncovering}, ~\cite{henry2011emergence}, ~\cite{mason2012collaborative}, ~\cite{wang2012cooperation}\\
~\cite{dandekar2013biased}, ~\cite{varga2019shorter},
~\cite{becker2019wisdom}, ~\cite{lucchini2020code}\\
~\cite{shirado2017locally}, ~\cite{wu2019large}, ~\cite{stewart2019information}, ~\cite{muchnik2013origins}\\
~\cite{wang2013impact}, ~\cite{lu2013retraction}, ~\cite{singh2013threshold}, ~\cite{wilkins2014network}\\ 
~\cite{gianetto2015network}, ~\cite{cuesta2015reputation}, ~\cite{ramos2015does}, ~\cite{yang2019identification}

\end{tabular} 

& 
\multicolumn{1}{l|}{
\begin{tabular}[c]{@{}l@{}}
~\cite{caliskan2017semantics},
~\cite{gerow2018measuring},
~\cite{sivak2019parents},
~\cite{sachan2012using}\\
~\cite{cresci2015linguistically}, ~\cite{nobata2016abusive},
~\cite{badjatiya2017deep},
~\cite{zhang2017regions}\\
~\cite{singer2017we},
~\cite{ma2018detect},
~\cite{bhatt2018combining},
~\cite{volkova2018misleading}\\
~\cite{khattar2019mvae}, ~\cite{ma2019detect},
~\cite{wang2019demographic},
~\cite{zhang2019reply}\\
~\cite{weerasinghe2020pod},
~\cite{zafarani2013connecting}, ~\cite{mukherjee2013spotting},
~\cite{yuan2013and}\\
~\cite{lucey2013assessing},
~\cite{ferraz2015rsc}\\ ~\cite{mu2016user},
~\cite{wang2018eann}, ~\cite{qiu2018deepinf},
~\cite{jiang2019deepurbanevent}\\
~\cite{ding2019modeling},
~\cite{zarrinkalam2019social},
~\cite{zhang2020general}, ~\cite{tang2020knowing}\\
~\cite{dutta2020deep},
~\cite{ye2019secret},
~\cite{stefanov2020predicting},
~\cite{baly-etal-2020-written}\\ ~\cite{da-san-martino-etal-2020-prta},
~\cite{wu-etal-2020-dtca},
~\cite{zhong2020integrating}, ~\cite{bansal-etal-2020-code}\\
~\cite{chowdhury2019youtoo},
~\cite{oprea-magdy-2019-exploring}, ~\cite{ma2019sentence},
~\cite{wan-etal-2019-fine}\\
~\cite{pan2019twitter}, ~\cite{wilson-mihalcea-2019-predicting},
~\cite{ma-etal-2018-rumor}, ~\cite{mishra2017learning}\\
~\cite{pavalanathan2017multidimensional}, ~\cite{sasaki-etal-2017-topics},
~\cite{volkova2016inferring},
~\cite{preoctiuc2015analysis}\\
~\cite{nguyen2015topic},
~\cite{li2014towards},
~\cite{yancheva2013automatic}\\
~\cite{diao2012finding},
~\cite{wang2012historical},
~\cite{garg2018word},
~\cite{almerekhi2020these}\\ 
~\cite{fu2013people}, ~\cite{djuric2015hate}, ~\cite{hughes2012quantitative}

\end{tabular}
}    
    
&
\multicolumn{1}{l}{
\begin{tabular}[c]{@{}l@{}}
~\cite{kosinski2013private},
~\cite{yang2011like},
~\cite{sachan2012using}, ~\cite{li2017deepcas}\\
~\cite{zhang2017regions}, 
~\cite{fan2019graph},
~\cite{wu2019dual},
~\cite{wang2012magnet}\\
~\cite{qiu2018deepinf}, ~\cite{zarrinkalam2019social},
~\cite{tang2020knowing},
~\cite{lu-li-2020-gcan}\\ 
~\cite{stefanov2020predicting},
~\cite{zhong2020integrating},
~\cite{pan2019twitter}\\

\end{tabular}
}


\\ \hline

\textbf{Anthropology}  
&
\begin{tabular}[c]{@{}l@{}}
~\cite{klingenstein2014civilizing},
~\cite{mehr2019universality}
\end{tabular}  
& 
\begin{tabular}[c]{@{}l@{}}
~\cite{schich2014network}, ~\cite{hart2017effects}, ~\cite{fowler2011correlated}\\ ~\cite{boardman2012social}, ~\cite{lulewicz2019social}, ~\cite{apicella2012social}, ~\cite{roberts2016wild}\\ ~\cite{hilger2017intelligence} 
\end{tabular} 

& 
\multicolumn{1}{l|}{
\begin{tabular}[c]{@{}l@{}}
~\cite{fetaya2020restoration}
\end{tabular}
} 
& 
\multicolumn{1}{l}{

} 

\\ \hline

\textbf{Psychology}    
& 
\begin{tabular}[c]{@{}l@{}}
~\cite{kramer2014experimental}, ~\cite{eichstaedt2018facebook},
~\cite{lim2018grass}, ~\cite{chang2020don}\\ ~\cite{frank2013happiness},
~\cite{golder2011diurnal},
~\cite{jaidka2020estimating},
~\cite{chen2018mood}\\
\end{tabular}

& 
\begin{tabular}[c]{@{}l@{}}
~\cite{turetsky2020psychological}, ~\cite{morelli2017empathy}, ~\cite{taruffi2017effects}\\
~\cite{kim2019social},
~\cite{ito2020influence}, ~\cite{schmalzle2017brain}
\end{tabular}

& 
\multicolumn{1}{l|}{
\begin{tabular}[c]{@{}l@{}}
~\cite{eichstaedt2018facebook},
~\cite{gaur2019knowledge},
~\cite{cao2017deepmood},
~\cite{jaidka2020estimating}\\
~\cite{mooijman2018moralization},
~\cite{yu2020identifying}, ~\cite{kern2019social}
\end{tabular}
}

& 
\multicolumn{1}{l}{
\begin{tabular}[c]{@{}l@{}}
~\cite{yu2020identifying}
\end{tabular}
}

\\ \hline

\textbf{Politics}      
& 
\begin{tabular}[c]{@{}l@{}}
~\cite{lupia2020does},
~\cite{jordan2019examining},
~\cite{bovet2018validation}, ~\cite{preoctiuc2017beyond}\\ ~\cite{alizadeh2020content},
~\cite{burfoot2011collective}, ~\cite{rule2015lexical}, ~\cite{DBLP:conf/acl/JohnsonG18}\\
~\cite{DBLP:conf/acl/JohnsonJG17}, ~\cite{DBLP:conf/acl/LamposPC13},
~\cite{DBLP:conf/acl/OConnorSS13}, ~\cite{badawy2019falls}\\
~\cite{brady2017emotion}

\end{tabular}        
&
\begin{tabular}[c]{@{}l@{}}
~\cite{grinberg2019fake}, ~\cite{farrell2016corporate},
~\cite{brady2017emotion}, ~\cite{bail2018exposure} \\
~\cite{bovet2018validation}, ~\cite{bovet2019influence}, ~\cite{pado2019sides}, ~\cite{volkova2014inferring}\\
~\cite{burfoot2011collective}, ~\cite{DBLP:conf/acl/JohnsonJG17},~\cite{rule2015lexical},
~\cite{badawy2019falls}\\
\end{tabular}
& 
\multicolumn{1}{l|}{
\begin{tabular}[c]{@{}l@{}}
~\cite{farrell2016corporate},
~\cite{Silva2020Facebook},
~\cite{stefanov2020predicting}, ~\cite{da-san-martino-etal-2020-prta}\\
~\cite{li2019encoding},
~\cite{preoctiuc2017beyond},
~\cite{DBLP:conf/acl/TsurCL15}, ~\cite{DBLP:conf/acl/NguyenBRM15}\\
~\cite{DBLP:conf/acl/IyyerEBR14},
~\cite{DBLP:conf/acl/OConnorSS13}, ~\cite{2020What}

\end{tabular}
}    
& 
\multicolumn{1}{l}{
\begin{tabular}[c]{@{}l@{}}
~\cite{stefanov2020predicting}, ~\cite{li2019encoding}\\
\end{tabular}
}       

\\ \hline

\textbf{Economics}     
& 
\begin{tabular}[c]{@{}l@{}}
~\cite{alanyali2013quantifying}, ~\cite{xie2013semantic}

\end{tabular}
&
\begin{tabular}[c]{@{}l@{}}
~\cite{battiston2016price}, ~\cite{anderson2017skill},
~\cite{bonaccorsi2020economic}, ~\cite{bardoscia2017pathways}\\
~\cite{porfirio2018economic}, ~\cite{ren2020bridging}, ~\cite{atalay2011network}, ~\cite{arinaminpathy2012size}\\ 
~\cite{haldane2011systemic}, ~\cite{atalay2011network}, ~\cite{pozzi2013spread}, ~\cite{squartini2013early}\\
~\cite{thurner2013debtrank}, ~\cite{cimini2015systemic}, ~\cite{ciampaglia2015production}, ~\cite{filip2016dynamics}\\ ~\cite{garcia2019stochastic}

\end{tabular} 

& 
\multicolumn{1}{l|}{
\begin{tabular}[c]{@{}l@{}}
~\cite{2014Quantifying}, ~\cite{DBLP:conf/www/YangNSD20},
~\cite{DBLP:conf/acl/CohenX18}, ~\cite{nguyen2015topic}\\
~\cite{xie2013semantic},
~\cite{liu2016repeat}, ~\cite{chakraborty2016predicting}

\end{tabular}
}

& 
\multicolumn{1}{l}{
\begin{tabular}[c]{@{}l@{}}
~\cite{zhong2020financial}
\end{tabular}
}
    
\\ \hline

\textbf{Linguistics}   
& 
\begin{tabular}[c]{@{}l@{}}
~\cite{michel2011quantitative}, 
~\cite{boyd2020narrative},
~\cite{piantadosi2011word},
~\cite{yang2013ontogeny}\\
~\cite{kao2014nonliteral},
~\cite{futrell2015large},
~\cite{dodds2015human},
~\cite{jordan2019examining}\\
~\cite{hovy2015user},
~\cite{hube2018detecting},
 ~\cite{danescu2012you},
~\cite{danescu-niculescu-mizil-etal-2013-computational}\\
~\cite{tan-etal-2014-effect}, ~\cite{bouchard2013automated}, ~\cite{hughes2012quantitative}, ~\cite{jackson2019emotion}
\end{tabular}

& 

\begin{tabular}[c]{@{}l@{}}
~\cite{youn2016universal}, ~\cite{sizemore2018knowledge},
~\cite{ronen2014links}\\
\end{tabular}  


& 
\multicolumn{1}{l|}{
\begin{tabular}[c]{@{}l@{}}
~\cite{roy2015predicting},
~\cite{kulkarni2015statistically},
~\cite{bokanyi2016race},
~\cite{gonen2020simple}\\
~\cite{doyle2016robust}, ~\cite{hughes2012quantitative}, ~\cite{huth2016natural}
\end{tabular}
}

&  

\multicolumn{1}{l}{}

\\ \hline

\textbf{Communication} 
&  
\begin{tabular}[c]{@{}l@{}}
~\cite{jenders2013analyzing}, ~\cite{arous2020opencrowd},
~\cite{sheshadri2019public}, ~\cite{green2020elusive}\\ ~\cite{brady2017emotion}

\end{tabular}
& 
\begin{tabular}[c]{@{}l@{}}
~\cite{brady2017emotion}, ~\cite{guille2012predictive}, ~\cite{pei2014searching}, ~\cite{quattrociocchi2014opinion}\\
~\cite{gao2014quantifying}, ~\cite{lima2015disease},
~\cite{gomez2016explosive}, ~\cite{luo2017inferring}\\
~\cite{shao2018spread}, ~\cite{bovet2019influence},
~\cite{zhang2019identifying}, ~\cite{gomez2019clustering}\\
~\cite{vosoughi2018spread}, ~\cite{Farrell2016Network}
\end{tabular}     

& 
\multicolumn{1}{l|}{
\begin{tabular}[c]{@{}l@{}}
~\cite{2011Topic}, ~\cite{Farrell2016Network}, ~\cite{sheshadri2019public}, ~\cite{DBLP:conf/acl/TsurCL15}\\

\end{tabular}
}      
& 
\multicolumn{1}{l}{
\begin{tabular}[c]{@{}l@{}}
~\cite{2011Topic}
\end{tabular}
}  

\\ \hline

\textbf{Geography}     
& 
\begin{tabular}[c]{@{}l@{}}
~\cite{ikawa2012location}, ~\cite{ryoo2014inferring},
~\cite{kim2014socroutes},
~\cite{wing2011simple}
\end{tabular}

&
\begin{tabular}[c]{@{}l@{}}
~\cite{halu2016data}, ~\cite{ganin2017resilience}, ~\cite{cho2011friendship}, ~\cite{li2012towards}\\
~\cite{yang2012socio}, ~\cite{bao2017planning}, ~\cite{expert2011uncovering}, ~\cite{sun2013understanding}\\
~\cite{deville2014dynamic}, ~\cite{santi2014quantifying},
~\cite{sekara2016fundamental},
~\cite{bonaccorsi2020economic}\\
~\cite{karduni2016protocol}, ~\cite{vazifeh2018addressing},
~\cite{kujala2018collection}, ~\cite{riascos2020networks}\\ ~\cite{wesolowski2012quantifying},
~\cite{barrington2020global}, ~\cite{wang2011human},
~\cite{liu2016rebalancing}\\ ~\cite{yadav2020resilience}, ~\cite{bettencourt2013origins}, ~\cite{barthelemy2013self}\\ ~\cite{daqing2014spatial}, ~\cite{su2014robustness}

\end{tabular}

& 
\multicolumn{1}{l|}{
\begin{tabular}[c]{@{}l@{}}
~\cite{ahmed2013hierarchical}, ~\cite{zhang2017regions},
~\cite{yuan2013and},
~\cite{wang2014travel}\\
~\cite{zhang2016gmove}, ~\cite{zhang2017triovecevent},
~\cite{miura2017unifying},
~\cite{hong2012discovering}

\end{tabular}
}  

& 
\multicolumn{1}{l}{
\begin{tabular}[c]{@{}l@{}}
~\cite{zhang2017regions}, ~\cite{yang2019revisiting},
~\cite{wang2020traffic},
~\cite{deng2016latent}\\
~\cite{li2018multi},
~\cite{sun2018exploring},
~\cite{wang2018you}, ~\cite{pan2019urban}\\
~\cite{miura2017unifying},
~\cite{yuan2012discovering},

\end{tabular}
}       


\\ \hline

\textbf{Environment}   
& 
\begin{tabular}[c]{@{}l@{}}
~\cite{kryvasheyeu2016rapid},
~\cite{ghosh2018class}
\end{tabular}

& 
\begin{tabular}[c]{@{}l@{}}
~\cite{reino2017networks},
~\cite{farrell2016corporate}, ~\cite{barnes2016social}\\
~\cite{camara2019indigenous},
~\cite{Farrell2016Network}, ~\cite{hsieh2015inferring}\\
~\cite{zheng2013u}, ~\cite{mcdonald2011urban}, ~\cite{dalin2012evolution}, 

\end{tabular} 


& 
\multicolumn{1}{l|}{
\begin{tabular}[c]{@{}l@{}}
~\cite{Farrell2016Network}, ~\cite{DBLP:conf/acl/JiangSWY19}

\end{tabular}
}    
& 
\multicolumn{1}{l}{
\begin{tabular}[c]{@{}l@{}}
~\cite{shang2014inferring}
\end{tabular}
}       

\\ \hline
\end{tabular}
}
\caption{Distribution of CSS applications}
\label{tab:domain-distribution-table}
\end{table*}

\mbox{}
\clearpage
\clearpage
\zihao{5}
\noindent
\noindent
\noindent
  \end{document}